\documentclass[11pt]{article}

\usepackage[top=2cm,bottom=2cm,outer=2.5cm,inner=2.5cm,marginparwidth=2.5cm]{geometry}
\usepackage{fvextra} 
\usepackage{authblk} 

\ifLuaTeX
\usepackage[unicode]{hyperref}
\usepackage{pdftexcmds} 
\makeatletter
\let\pdfstrcmp\pdf@strcmp
\makeatother
\else
\usepackage{hyperref}
\fi

\usepackage{amsmath}
\usepackage{amsfonts}
\usepackage{amssymb}
\usepackage{empheq} 
\usepackage{romanbar} 
\usepackage{dsfont} 
\usepackage{slashed}
\usepackage{cite}

\usepackage{lmodern} 
\usepackage[T1]{fontenc}
\usepackage{caption}
\usepackage{subcaption} 
\usepackage{titlesec} 
\usepackage{fancyhdr} 
\usepackage{abstract} 
\usepackage[super]{nth}
\usepackage{tocloft} 
\usepackage{bbm}

\usepackage[dvipsnames]{xcolor}

\hypersetup{
pdfstartview={FitH},
pdftitle={Defect RG flows},
pdfauthor={J. Belton, N. Drukker, B. Sahoo},
pdfsubject={},
pdfcreator={},
pdfkeywords={},
colorlinks=true,
citecolor=MidnightBlue,
linkcolor=MidnightBlue,
urlcolor=MidnightBlue
}

\numberwithin{equation}{section}

\usepackage{mathtools}

\DeclareMathOperator{\Tr}{Tr}
\DeclareMathOperator{\tr}{tr}
\DeclareMathOperator{\diag}{diag}

\DeclareMathOperator{\Gr}{Gr}

\def\bR {\mathbb{R}}
\def\bZ {\mathbb{Z}}

\def\cN{{\mathcal{N}}}
\def\cO{{\mathcal{O}}}

\newcommand{\sof}{\mathfrak{so}}

\newcommand{\bea}{\begin{eqnarray}}
\newcommand{\eea}{\end{eqnarray}}
\newcommand{\beq}{\begin{equation}}
\newcommand{\eeq}{\end{equation}}
\newcommand{\bal}{\begin{equation}\begin{aligned}{}}
\newcommand{\eal}{\end{aligned} \end{equation}}

\newcommand{\cC}{{\cal C}}
\newcommand{\cD}{{\cal D}}
\newcommand{\bD}{\mathbb D}
\newcommand{\bt}{{\mathbbm{t}}}

\title{
\vspace{1 cm}
Flowing with Displacements and Tilts:\\
Surface Operators in $O(N)$ Models
\vspace{5mm}}
\author{Jake~Belton%
\thanks{\href{mailto:jake.belton@kcl.ac.uk}{jake.belton@kcl.ac.uk}}}
\author{Nadav~Drukker%
\thanks{\href{mailto:nadav.drukker@gmail.com}{nadav.drukker@gmail.com}}}
\author{Biswajit~Sahoo%
\thanks{\href{mailto:biswajit.sahoo@kcl.ac.uk}{biswajit.sahoo@kcl.ac.uk}}}
\affil{\it Department of Mathematics, King's College London,\protect\\London, 
WC2R 2LS, United Kingdom\vspace{4pt}}
\date{}

\begin{document}
\maketitle
  \begin{abstract}
Defect conformal field theories have special operators of protected dimension 
known as displacements and tilts. They arise due to the breaking of global symmetries 
by the defect and the normalisations of their two-point functions are characteristics 
of the defect. In the case of surface defects, these normalisations are related to 
some of the anomaly coefficients in the surface effective action. To study these 
operators and their flows between different defect renormalization group fixed 
points we present an elegant approach using conformal perturbation theory that easily 
reproduces the known examples from the critical Wilson--Fisher $O(N)$ model in 
$4-\varepsilon$ dimensions and allows us to construct new ones in other multiscalar theories. In 
all the systems that we study the flows are short and under full control, as is the 
change of the displacement and tilt normalizations. We point out some novel 
features like the existence of vortices when the defect conformal manifold is 
not simply connected. 
In addition to regular human labour, this work relied heavily on generative AI; 
see full disclosure in methodology section.
\end{abstract}
\vfill
\eject
\tableofcontents

\section{Introduction}
\label{sec:intro}

Conserved stress tensors and symmetry currents are of paramount 
importance in local conformal field theories. When introducing defects 
into a theory, the defect dynamics are themselves often non-local, 
as they can be mediated via bulk interactions. Still, instead 
of the conserved currents, they have distinguished operators 
known as displacements $\bD$ and tilts $\bt$, for broken spacetime 
and internal symmetries, respectively. 

The displacements represent a small deformation of a defect away from 
a symmetric configuration (plane or sphere). The tilts arise from 
localised deviation from a uniform coupling of a defect to operators 
charged under a global symmetry. For the formal definitions, see 
the top of Section~\ref{sec:tD-in-ST}. 
For a $p$ dimensional defect, they have 
protected dimensions $\Delta_\bD=p+1$ and $\Delta_\bt=p$. 
The normalizations of these operators 
are physically meaningful quantities characteristic of the 
particular defects%
\footnote{Here $r,s$ are indices for the directions transverse to the 
defect and $\hat\imath, \check\jmath$ for the bifundamental arising 
in the breaking of a global symmetry $O(N)\to O(n)\times O(m)$.}
\begin{equation}
  \label{eq:CD-Ct}
  \langle\bD_r(\tau)\bD_s(0)\rangle
  =
  \frac{C_{\bD}\delta_{rs}}{|\tau|^{2(p+1)}},
\qquad
  \langle\bt_{\hat\imath\check\jmath}(\tau)\bt_{\hat k\check l}(0)\rangle
  =
  \frac{C_{\bt}\delta_{\hat\imath\hat k}\delta_{\check\jmath\check l}}{|\tau|^{2p}}.
\end{equation}

The fact that these operators arise from broken symmetries means 
that correlators with and without them are related by Ward identities. 
For a single extra insertion, these relations were first derived 
in~\cite{Billo:2016cpy, Padayasi:2021sik}. Nonlinear relations 
for correlators with multiple tilts were first derived 
in~\cite{Drukker:2022pxk} and related to 
the curvature of the defect conformal manifold---the global 
symmetry breaking coset. Those relations were widely expanded 
in \cite{Gabai:2025zcs, Belton:2025hbu, Gabai:2025hwf, 
Kong:2025sbk, Girault:2025kzt, Belton:2025ief, Drukker:2025dfm, 
whale}, 
to include displacements, correlators involving other operators 
and bulk--defect correlators.

For a surface operator, the constants $C_\bD$ and $C_\bt$ are also 
related to defect anomalies. In particular, $C_\bD$ is related 
to an anomaly 
term arising from nontrivial embeddings, so roughly the extrinsic 
curvature, see, e.g.~\cite{Bianchi:2015liz, 
Herzog:2017xha, Herzog:2021spv}. 
$C_\bt$ is related to anomalies from nonuniform 
global symmetry breaking, see e.g.~\cite{Drukker:2020atp, 
Jensen:2017eof, Choi:2025ebk, Wen:2025xka,
Copetti:2025sym,Drukker:2025dfm, whale}.

The purpose of this paper is to explore the behaviour of 
$\bD$, $\bt$, $C_\bD$ and $C_\bt$ under renormalization group 
(RG) flows.
Anomaly coefficients are sometimes monotonic under RG flows. 
In particular, surface defects and two-dimensional boundaries satisfy 
the $b$-theorem~\cite{Jensen:2015swa} (see also~\cite{Casini:2018nym,
Wang:2020xkc, Shachar:2024ubf}) for the anomaly coefficient 
associated to the Euler density. The anomaly coefficients 
related to $C_\bD$ and $C_\bt$, which are the focus of this 
paper are not monotonic, but since their values at the fixed 
points are of importance, it is natural to explore them 
also along the flows.

While defects with their protected operators exist in different 
dimensions, we focus on surface defects because there is 
a large family of such defects that are accessible 
perturbatively. 
Furthermore, it is known that the $O(N)$ model in three dimensions 
has rich universality classes of surface 
defects~\cite{Metlitski:2020cqy, Toldin:2021kun, Krishnan:2023cff}.
For the most part we employ conformal 
perturbation theory techniques valid for any theory in $d$ dimensions 
with $O(N)$ symmetry and operators of dimension close 
to two and three in different $O(N)$ representations. 
This includes the Wilson--Fisher (WF) critical $O(N)$ 
model in $d=4-\varepsilon$, for which surface operators were 
studied in~\cite{Trepanier:2023tvb, Raviv-Moshe:2023yvq, 
Giombi:2023dqs, Diatlyk:2024ngd}. Surface operators in 
closely related theories were also studied 
in~\cite{Anataichuk:2025zoq}.

We classify a broad family of perturbative surface defects and analyse 
their stability. We determine the displacement and tilt normalizations
at the fixed points and study their behaviour along RG flows. The
particular models presented in this work 
exhibit fixed-point collisions and defect conformal
manifolds with distinct types of tilt operators.

We first formulate the problem in terms of arbitrary $O(N)$ 
bulk theories with perturbative spectra.
The input is a small set of generic anomalous dimensions and three-point coefficients for operators near dimension two in specified $O(N)$ representations, together with the current and descendant data needed to track displacements.
This gives some more 
options for defects beyond those in the WF theory. It also allows 
for the flow analysis to be model independent. In 
Section~\ref{sec:SandT} we analyse the possible defect fixed 
points arising from deforming a bulk theory by one $O(N)$ singlet 
field and one traceless-symmetric tensor. 
In Section~\ref{sec:scalar-tensor-anti} we additionally turn on an 
antisymmetric field, or combining all fields together, a completely 
unconstrained real $N\times N$ deformation.

Another possible field of dimension close to two is 
in the fundamental of $O(N)$. This is natural around $d=6$ and 
was studied in \cite{Trepanier:2023tvb, Giombi:2023dqs}, 
so we do not repeat this 
example. We treat the case of defects not breaking $O(N)$ as 
a special class within the examples we study and do not focus 
on the large-$N$ limit, where they arise naturally as the 
Hubbard--Stratonovich fields. 

In Section~\ref{sec:protected-flow} we analyse $C_\bD$ and $C_\bt$ at the 
varied surface defects that we find. We also analyse their 
behaviour along the flows between the RG fixed points.

We then specialise to particular examples in Section~\ref{sec:examples}. 
We first redo the case of the WF $O(N)$ model reproducing many  
of the results of~\cite{Trepanier:2023tvb, Raviv-Moshe:2023yvq, 
Giombi:2023dqs, Diatlyk:2024ngd} using this unified framework. 
We then look at the long-range WF theory~\cite{Bianchi:2024eqm}, which is a 
nonlocal generalisation of the theory with one extra continuous parameter, 
parameterising the nonlocality. As its nonlocality parameter is varied, this 
model exhibits a family of fixed-point collisions. Our construction reproduces and 
generalises the results of~\cite{Bianchi:2024eqm}.

The next example is the chiral $O(N)\times O(2)$ WF 
theory~\cite{Osborn:2017ucf, Henriksson:2020fqi, Kawamura:1988zz}. 
This model has an antisymmetric field, allowing to realise the 
construction in Section~\ref{sec:scalar-tensor-anti}. This is a 
rather rich theory with a multitude of surface operators and we 
do not attempt to classify them all, just find relatively symmetric 
ones. Still, the resulting spaces of defects, or defect conformal 
manifolds, include $O(N)/(U(p)\times O(N-2p))$, which is homogeneous 
but not symmetric. This is manifested by this model having 
two different types of 
tilt operators with different values of $C_\bt$. 
Lastly we use the tricritical $O(N)$ model in $d=3-\varepsilon$ as an example of how to extend our framework beyond singlet, rank two and antisymmetric perturbations. Consistency requires including a rank four symmetric traceless perturbation which qualitatively changes the stability of the defects.

We conclude with a brief discussion.

\subsection{Methodology}
\label{sec:method}
This project started using traditional computational methods 
with the initial focus on 
the flow of $C_\bD$ and $C_\bt$ in conformal perturbation theory 
between surface operators in the WF theory. For simplicity 
we focused on the $O(2)$ model. We then decided to see 
whether generative AI (ChatGPT 5.3--5.6, Claude Opus 4.7--4.8 and Gemini 3.1) could speed up the 
research and allow us to generalise to $N>2$ and other models. 
Many of the calculations in Sections~\ref{sec:SandT}, 
\ref{sec:scalar-tensor-anti}, \ref{sec:examples} were first 
done using the AI tools.

All AI output was extensively 
tested by the authors on paper and in Mathematica, and manu\-script 
writing was performed by the authors. All the authors stand 
behind the results reported here and take responsibility for 
any possible errors.

\section{Scalar--tensor surface operators}
\label{sec:SandT}

\subsection{Bulk theories and notation}
\label{sec:bulk}

We start from an $O(N)$-symmetric bulk CFT in $d$-dimensions. 
We assume there is an $O(N)$ singlet operator $S$ and a symmetric 
traceless operator $T_{ij}$ both of dimension close to two and 
that at least one of them is weakly relevant. 
This is valid for many variants of the Wilson--Fisher 
theory in $d=4-\varepsilon$, as presented in Section~\ref{sec:examples}. 
This also holds for the tricritical WF theory in $d\lesssim3$, 
presented in Section~\ref{sec:3-ep}, though that theory 
also has a rank four tensor of a similar dimension so requires 
the separate analysis there. 

To track the displacement operator, we also look at some operators 
of dimension close to three, specifically those that are vectors of 
the transverse $O(d-2)$ symmetry. In the bulk these may be 
the conserved current $j_r$ (for $d\simeq4$) and the transverse 
descendants of $S$ and $T_{ij}$, which we denote as $V_r$ and
$U_{ij,r}$ respectively.

The bulk operator content that we study is summarised 
in Table~\ref{tab:uv-field-content}.
We denote their dimension as
\begin{equation}
  \Delta_S=2+\gamma_S,
  \qquad
  \Delta_T=2+\gamma_T,
  \qquad
  \Delta_V=3+\gamma_S,
  \qquad
  \Delta_U=3+\gamma_T,
  \qquad
  \Delta_j=d-1 .
\label{eq:uv-dimensions}
\end{equation}

\begin{table}[t]
\centering
\renewcommand{\arraystretch}{1.25}
\begin{tabular}{llll}
\hline
operator & $O(N)$ reps & dimension & WF dimension \\
\hline
$S$ & $\mathbf1$ & $2+\gamma_S$
& $2-\frac{6\varepsilon}{N+8}+O(\varepsilon^2)$ \\
$T_{ij}$ & $\mathbf T_N$ & $2+\gamma_T$
& $2-\frac{N+6}{N+8}\varepsilon+O(\varepsilon^2)$ \\
$V_r$ & $\mathbf1$ & $3+\gamma_S$
& $3-\frac{6\varepsilon}{N+8}+O(\varepsilon^2)$ \\
$U_{ij,r}$ & $\mathbf T_N$ & $3+\gamma_T$
& $3-\frac{N+6}{N+8}\varepsilon+O(\varepsilon^2)$ \\
$j_{ij,r}$ & $\mathbf A_N$ & $d-1$
& $3-\varepsilon$ \\
\hline
\end{tabular}
\caption{Bulk operators used in the surface operator 
construction. $\mathbf1$ is the singlet, 
$\mathbf T_N$ is the traceless symmetric 
and $\mathbf A_N$ is the adjoint antisymmetric
of $O(N)$. The final column gives the dimensions 
of the operators 
in the Wilson--Fisher $O(N)$ model, discussed in 
Section~\ref{sec:WFON}.
\label{tab:uv-field-content}}
\end{table}

All operators, other than the current, are taken 
to be unit normalised. In particular
\begin{equation}
  \langle S(x_1)S(x_2)\rangle
  =\frac{1}{x_{12}^{2\Delta_S}},
  \qquad
  \langle T_{ij}(x_1)T_{kl}(x_2)\rangle
  =\frac{\Pi_{ij,kl}}{x_{12}^{2\Delta_T}},
  \qquad
  \langle j_{ij,r}(x_1)j_{kl,s}(x_2)\rangle
  =C_j\frac{P_{ij,kl}\delta_{rs}}{x_{12}^{2d-2}}.
  \label{eq:uv-two-point-normalization}
\end{equation}
For the symmetric traceless and adjoint representations 
one has the projectors
\begin{equation}
  \Pi_{ij,kl}
  =
  \frac12\left(\delta_{ik}\delta_{jl}+\delta_{il}\delta_{jk}\right)
  -\frac1N\delta_{ij}\delta_{kl},
  \qquad
  P_{ij,kl}
  =
  \frac12\left(\delta_{ik}\delta_{jl}-\delta_{il}\delta_{jk}\right).
\label{eq:st-and-adj-projectors}
\end{equation}
They satisfy
\begin{equation}
\begin{aligned}
  \Pi_{ij,kl}&=\Pi_{kl,ij},
  &\qquad
  \Pi_{ij,kl}\Pi_{kl,mn}&=\Pi_{ij,mn},
  &\qquad
  \delta_{ij}\Pi_{ij,kl}&=0, \\
  P_{ij,kl}&=P_{kl,ij},
  &
  P_{ij,kl}P_{kl,mn}&=P_{ij,mn},
  &
  P_{ji,kl}&=-P_{ij,kl}.
  \label{eq:projector-identities}
\end{aligned}
\end{equation}
The cubic symmetric traceless tensor is
\begin{equation}
  {\cal I}_{ij,kl,mn}
  =
  \Pi_{ij,ab}\Pi_{kl,bc}\Pi_{mn,ca},
  \qquad
  M_{ij}N_{kl}K_{mn}{\cal I}_{ij,kl,mn}
  =
  \Tr(MNK),
\label{eq:st-cubic-invariant}
\end{equation}
for symmetric traceless matrices $M,N,K$.

The descendants are also unit normalised (with an extra $\delta_{rs}$), 
so
\begin{equation}
  V_r=\frac{1}{\sqrt{2\Delta_S}}\partial_rS
  =
  \frac12\partial_rS+O(\varepsilon),
  \qquad
  U_{ij,r}=\frac{1}{\sqrt{2\Delta_T}}\partial_rT_{ij}
  =
  \frac12\partial_rT_{ij}+O(\varepsilon).
\label{eq:uv-descendant-normalization}
\end{equation}

We take arbitrary positive structure constants
\begin{equation}
  C_{SSS},
  \quad
  C_{STT},
  \quad
  C_{Sjj},
  \quad
  C_{TTT},
\label{eq:structure-constants}
\end{equation}
and $C_{SST}=C_{SSj}=C_{STj}=0$ by group theory. 
The normalisations are such that the tensor three-point functions are
\begin{equation}
\begin{aligned}
  \langle S(x_1)T_{ij}(x_2)T_{kl}(x_3)\rangle
  =
  \frac{C_{STT}\Pi_{ij,kl}}
  {x_{12}^{\Delta_S}x_{13}^{\Delta_S}x_{23}^{2\Delta_T-\Delta_S}},
\qquad
  \langle T_{ij}(x_1)T_{kl}(x_2)T_{mn}(x_3)\rangle
  =
  \frac{C_{TTT}{\cal I}_{ij,kl,mn}}
  {x_{12}^{\Delta_T}x_{13}^{\Delta_T}x_{23}^{\Delta_T}}.
  \label{eq:scalar-three-point-definitions}
\end{aligned}
\end{equation}
Then to leading order 
in small $\gamma$, for the descendants
\begin{equation}
  C_{VVS}=\frac12 C_{SSS},
  \qquad
  C_{UUS}=C_{VUT}=\frac12 C_{STT},
  \qquad
  C_{UUT}=\frac12 C_{TTT}.
\label{eq:descendant-structure-constants}
\end{equation}
Their three-point functions are normalised as in 
\eqref{eq:scalar-three-point-definitions} with an extra 
$\delta_{rs}$ on the transverse-vector indices.

The structure constants involving the current are defined via
\begin{equation}
\begin{aligned}
  \langle S(x_1)j_{ij,r}(x_2)j_{kl,s}(x_3)\rangle
  &=
  \frac{C_{Sjj}P_{ij,kl}\delta_{rs}}
  {x_{12}^{\Delta_S}x_{13}^{\Delta_S}x_{23}^{2d-2-\Delta_S}}, \\
  \langle T_{ij}(x_1)j_{kl,r}(x_2)U_{mn,s}(x_3)\rangle
  &=
  \frac{C_{TjU}\Pi_{ij,ab}P_{kl,ac}\Pi_{mn,cb}\delta_{rs}}
  {x_{12}^{d-2}x_{13}^{2\Delta_T+2-d}x_{23}^{d}},
\\
  \langle T_{ij}(x_1)j_{kl,r}(x_2)j_{mn,s}(x_3)\rangle
  &=
  \frac{C_{Tjj}\Pi_{ij,ab}P_{kl,ac}P_{mn,cb}\delta_{rs}}
  {x_{12}^{\Delta_T}x_{13}^{\Delta_T}x_{23}^{2d-2-\Delta_T}}.
  \label{eq:current-three-point-definitions}
\end{aligned}
\end{equation}
The remaining three point functions of a single operator of 
dimension near two and two transverse vectors vanish
\begin{equation}
  C_{SVU}=
  C_{SVj}=
  C_{SUj}=
  C_{TVV}=
  C_{TVj}=0.
\label{eq:vector-structure-vanishings}
\end{equation}

$C_{Sjj}$ and $C_{Tjj}$ are further data, which we postpone to 
Section~\ref{sec:3-ep}, 
but $C_{TjU}$ is fixed by a Ward identity and a choice of 
Lie algebra generator convention, which we take as
\begin{equation}
  (E_{ij})_k{}^l
  =\delta_{ik}\delta_j{}^l-\delta_{jk}\delta_i{}^l.
\label{eq:adjoint-generator-convention}
\end{equation}
this fixes the current Ward identity
\begin{equation}
\begin{aligned}
\partial_\mu
\left\langle j_{kl}^{\mu}(x)T_{ij}(x_1)T_{mn}(x_2)\right\rangle
&=
-\delta^{(d)}(x-x_1)
\big((E_{kl})_i{}^a\delta_j^b+(E_{kl})_j{}^a\delta_i^b\big)
\left\langle T_{ab}(x_1)T_{mn}(x_2)\right\rangle
\\
&\quad
-\delta^{(d)}(x-x_2)
\big((E_{kl})_m{}^a\delta_n^b+(E_{kl})_n{}^a\delta_m^b\big)
\left\langle T_{ij}(x_1)T_{ab}(x_2)\right\rangle .
\end{aligned}
\end{equation}
Together with the normalizations in 
\eqref{eq:uv-two-point-normalization} 
and \eqref{eq:uv-descendant-normalization}, this gives
\begin{equation}
  \frac{C_{TjU}^2}{2C_j}=\frac{d-2}{\Delta_T}=1+O(\varepsilon).
\label{eq:ward-CjU-ratio-abstract}
\end{equation}
Equivalently, we can choose the sign
\begin{equation}
  C_{TjU}=\sqrt{(d-2)C_j}+O(\varepsilon).
\label{eq:ward-CjU-sqrt}
\end{equation}

\subsection{Surface perturbation and fixed points}
\label{sec:fixed-points}

Our main objects of study are flat surface operators achieved 
by perturbing the bulk theory by the scalar and symmetric 
traceless operators
\begin{equation}
  S_{\rm def}
  =S_{\rm UV}
  +\mu^{-\gamma_S}h^S 
  \int_{\bR^2} d^2\tau\,S
  +\mu^{-\gamma_T}h^T_{ij}
  \int_{\bR^2} d^2\tau\,T_{ij}.
\label{eq:surface-action}
\end{equation}
Here $h^S$ and $h^T_{ij}$ are dimensionless renormalized couplings. 
This closely mirrors the analysis in~\cite{Trepanier:2023tvb, 
Raviv-Moshe:2023yvq, Giombi:2023dqs, Diatlyk:2024ngd}, where 
the focus was solely on the WF model. Some of these papers 
have slightly different notations and some also use 
$h_{ij}=h^S\delta_{ij}+h^T_{ij}$.

The beta functions take their usual form in conformal perturbation 
theory in two dimensions
\begin{align}
  \beta^S
  &=\mu\frac{d h^S}{d\mu}
  =\gamma_S h^S
  +\pi C_{SSS}(h^S)^2
  +\pi C_{STT}h^T_{ij}h^T_{ij}
  +O(h^3,\varepsilon h^2),
\label{eq:singlet-h-beta-functions}
\\
  \beta^T_{ij}
  &=\mu\frac{d h^T_{ij}}{d\mu}
  =\gamma_T h^T_{ij}
  +2\pi C_{STT}h^S h^T_{ij}
  +\pi C_{TTT}\left(h^T_{ik}h^T_{kj}
  -\frac{\delta_{ij}}{N}h^T_{kl}h^T_{kl}\right)
  +O(h^3,\varepsilon h^2).
\label{eq:tensor-h-beta-functions}
\end{align}
The contractions of the symmetric traceless tensors above follow by using
the projection tensors in \eqref{eq:st-and-adj-projectors}, \eqref{eq:st-cubic-invariant}
\begin{equation}
  h^T_{ij}h^T_{kl}\Pi_{ij,kl}
  =
  h^T_{ij}h^T_{ij},
  \qquad
  h^T_{kl}h^T_{mn}{\cal I}_{ij,kl,mn}
  =
  h^T_{ik}h^T_{kj}
  -
  \frac{\delta_{ij}}{N}h^T_{kl}h^T_{kl}.
  \label{eq:h-tensor-contractions}
\end{equation}

The fixed points are the zeros of \eqref{eq:singlet-h-beta-functions} and \eqref{eq:tensor-h-beta-functions}. 
Given that 
there are multiple fixed points, we label them as $\cD_n$
rather than the traditional asterisk.

The trivial fixed point is
\begin{equation}
  \cD_0:\qquad
  h^{S,\cD_0}=0,
  \qquad
  h^{T,\cD_0}_{ij}=0,
  \label{eq:D0-h-couplings}
\end{equation}
and the nontrivial $O(N)$ symmetric fixed point exists as long as $C_{SSS}\neq0$ and $\gamma_S<0$ and is
\begin{equation}
  \cD_N:\qquad
  h^{T,\cD_N}_{ij}=0,
  \qquad
  h^{S,\cD_N}=-\frac{\gamma_S}{\pi C_{SSS}}.
  \label{eq:DN-h-couplings}
\end{equation}

To analyse fixed points with nonzero $h^T_{ij}$, 
let us first assume $C_{TTT}\neq0$. Then diagonalizing 
$h^T_{ij}$, we find from \eqref{eq:tensor-h-beta-functions} 
the equation for the eigenvalues (with no sum over repeated indices)
\begin{equation}
\frac{\gamma_T+2\pi C_{STT}h^S}{\pi C_{TTT}}h^T_{ii}
+
(h^T_{ii})^2-\frac{1}{N}\sum_{j=1}^N (h^T_{jj})^2
=0.
\label{eq:tensor-h-CTTT}
\end{equation}
Treating $\sum (h^T_{jj})^2$ as a constant, this is a quadratic equation 
for $h_{ii}$, so it has two possible roots with degeneracies 
$n$ and $m=N-n$. So we conclude that all the fixed points must have 
residual symmetry $O(n)\times O(m)$.

Let us denote the two eigenvalues by $h_n^{\cD_n}$ and $h_m^{\cD_n}$,
where the subscript denotes the degeneracy. As $n$ runs over
$0\leq n\leq N$, it enumerates the fixed points, with
$\cD_0$ the trivial point and $\cD_N$
the $O(N)$-symmetric point. The distinction between 
$\cD_n$ and $\cD_{N-n}$ is fixed in
\eqref{eq:hS-roots-sigma-gen-generic} below.
A diagonal realisation is
\begin{equation}
  h^{T,\cD_n}_{ij}
  =\diag\big(
  \underbrace{h_n^{\cD_n},\cdots,h_n^{\cD_n}}_{n},
  \underbrace{h_m^{\cD_n},\cdots,h_m^{\cD_n}}_{m}\big),
  \qquad
  n h_n^{\cD_n}+m h_m^{\cD_n}=0.
\label{eq:hn-hm-ansatz}
\end{equation}

Let us define
\begin{equation}
  \nu=2n-N=n-m.
  \label{eq:nu-definition}
\end{equation}
Assuming $n\neq m$, the tensor equation $\beta^T_{ij}=0$
\eqref{eq:tensor-h-CTTT} gives
\begin{equation}
  h_n^{\cD_n}
  =
  \frac{m}{\nu}
  \frac{\gamma_T+2\pi C_{STT}h^{S,\cD_n}}{\pi C_{TTT}},
  \qquad
  h_m^{\cD_n}
  =
  -\frac{n}{\nu}
  \frac{\gamma_T+2\pi C_{STT}h^{S,\cD_n}}{\pi C_{TTT}}.
  \label{eq:hn-hm-generic}
\end{equation}
Then
\begin{equation}
\big|h^{T,\cD_n}\big|^2
\equiv
  h^{T,\cD_n}_{ij}h^{T,\cD_n}_{ij}
  =\frac{nmN}{\nu^2}
  \left[\frac{\gamma_T+2\pi C_{STT}h^{S,\cD_n}}{\pi C_{TTT}}\right]^2.
\label{eq:hTnorm-generic}
\end{equation}
Substituting this into $\beta^S=0$ \eqref{eq:singlet-h-beta-functions}
gives a quadratic equation. Defining the rescaled discriminant
\begin{equation}
\sigma_n^2
  =\frac{9\nu^2}{N^2}
  +\frac{36nm}{N}
  \frac{C_{STT}}{C_{TTT}^2}\,
  \frac{\gamma_T}{\gamma_S^2}
  \left(2\gamma_SC_{STT}-\gamma_TC_{SSS}\right)
  \label{eq:discriminant}
\end{equation}
where $\sigma_n$ is chosen nonnegative when the roots are real. The
quadratic equation for $h^{S,\cD_n}$ has the roots
\begin{equation}
  h^{S,\cD_n}
  =-\frac{1}{2\pi}\frac{\nu^2\gamma_SC_{TTT}^2
    +4nmN\gamma_TC_{STT}^2
    +\frac{N\nu}{3}\gamma_SC_{TTT}^2\sigma_n}
    {\nu^2C_{TTT}^2C_{SSS}
      +4nmNC_{STT}^3}.
  \label{eq:hS-roots-sigma-gen-generic}
\end{equation}
By choosing the positive root of $\sigma_n$, this expression 
uniquely distinguishes $\cD_n$ from $\cD_m$ (when $\sigma_n^2>0$), 
the two different solutions with $O(n)\times O(m)$ symmetry. 
To see that, note that $\nu^2$, $nm$ and $\sigma_n$ are 
invariant under $n\leftrightarrow m$, but $\nu$ changes sign. 
So the only term in \eqref{eq:hS-roots-sigma-gen-generic} 
that changes sign is the last one in the numerator, proportional 
to $\sigma_n$. So for an $(n,m)$ pair we keep $\sigma_n$ fixed 
but allow the two values of $h^{S,\cD_n}$ to depend on the sign 
of $\nu$, thus fixing $\cD_n$, except when 
$\sigma_n=0$, where the two roots and
hence $\cD_n$ and $\cD_{N-n}$ coincide. The case $n=N/2$ 
($\nu=0$, $N$ even) is the separate point treated in 
\eqref{eq:hT-N/2},
\eqref{eq:hT-half}.

To have a real fixed point, $\sigma_n^2$
\eqref{eq:discriminant} must of course be nonnegative. 
The fixed point reality and stability conditions are discussed 
in detail in Section~\ref{sec:real-and-stable} below. 
See also~\cite{Pannell:2024hbu}.

Cases when $\sigma_n=0$ require special care, as the next order 
contributions to $\sigma_n^2$ are generically of order $\gamma^3$. 
This introduces half-integer powers of the couplings into the 
expressions for the dimensions. In the case of the WF defect, 
this was noted already 
in \cite{deSabbata:2024xwn} and mentioned in 
Section~\ref{sec:WF-dim2} below. 
Further examples arise in the long-range 
WF theory, see Section~\ref{sec:near2-LR}.

When $N$ is even and $n=m=N/2$, the symmetry is really 
$\bigl(O(n)\times O(n)\bigr)\rtimes\bZ_2$,
where the $\bZ_2$ exchanges the two equal blocks.
In this case the part in \eqref{eq:tensor-h-beta-functions} 
quadratic in $h^T$ vanishes, so a nonzero tensor coupling requires
\begin{equation}
  h^{S,\cD_{N/2}}=-\frac{\gamma_T}{2\pi C_{STT}},
  \label{eq:n-half-hS}
\end{equation}
assuming $C_{STT}\neq0$.
The singlet beta function equation \eqref{eq:singlet-h-beta-functions} 
now constrains $h_{N/2}^{\cD_{N/2}}\equiv h_n^{\cD_{N/2}} = - h_m^{\cD_{N/2}}$
\begin{equation}
NC_{STT}(h^{\cD_{N/2}}_{N/2})^2
  =
  -\frac{\gamma_S}{\pi} h^{S,\cD_{N/2}}
  -C_{SSS}(h^{S,\cD_{N/2}})^2
  =
  \frac{\gamma_T}{4\pi^2 C_{STT}^2}
  (2\gamma_SC_{STT}-\gamma_TC_{SSS})
\label{eq:hT-N/2}
\end{equation}
or
\begin{equation}
h^{\cD_{N/2}}_{N/2}
  =
  \frac{\sqrt{\gamma_T(2\gamma_SC_{STT}-\gamma_TC_{SSS})}}
  {2\pi\sqrt{NC_{STT}^3}},
\label{eq:hT-half}
\end{equation}
where we have chosen the positive root.
The other sign is related to it by the $\mathbb{Z}_2$ exchange of the two blocks.
Under this symmetry, $h^T\rightarrow-h^T$ while $h^S$ is unchanged, so the two roots describe equivalent defects.
Of course, this fixed point is real only for 
$\gamma_T(2\gamma_SC_{STT}-\gamma_TC_{SSS})\geq0$.

Very similar equations arise when $C_{TTT}=0$. The tensor 
equation \eqref{eq:tensor-h-beta-functions} factorises
\begin{equation}
(\gamma_T+2\pi C_{STT}h^S)h^T_{ij}=0\,,
\end{equation}
so as in \eqref{eq:n-half-hS}, 
the non $O(N)$-invariant fixed points must have 
\begin{equation}
  h^{S,\cD_n}=-\frac{\gamma_T}{2\pi C_{STT}}.
\label{eq:hS-O2}
\end{equation}
Up to $O(N)$ rotations, one can always diagonalise $h^T_{ij}$, so 
\eqref{eq:singlet-h-beta-functions} enforces, in analogy to 
\eqref{eq:hT-N/2} (with no sum over repeated 
indices)
\begin{equation}
\sum_{i=1}^N(h^T_{ii})^2
  =
  \frac{\gamma_T}{4\pi^2 C_{STT}^3}
  (2\gamma_SC_{STT}-\gamma_TC_{SSS})
\label{eq:hT-O2}
\end{equation}
This requirement and the tracelessness of $h^T_{ij}$ 
are two conditions on the $N$ eigenvalues of 
$h^T$, so for $N>2$, there are continuous families of defects not 
related by symmetry. Those are presumably lifted at higher order 
in conformal perturbation theory, but we have not verified that.

One case when $C_{TTT}$ is necessarily zero is the $O(2)$ model, 
because there are no possible singlet contractions. 
More specifically, the tensor ${\cal I}_{ij,kl,mn}$ of 
\eqref{eq:st-cubic-invariant} vanishes. In that 
case, though, tracelessness and \eqref{eq:hT-O2} fully fix $h^T$, 
up to global $O(2)$ rotations, so it really matches the 
usual $n=N/2$ case.

It is worth mentioning here that these expressions are significantly 
more involved than in the WF $O(N)$ theory, where 
$C_{SSS}=C_{STT}$. We keep the general form and specialise 
to WF only in Section~\ref{sec:WFON}.

\subsection{Defect conformal manifold}
\label{sec:grassmannian}

In the case of the symmetry breaking defects, the choice 
of a diagonal $h^T_{ij}$ in \eqref{eq:hn-hm-ansatz} is one 
of many possible choices. The space of allowed solutions 
(a.k.a defect conformal manifold) is the 
quotient of the original symmetry group by 
the symmetry preserved by the defect, which in this case is  the
real Grassmannian
\begin{equation}
\frac{O(N)}{O(n)\times O(m)}=\Gr(n,\bR^N).
\label{eq:grassmannian}
\end{equation}
This is an $nm$ dimensional manifold and deformations of the defect in those directions are realised by the protected 
tilt operators $\bt$ in the bifundamental representation.

In addition to defects with uniform profiles (a point on the 
Grassmannian), one can have the defect with different points 
along the surface mapping to different points on the coset. 
Generally those are not conformal, though kinks 
may be~\cite{Drukker:2022beq,Trepanier:2023tvb}. 
Another conformal possibility is when the coset is not 
simply connected, as is the case for the Grassmannian 
with $N\geq3$ since $\pi_1(\Gr(n,\bR^N))=\bZ_2$. 
Then there are configurations on the two dimensional 
defect where encircling the origin, the choice of 
coset representative transverses the noncotractible cycle 
on the defect conformal manifold. The IR limit of such 
a configuration is a 
$\bZ_2$ charged local vortex operators.

For $N=2$ and $n=1$, the defect conformal manifold is simply 
$\Gr(1,\bR^2)=S^1$ with fundamental group $\bZ$, so in this case 
the vortices are integer valued.

It would be interesting to study these vortices, which can be thought 
of as defects within defects~\cite{Antunes:2021qpy, Drukker:2022beq, Sun:2024qhv, 
Shimamori:2024yms}.

\subsection{Stability matrix and near dimension two operators}
\label{sec:stability-matrix}

The mixing matrix of nearly degenerate operators 
is determined by linearizing the beta functions
\eqref{eq:singlet-h-beta-functions}, \eqref{eq:tensor-h-beta-functions} 
at a fixed point. For operators of dimension close to two 
this is the usual stability matrix at $\cD$
\begin{equation}
  \Gamma^{\cD}
  =
  \left.
  \frac{\partial(\beta^S,\beta^T_{ij})}{\partial(h^S,h^T_{kl})}
  \right|_{\cD}.
  \label{eq:Gamma-definition}
\end{equation}
If $\gamma^{\cD}$ is an eigenvalue of $\Gamma^{\cD}$, the corresponding
near-two dimension is
\begin{equation}
  2+\gamma^{\cD}.
  \label{eq:operator-dimension-from-Gamma}
\end{equation}

At the UV fixed point $\cD_0$, \eqref{eq:Gamma-definition} gives back the UV
dimensions in \eqref{eq:uv-dimensions}.
\begin{equation}
  \gamma_S^{\cD_0}=\gamma_S,
  \qquad
  \gamma_T^{\cD_0}=\gamma_T.
  \label{eq:D0-stability-dimensions}
\end{equation}
At the fully symmetric fixed point \eqref{eq:DN-h-couplings}, the singlet
and symmetric traceless operators have anomalous 
dimensions
\begin{equation}
  \gamma_S^{\cD_N}
  =
  -\gamma_S,
  \qquad
  \gamma_T^{\cD_N}
  =
  \gamma_T-\frac{2C_{STT}}{C_{SSS}}\gamma_S.
  \label{eq:DN-stability-dimensions}
\end{equation}

For the generic $\cD_n$ fixed point with $C_{TTT}\neq0$, 
$n\neq m$, the residual symmetry is $O(n)\times O(m)$. The 
traceless symmetric $T_{ij}$ decomposes into 
two traceless symmetrics $T_n$ and $T_m$, a bifundamental tilt $\bt$ and a 
singlet. The operators or dimensions close to two, their representations 
and dimensions are in Table~\ref{tab:Dn-near-two-dimensions}.

\begin{table}[t]
\centering
\renewcommand{\arraystretch}{1.25}
\begin{tabular}{llll}
\hline
operator & $O(n)\times O(m)$ reps & generic dimension & WF dimension \\
\hline
$T_n$ & $(\mathbf T_n,\mathbf1)$
& see~\eqref{eq:tensor-gammas-generic}
& $2+\frac{2\sigma_n-\nu}{N+8}\varepsilon$ \\
$T_m$ & $(\mathbf1,\mathbf T_m)$
& see~\eqref{eq:tensor-gammas-generic}
& $2-\frac{2\sigma_n-\nu}{N+8}\varepsilon$ \\
$\bt$ & $(\mathbf n,\mathbf m)$
& $2$
& $2$ \\
$\cO_\pm$ & $(\mathbf1,\mathbf1)$
& see~\eqref{eq:mixed-singlets-gamma-generic}
& $2+\frac{N\pm\sqrt{N^2+16\sigma_n^2-8\nu\sigma_n}}{2(N+8)}\varepsilon$\\
\hline
\end{tabular}
\caption{Near-two operators at a generic $O(n)\times O(m)$ fixed 
point with $C_{TTT}\neq0$ and $n\neq m$. Here $m=N-n$, 
and the last column gives the dimensions in the WF $O(N)$ 
model, where $\nu$, $\sigma_n$ are defined in 
\eqref{eq:WF-fixed-point-shorthands}. Also,
$\mathbf1$ is the singlet, $\mathbf T_n$ and $\mathbf T_m$
are traceless symmetric and $\mathbf n$ and $\mathbf m$ are
fundamental representations. For $d\simeq3$, the current is also 
of dimension close to two, but is treated only in Section~\ref{sec:3-ep}.}
\label{tab:Dn-near-two-dimensions}
\end{table}

$T_n$, $T_m$ and $\bt$ do not mix with other operators, but their 
dimensions are shifted and can be read from diagonal elements in 
the stability matrix \eqref{eq:Gamma-definition}.
Their anomalous dimensions (away from 2) are
\begin{equation}
\begin{aligned}
  \gamma_{T_n}^{\cD_n}
  &=\frac{N}{\nu}
  \big(\gamma_T+2\pi C_{STT}h^{S,\cD_n}\big), 
\\
  \gamma_{T_m}^{\cD_n}
  &=-\frac{N}{\nu}
  \big(\gamma_T+2\pi C_{STT}h^{S,\cD_n}\big), 
  \\
  \gamma_\bt^{\cD_n}
  &=0.
\label{eq:tensor-gammas-generic}
\end{aligned}
\end{equation}

The singlet component of $T_{ij}$, which is parallel to the 
diagonal coupling $h^{T,\cD_n}_{ij}$ \eqref{eq:hn-hm-ansatz} 
mixes with the $O(N)$ singlet $S$. The resulting two
operators are denoted $\cO_\pm$. 
Their anomalous-dimension mixing matrix, in the original basis, 
is
\begin{equation}
  \Gamma_{\rm singlet}^{\cD_n}
  =
  \begin{pmatrix}
    \gamma_S+2\pi C_{SSS}h^{S,\cD_n}
    &2\pi C_{STT}\big|h^{T,\cD_n}\big|
    \\[.5em]
    2\pi C_{STT}\big|h^{T,\cD_n}\big|
    &-\gamma_T-2\pi C_{STT}h^{S,\cD_n}
  \end{pmatrix},
\label{eq:mixed-singlet-Gamma-generic}
\end{equation}
with $h^{S,\cD_n}$ in \eqref{eq:hS-roots-sigma-gen-generic} and 
$\big|h^{T,\cD_n}\big|$ in \eqref{eq:hTnorm-generic}. 
Explicitly, the eigenvalues are
\begin{equation}
\begin{aligned}
  \gamma_\pm^{\cD_n}
  &=
  \frac12\bigg[\gamma_S-\gamma_T
    +2\pi (C_{SSS}-C_{STT})h^{S,\cD_n} \\
  &\qquad
    \pm\sqrt{\left(\gamma_S+\gamma_T
        +2\pi (C_{SSS}+C_{STT})h^{S,\cD_n}\right)^2
      +\frac{16nmN}{\nu^2}\frac{C_{STT}^2}{C_{TTT}^2}
      \left(\gamma_T+2\pi C_{STT}h^{S,\cD_n}\right)^2}
  \bigg].
  \label{eq:mixed-singlets-gamma-generic}
\end{aligned}
\end{equation}

When $N$ is even and $n=m=N/2$, the fixed point couplings are instead
\eqref{eq:n-half-hS} and \eqref{eq:hT-half}. 
The tensor anomalous dimensions are then
\begin{equation}
  \gamma_{T_n}^{\cD_{N/2}}
  =
  2\pi C_{TTT}h_{N/2}^{\cD_{N/2}},
  \qquad
  \gamma_{T_m}^{\cD_{N/2}}
  =
  -2\pi C_{TTT}h_{N/2}^{\cD_{N/2}},
  \qquad
  \gamma_\bt^{\cD_{N/2}}
  =
  0.
\label{eq:n-half-tensor-gammas-generic}
\end{equation}
The two singlets are again $S$ and the component of $T_{ij}$ parallel 
to the diagonal tensor in \eqref{eq:hn-hm-ansatz}. Their mixing matrix 
is again \eqref{eq:mixed-singlet-Gamma-generic}, plugging in 
\eqref{eq:n-half-hS} it simplifies such that the bottom right entry 
vanishes. Likewise, the eigenvalues 
\eqref{eq:mixed-singlets-gamma-generic} simplify to
\begin{equation}
  \gamma_\pm^{\cD_{N/2}}
  =\frac12\left[\gamma_S-\frac{C_{SSS}}{C_{STT}}\gamma_T
  \pm\sqrt{\left(\gamma_S-\frac{C_{SSS}}{C_{STT}}\gamma_T\right)^2
  +8\gamma_S\gamma_T
  -\frac{4C_{SSS}}{C_{STT}}\gamma_T^2}
  \right].
  \label{eq:n-half-mixed-singlets-gamma-generic}
\end{equation}

\subsubsection{Reality and stability analysis}
\label{sec:real-and-stable}
The stability analysis of such defects was performed in~\cite{Pannell:2024hbu}, 
generalising Michel's theorem~\cite{Michel:1983in}.
At this order the beta functions \eqref{eq:singlet-h-beta-functions}, 
\eqref{eq:tensor-h-beta-functions} are gradient, 
$\beta^I=\partial A/\partial h^I$, with
\begin{equation}
  A=\frac{1}{2}\gamma_I (h^I)^2
  +\frac{\pi}{3}C_{IJK}\,h^I h^J h^K,
  \qquad I\in\{S,T_{ij}\}.
  \label{eq:A-function}
\end{equation}
The function $A$ decreases monotonically towards the IR and stable fixed points 
minimise $A$. Furthermore, the theorem states that if the quadratic part of 
$A$ is negative definite, meaning  $\gamma_S, \gamma_T<0$, 
there is at most one totally stable defect and, if it exists, it preserves
$O(N)$ symmetry, so is $\cD_N$. 
Since the potential is cubic, it certainly is not a global minimum.

Looking at \eqref{eq:DN-stability-dimensions}, we see 
that the symmetric fixed point $\cD_N$ is stable 
iff $\gamma_TC_{SSS}>2\gamma_SC_{STT}$. This is clearly 
correct if $\gamma_T>0$ (assuming $\gamma_S<0$, or else there 
would be no flow).

The situation is more interesting when both $\gamma_T$ and $\gamma_S$ 
are negative. In that case, if $\gamma_TC_{SSS}>2\gamma_SC_{STT}$, 
then it is easy to see that all of $\cD_n$ with $n\neq m$ 
are real saddle points. We note that the discriminant in 
\eqref{eq:discriminant} can be rewritten in terms of $\gamma_T^{\cD_N}$ 
\eqref{eq:DN-stability-dimensions} as
\begin{equation}
  \sigma_n^2 = \frac{9\nu^2}{N^2} - \frac{36 nm}{N} 
  \frac{C_{STT}C_{SSS}}{C_{TTT}^2} \frac{\gamma_T}{\gamma_S^2} 
  \gamma_T^{\cD_N}.
  \label{eq:discriminant-rewritten}
\end{equation}
Given that $\gamma_T < 0$, the sign of the second term depends entirely on 
$\gamma_T^{\cD_N}$. If $\cD_N$ is stable, then the second term in 
\eqref{eq:discriminant-rewritten} is non-negative and 
$\sigma_n^2 \geq 9\nu^2/N^2$ for all $n \neq m$. 

Also for $n=m$ in the even $N$ case, the reality condition of this 
fixed point 
\eqref{eq:hT-half} is exactly the same as the $\cD_N$ stability condition.
This is because the $n=m=N/2$ branch meets the symmetric fixed 
point $\cD_N$ exactly when 
$h_{N/2}$ coupling in \eqref{eq:hT-N/2} vanishes, which is 
the same as $\gamma_T^{\cD_N}=0$.
For the ordinary WF theory this
happens at $N=6$: $\cD_6^{\rm WF}$ meets
$\cD_3^{\rm WF}$. Resolving this degeneracy requires going to 
higher orders in perturbation theory, as done 
in~\cite{deSabbata:2024xwn}. We note further such examples in 
the long-range WF theory in Section~\ref{sec:near2-LR}.

For $\gamma_S<0<\gamma_T$, the symmetric point $\cD_N$ is stable 
but a possible $n=m$ saddle is complex. There may be real saddle 
points for other values of $n$. 
For $\gamma_T<0<\gamma_S$, the fixed point $\cD_N$ exists only as 
a formal solution, as $S$ is irrelevant. 
The $n=m$ case is complex, but there could be real fixed points 
for small or large $n$, but again, they are unstable.

\subsection{Near dimension-three operators}
\label{sec:dimension-three-operators}

We now consider the operators arising from the UV descendant fields 
$V_r$ and $U_{ij,r}$ in
\eqref{eq:uv-descendant-normalization}, together for $d\simeq4$, 
with the currents $j_{ij}$.

Under $O(N)\to O(n)\times O(m)$ the transverse 
vector operators decompose as
\begin{equation}
\begin{aligned}
  &V_r:&
  \mathbf1
  &\to
  (\mathbf1,\mathbf1),
  \\
  &U_{ij,r}:&\quad
  \mathbf T_N
  &\to
  (\mathbf T_n,\mathbf1)
  \oplus
  (\mathbf1,\mathbf T_m)
  \oplus
  (\mathbf n,\mathbf m)
  \oplus
  (\mathbf1,\mathbf1),
  \\
  &j_{ij,r}:&
  \mathbf A_N
  &\to
  (\mathbf A_n,\mathbf1)
  \oplus
  (\mathbf1,\mathbf A_m)
  \oplus
  (\mathbf n,\mathbf m).
  \label{eq:near-three-rep-decomposition}
\end{aligned}
\end{equation}
Here $\mathbf A_n$ is the antisymmetric representation of $O(n)$.
The operators of dimension near three at the $\cD_n$ fixed point are shown in
Table~\ref{tab:Dn-near-three-dimensions}. 
For the broken symmetry group, we use the indices 
$\hat\imath,\hat\jmath=1,\cdots,n$ to label the $O(n)$ 
factor, and $\check\imath,\check\jmath=n+1,\cdots,N$ to 
label the $O(m)$ factor. The transverse indices remain $r,s$. 

\begin{table}[t]
\centering
\renewcommand{\arraystretch}{1.25}
\begin{tabular}{llll}
\hline
operator & $O(n)\times O(m)$ reps & generic dimension & WF dimension \\
\hline
$\bD_r$ & $(\mathbf1,\mathbf1)$
& $3$
& $3$ \\
$V'_r$ & $(\mathbf1,\mathbf1)$
& see~\eqref{eq:Vprime-gamma-generic}
& $3-\frac{6\varepsilon}{N+8}$ \\
$U_{n,\hat\imath\hat\jmath,r}$ & $(\mathbf T_n,\mathbf1)$
& see~\eqref{eq:Un-Um-dimensions-generic}
& $3+\frac{\sigma_n-n-3}{N+8}\varepsilon$ \\
$U_{m,\check\imath\check\jmath,r}$ & $(\mathbf1,\mathbf T_m)$
& see~\eqref{eq:Un-Um-dimensions-generic}
& $3-\frac{m+3+\sigma_n}{N+8}\varepsilon$ \\
$j_{n,\hat\imath\hat\jmath,r}$ & $(\mathbf A_n,\mathbf1)$
& $3-\varepsilon$
& $3-\varepsilon$ \\
$j_{m,\check\imath\check\jmath,r}$ & $(\mathbf1,\mathbf A_m)$
& $3-\varepsilon$
& $3-\varepsilon$ \\
$u_{\pm,\hat\imath\check\jmath,r}$ & $(\mathbf n,\mathbf m)$
& see~\eqref{eq:Wpm-gamma-generic}
& $3-\frac{3N+22\pm\sqrt{(N+10)^2+4\left(2\sigma_n-\nu\right)^2}}{4(N+8)}\varepsilon$ \\
\hline
\end{tabular}
\caption{Near-three operators at a generic $O(n)\times O(m)$ fixed point with
$C_{TTT}\neq0$ and $n\neq m$. Here $m=N-n$, and the WF quantities
$\nu,\sigma_n$ are defined in \eqref{eq:WF-fixed-point-shorthands}. 
For $d\not\simeq4$ the current dimension is far from three and it does 
not mix in this sector.}
\label{tab:Dn-near-three-dimensions}
\end{table}

We do not treat the symmetric fixed point $\cD_N$ separately, it 
can be inferred by setting $n\to N$ below.

For a fixed point $\cD_n$ with couplings 
\eqref{eq:hn-hm-ansatz}, the operators whose UV limits are 
the descendant of $S$ and the descendant of the
component of $T_{ij}$ parallel to $h^{T,\cD_n}_{ij}$ mix. 
In the basis
\begin{equation}
  \left(
    V_r^{\cD_n},
    \frac{h^{T,\cD_n}_{ij}U_{ij,r}^{\cD_n}}
    {\big|h^{T,\cD_n}\big|}
  \right),
  \label{eq:vector-singlet-basis}
\end{equation}
their anomalous dimensions are the eigenvalues of
\begin{equation}
  \Gamma_{\rm vector}^{\cD_n}
  =
  \begin{pmatrix}
    \displaystyle
    \gamma_S+\pi C_{SSS}h^{S,\cD_n}
    &
    \displaystyle
    \pi C_{STT}\big|h^{T,\cD_n}\big|
    \\[.5em]
    \displaystyle
    \pi C_{STT}\big|h^{T,\cD_n}\big|
    &
    \displaystyle
    -\pi C_{STT}h^{S,\cD_n}
  \end{pmatrix}.
  \label{eq:Gamma-vector-singlet-generic}
\end{equation}
The determinant of \eqref{eq:Gamma-vector-singlet-generic} vanishes by the
$\beta^S=0$ equation \eqref{eq:singlet-h-beta-functions}. The protected
operator is the displacement,
\begin{equation}
  \Delta_{\bD}^{\cD_n}=3.
  \label{eq:displacement-dimension-generic}
\end{equation}
The other singlet vector operator is denoted $V'_r{}^{\cD_n}$ and has
\begin{equation}
  \gamma_{V'}^{\cD_n}
  =
  \gamma_S+\pi(C_{SSS}-C_{STT})h^{S,\cD_n}.
  \label{eq:Vprime-gamma-generic}
\end{equation}

The components of $U_{ij,r}^{\cD_n}$ not 
in \eqref{eq:vector-singlet-basis}
decompose into an $O(n)$ traceless symmetric tensor, an $O(m)$ traceless
symmetric tensor and bifundamentals. The traceless symmetrics do 
not mix and their anomalous dimensions are simply
\begin{equation}
\begin{aligned}
  \gamma_{U_n}^{\cD_n}
  &=
  \gamma_T+\pi C_{STT}h^{S,\cD_n}+\pi C_{TTT}h_n^{\cD_n},
\\
  \gamma_{U_{m}}^{\cD_n}
  &=
  \gamma_T+\pi C_{STT}h^{S,\cD_n}+\pi C_{TTT}h_m^{\cD_n}.
  \label{eq:Un-Um-dimensions-generic}
\end{aligned}
\end{equation}
The bifundamental part $U_{\hat\imath\check\jmath,r}^{\cD_n}$ 
mixes with the broken components of the current. Recall that the 
current two-point has the central charge $C_j$ in its normalisation 
\eqref{eq:uv-two-point-normalization}. For the mixing, we use the 
unit normalised operator, so in the basis
\begin{equation}
  \left(
    U_{\hat\imath\check\jmath,r}^{\cD_n},
    \frac{1}{\sqrt{C_j}}j_{\hat\imath\check\jmath,r}^{\cD_n}
  \right),
  \label{eq:bifundamental-vector-basis}
\end{equation}
where $C_j$ is defined in \eqref{eq:uv-two-point-normalization}, 
the anomalous-dimension matrix is
\begin{equation}
  \Gamma_{\rm bifund}^{\cD_n}
  =
  \begin{pmatrix}
    \displaystyle
    \gamma_T+\pi C_{STT}h^{S,\cD_n}+\frac{\pi C_{TTT}}{2}
    \big(h_n^{\cD_n}+h_m^{\cD_n}\big)
    &\displaystyle
    \pi \frac{C_{TjU}}{\sqrt{C_j}}
    \big(h_n^{\cD_n}-h_m^{\cD_n}\big)
    \\[.8em]\displaystyle
    \pi \frac{C_{TjU}}{\sqrt{C_j}}
    \big(h_n^{\cD_n}-h_m^{\cD_n}\big)
    &
    -\varepsilon
  \end{pmatrix}.
  \label{eq:bifundamental-vector-Gamma-generic}
\end{equation}
Recall that $C_{TjU}=\sqrt{(d-2)C_j}$ \eqref{eq:ward-CjU-sqrt}, 
so we find the eigenvectors
$u_{\pm,\hat\imath\check\jmath,r}^{\cD_n}$, with dimensions
\begin{equation}
\begin{aligned}
  \gamma_{u_\pm}^{\cD_n}
  &=\frac12\Bigg[
    \gamma_T-\varepsilon
    +\pi C_{STT}h^{S,\cD_n}
    +\frac{\pi C_{TTT}}{2}
    \big(h_n^{\cD_n}+h_m^{\cD_n}\big)
  \\
  &\quad
    \pm\sqrt{\left(\gamma_T+\varepsilon
    +\pi C_{STT}h^{S,\cD_n}
    +\frac{\pi C_{TTT}}{2}\big(h_n^{\cD_n}+h_m^{\cD_n}\big)\right)^2
    +4\pi^2(d-2)\big(h_n^{\cD_n}-h_m^{\cD_n}\big)^2}
  \Bigg].
\end{aligned}
  \label{eq:Wpm-gamma-generic}
\end{equation}

The components of $j_r$ that are adjoints of $O(n)$ and $O(m)$ 
remain conserved currents of the residual symmetry with
\begin{equation}
  \Delta_{j_{n,r}}^{\cD_n}
  =3-\varepsilon,
  \qquad
  \Delta_{j_{m,r}}^{\cD_n}
  =3-\varepsilon.
  \label{eq:unbroken-current-dimensions-generic}
\end{equation}

When $N$ is even and $n=m=N/2$, use instead 
\eqref{eq:n-half-hS} and \eqref{eq:hT-half}. 
The matrix \eqref{eq:Gamma-vector-singlet-generic} is 
as before, just with the couplings in \eqref{eq:n-half-hS} 
and \eqref{eq:hT-half}. The dimension of $U_n$ and $U_m$ 
also remain as in \eqref{eq:Un-Um-dimensions-generic}.
The bifundamental matrix in \eqref{eq:bifundamental-vector-Gamma-generic}
reduces to
\begin{equation}
  \Gamma_{\rm bifund}^{\cD_n}
  =
  \begin{pmatrix}
    \displaystyle
    \gamma_T/2
    &
    2\sqrt{2}\,\pi h_n^{\cD_n}
    \\[.5em]
    2\sqrt{2}\,\pi h_n^{\cD_n}
    &
    -\varepsilon
  \end{pmatrix},
  \label{eq:n-half-bifundamental-vector-Gamma-generic}
\end{equation}
and the corresponding dimensions are
\begin{equation}
\begin{aligned}
  \gamma_{u_\pm}^{\cD_n}
  &=
  \frac14\left[\gamma_T-2\varepsilon
    \pm\sqrt{\left(\gamma_T+2\varepsilon\right)^2
    +64\pi^2(d-2)(h_n^{\cD_n})^2}\right].
\end{aligned}
  \label{eq:n-half-Wpm-gamma-generic}
\end{equation}

\subsection{Deformations of a defect}
\label{sec:deformations-of-the-defect}

The analysis above treats flows from a bulk theory (or 
trivial defect) to nontrivial defects. We now apply the same 
machinery for flows starting at a nontrivial defect. If the 
defect is exactly one of the fixed points that arises by 
deformations of the bulk, the beta function equations are 
just the expansion of the equations in 
\eqref{eq:singlet-h-beta-functions}, 
\eqref{eq:tensor-h-beta-functions} around nonzero $h^S$ and 
$h^T$.

Instead of doing this trivial exercise, let us consider a 
general defect whose low-lying spectrum is reminiscent of 
that of $\cD_n$, but possibly not identical. 
In particular, it still has internal symmetry $O(n)\times O(m)$ 
and the same set of near dimension two operators in 
Table~\ref{tab:Dn-near-two-dimensions}
\begin{itemize}
\item 
A pair of singlets $\cO_\pm$ with anomalous dimensions $\gamma_\pm$.
\item
$T_n$ in the $(\mathbf{T}_n,\mathbf{1})$ representation 
with anomalous dimension $\gamma_n$.
\item
$T_m$ transforming in $(\mathbf{1},\mathbf{T}_m)$ with 
anomalous dimension $\gamma_m$.
\item
Finally an operator $t$ in the bifundamental $(\mathbf{n},\mathbf{m})$ 
with anomalous dimension $\gamma_t$. As we do not assume that 
it arises from symmetry breaking, it may not be a tilt and
$\gamma_t$ may be different from zero.
\end{itemize}

We assume that the anomalous dimensions are all small and of the same scale (as in the $\varepsilon$ expansion).
Deforming the defect by these operators, we use conformal 
perturbation theory in the anomalous dimensions to find further defects.
The most general deformed action is 
\begin{equation}
\begin{aligned}
  S_g^{\cD_n}
  &=
  S^{\cD_n}
  +
  \mu^{-\gamma_+^{\cD_n}}g_+
  \int_{\cD_n}d^2\tau\,\cO_+
  +
  \mu^{-\gamma_-^{\cD_n}}g_-
  \int_{\cD_n}d^2\tau\,\cO_-
  \\ &\quad
  +
  \mu^{-\gamma_n^{\cD_n}}g_{n;\hat\imath\hat\jmath} 
  \int_{\cD_n}d^2\tau\,T_{n;\hat\imath\hat\jmath}
  +
  \mu^{-\gamma_m^{\cD_n}}g_{m;\check\imath\check\jmath} 
  \int_{\cD_n}d^2\tau\,T_{m;\check\imath\check\jmath}
  +
  \mu^{-\gamma_t^{\cD_n}}g_{t;\hat\imath\check\jmath} \int_{\cD_n}d^2\tau\, t_{\hat\imath\check\jmath}.
  \label{eq:dn-two-singlet-deformation}
\end{aligned}
\end{equation}
Deforming by any $T_n$ or $T_m$ breaks the $O(n)\times O(m)$ symmetry, 
as do deformations by $t$, if it is not a protected tilt. One 
can look at the resulting defects and in particular try to find 
ones where the spectrum reorganises into $O(n')\times O(m')$ 
representations (as would happen by flowing between $\cD_n$ and 
$\cD_{n'}$). But for simplicity, we focus on deformations 
which do not break the symmetry and set $g_{n;\hat\imath\hat\jmath}=g_{m;\check\imath\check\jmath}=g_{t;\hat\imath\check\jmath}=0$. 
In order for the remaining perturbation to trigger an RG flow, we make the assumption that at least one of the singlet operators $\cO_\pm$ is relevant.

The beta functions for $g_\pm$, 
to quadratic order in conformal perturbation theory, is
\begin{equation}
\begin{aligned}
  \beta_+^{\cD_n}
  &=
  \gamma_+^{\cD_n}g_+
  +\pi\left(
    C_{+++}^{\cD_n}g_+^2
    +2C_{++-}^{\cD_n}g_+g_-
    +C_{+--}^{\cD_n}g_-^2
  \right), \\
  \beta_-^{\cD_n}
  &=
  \gamma_-^{\cD_n}g_-
  +\pi\left(
    C_{++-}^{\cD_n}g_+^2
    +2C_{+--}^{\cD_n}g_+g_-
    +C_{---}^{\cD_n}g_-^2
  \right).
  \label{eq:dn-two-singlet-beta-functions}
\end{aligned}
\end{equation}
Note that even if only one of the singlet operators is relevant, say $\cO_-$, we still need to perturb by the irrelevant operator $\cO_+$. 
If we didn't, we would be setting $g_+ = 0$ and the beta function for $g_+$ then becomes $\beta_+^{\cD_n} = \pi C_{+--}^{\cD_n}g_-^2$. 
If $C_{+--}^{\cD_n} \neq 0$, then the deformation by $\cO_-$ sources $\cO_+$ at this order.

To look for a non-trivial fixed point $g_\pm^*$, we take $g_+^*\neq 0$ 
which allows to rewrite $\beta_+^{\cD_n}=0$ as
\begin{equation}
  g_+^*
  =
  -\frac{\gamma_+^{\cD_n}}{\pi (C_{+++}^{\cD_n}
  +2C_{++-}^{\cD_n}\rho
  +C_{+--}^{\cD_n}\rho^2)},
  \label{eq:dn-generic-endpoint-couplings}
\end{equation}
with
\begin{equation}
  \rho=\frac{g_-^*}{g_+^*}.
  \label{eq:dn-rho-definition}
\end{equation}
Substituting this into the equation $\beta_-^{\cD_n}=0$, we find
\begin{equation}
\begin{aligned}
  0={}&
  -C_{++-}^{\cD_n}\gamma_+^{\cD_n}
  +(C_{+++}^{\cD_n}\gamma_-^{\cD_n}
  -2C_{+--}^{\cD_n}\gamma_+^{\cD_n})\rho
  +(2C_{++-}^{\cD_n}\gamma_-^{\cD_n}
  -C_{---}^{\cD_n}\gamma_+^{\cD_n})\rho^2
  +C_{+--}^{\cD_n}\gamma_-^{\cD_n}\rho^3 .
  \label{eq:dn-rho-fixed-point-cubic}
\end{aligned}
\end{equation}
Solving this equation for $\rho$ fully determines the couplings 
at any other fixed point. In general, there are three distinct 
roots which would correspond to distinct fixed points with at least $O(n)\times O(m)$ symmetry.

We may ask under what conditions is the internal symmetry of an IR fixed point enhanced to full $O(N)$ symmetry. This imposes constraints on the operator spectrum.
One condition is that the near dimension two spectrum must reorganise into an $O(N)$ singlet and a traceless symmetric multiplet.
Since we have the decomposition
\begin{equation}
    \mathbf{T}_N \rightarrow 
    (\mathbf{1},\mathbf{1})
    \oplus(\mathbf{T}_n,\mathbf{1})
    \oplus(\mathbf{1},\mathbf{T}_m)
    \oplus(\mathbf{n},\mathbf{m})\,,
\end{equation}
we see that the operators $T_n$, $T_m$, $t$ and one linear 
combination of $\cO_\pm$, which we denote $T_\parallel$ 
must become part of the $O(N)$ traceless symmetric multiplet. 
We parametrise $T_\parallel$ and the orthogonal singlet $S$ 
using a parameter $q$ as
\begin{equation}
  T_\parallel
  =
  \frac{\cO_+ + q\cO_-}{\sqrt{1+q^2}},
  \qquad
  S
  =
  \frac{-q\cO_+ + \cO_-}{\sqrt{1+q^2}}.
  \label{eq:dn-endpoint-S-T-basis}
\end{equation}
$T_\parallel$ and $S$ are eigenvectors of the stability matrix
\begin{equation}
  M(\rho)
  =
  \begin{pmatrix}
    M_{++} & M_{+-} \\
    M_{+-} & M_{--}
  \end{pmatrix},
  \label{eq:dn-endpoint-singlet-stability-matrix}
\end{equation}
which in the $\cO_\pm$ basis at $g_\pm=g_\pm^*$ has the components
\begin{equation}
\begin{aligned}
  M_{++}
  &=
  \gamma_+^{\cD_n}
  +2\pi\left(
    C_{+++}^{\cD_n}g_+^*
    +C_{++-}^{\cD_n}g_-^*
  \right), \\
  M_{+-}
  &=
  2\pi\left(
    C_{++-}^{\cD_n}g_+^*
    +C_{+--}^{\cD_n}g_-^*
  \right), \\
  M_{--}
  &=
  \gamma_-^{\cD_n}
  +2\pi\left(
    C_{+--}^{\cD_n}g_+^*
    +C_{---}^{\cD_n}g_-^*
  \right).
  \label{eq:dn-endpoint-singlet-stability-entries}
\end{aligned}
\end{equation}

By treating the non-singlet operators as probes, their linearized beta functions gives the eigenvalues at the fixed point
\begin{equation}
\begin{aligned}
  \gamma_n^{\rm IR}
  &=
  \gamma_{n}^{\cD_n}
  +2\pi\left(
    C_{+nn}^{\cD_n}g_+^*
    +C_{-nn}^{\cD_n}g_-^*
  \right), 
\\
  \gamma_m^{\rm IR}
  &=
  \gamma_{m}^{\cD_n}
  +2\pi\left(
    C_{+mm}^{\cD_n}g_+^*
    +C_{-mm}^{\cD_n}g_-^*
  \right), 
\\
  \gamma_\bt^{\rm IR}
  &=
  \gamma_\bt^{\cD_n}
  +2\pi\left(
    C_{+tt}^{\cD_n}g_+^*
    +C_{-tt}^{\cD_n}g_-^*
  \right).
  \label{eq:dn-unturned-probe-eigenvalues}
\end{aligned}
\end{equation}
Imposing $O(N)$ spectrum degeneracy sets
\begin{equation}
\begin{gathered}
\gamma_n^{\rm IR}
=\gamma_m^{\rm IR}=\gamma_\bt^{\rm IR}
=M_{++}+qM_{+-}
=M_{--}+\frac{M_{+-}}{q}.
\label{eq:dn-dimension-recombination-conditions}
\end{gathered}
\end{equation}
After substituting the expressions for the stability matrix components 
\eqref{eq:dn-endpoint-singlet-stability-entries} and the 
anomalous dimensions \eqref{eq:dn-unturned-probe-eigenvalues}, 
these equations turn into polynomial constraints on the 
original defect data.

Having $O(N)$ invariant spectrum is not sufficient for $O(N)$ 
symmetry. 
It remains to impose $O(N)$ symmetry conditions on the structure constants at the IR fixed point. In particular, for the mixed structure constants involving both singlet and traceless symmetric operators, we require
\begin{equation}
\begin{gathered}
  C_{SST_\parallel}(q)=0, \\
  C_{ST_\parallel T_\parallel}(q)
  =
  C_{Snn}(q)
  =
  C_{Smm}(q)
  =
  C_{Stt}(q).
  \label{eq:dn-endpoint-S-cubic-selection}
\end{gathered}
\end{equation}
For the three tensor structure constants, enhanced $O(N)$ symmetry 
requires all components to arise as projections of the unique $O(N)$ invariant traceless symmetric cubic coefficient $C_{TTT}$.
This gives%
\footnote{If any of the denominators vanish, one should impose 
that the numerator vanishes as well.}
\begin{equation}
\begin{gathered}
  C_{nnn}^{\cD_n}
  =C_{mmm}^{\cD_n}
  =2C_{ntt}^{\cD_n}
  =2C_{mtt}^{\cD_n}
  =\frac{C_{T_\parallel nn}(q)}{\alpha}
  =\frac{C_{T_\parallel mm}(q)}{\beta}
  =\frac{2C_{T_\parallel tt}(q)}{\alpha+\beta}
  =\frac{C_{T_\parallel T_\parallel T_\parallel}(q)}{n \alpha^3 +m \beta^3}.
\label{eq:dn-endpoint-tensor-cubic-ratios}
\end{gathered}
\end{equation}
Here, $\alpha$ and $\beta$ are the diagonal entries of the component of the $O(N)$ traceless symmetric tensor which transforms as a singlet under $O(n)\times O(m)$: $\alpha$ is the entry of the $O(n)$ block and $\beta$ is the entry of the $O(m)$ block. 
Their values are fixed by tracelessness, $n\alpha +m\beta=0$, and unit normalisation, $n\alpha^2+m\beta^2=1$, and are given by
\begin{equation}
  \alpha=\sqrt{\frac{m}{nN}},
  \qquad
  \beta=-\sqrt{\frac{n}{mN}}.
  \label{eq:dn-endpoint-tensor-ratios}
\end{equation}
The relative factors appearing in \eqref{eq:dn-endpoint-tensor-cubic-ratios} come from the projections of $C_{TTT}$ onto the differing $O(n)\times O(m)$ invariant subspaces.

Equations \eqref{eq:dn-rho-fixed-point-cubic}, \eqref{eq:dn-dimension-recombination-conditions}, \eqref{eq:dn-endpoint-S-cubic-selection} and \eqref{eq:dn-endpoint-tensor-cubic-ratios} serve as 16 constraints on 19 UV CFT data and two auxiliary parameters, $\rho$ and $q$. 
This leaves five degrees of freedom, which is the expected number of IR degrees of freedom for a defect with full $O(N)$ symmetry: 
$\gamma_S^{\rm IR}$, $\gamma_T^{\rm IR}$, $C_{SSS}^{\rm IR}$, 
$C_{STT}^{\rm IR}$ and $C_{TTT}^{\rm IR}$.

Let us now examine the three roots of 
\eqref{eq:dn-rho-fixed-point-cubic} under the assumptions that 
all the $O(N)$ conditions above are satisfied. Obviously, as 
engineered, the IR fixed point has $O(N)$ symmetry. There is another 
root with $O(N)$ symmetry. In the language of 
Section~\ref{sec:fixed-points}, those are $\cD_N$ and $\cD_0$ 
respectively. The last fixed point is a mirror of the starting 
point $\cD_n$, the other point on the conformal manifold intersecting 
the $g_\pm$ plane.

A simple way to see this is to change variables from $g_\pm$ to 
$h^S$ and $h_n-h_m$, which describe the flow in the same 
two-dimensional plane. Specifically, after the $O(N)$ 
constraints above are imposed, the two-coupling
system is just the restriction of \eqref{eq:singlet-h-beta-functions}
and \eqref{eq:tensor-h-beta-functions} to
\begin{equation}
  h^T_{ij}=y\,\diag(\alpha\,\mathbf1_n,\beta\,\mathbf1_m).
\end{equation}
The projection of \eqref{eq:tensor-h-beta-functions} onto the $y$ 
coordinate is
\begin{equation}
  0
  =
  y\left[
  \gamma_T+2\pi C_{STT}h^S
  +\pi C_{TTT}(n\alpha^3+m\beta^3)y
  \right].
\end{equation}
$y=0$ gives the two $O(N)$ fixed points $\cD_0$ and
$\cD_N$ (depending on the value of $h^S$). The $y\neq0$ solution fixes $h^S$ linearly in
$y$, and substituting into $\beta^S=0$ gives a quadratic equation for
$y$. Since one root is the original $\cD_n$ point, the other root is
fixed by Vieta's formula and is the mirror point. Thus,
after translating the starting point to $g_\pm=0$, the three nonzero
roots of \eqref{eq:dn-rho-fixed-point-cubic} are $\cD_0$, $\cD_N$ and
the mirror $O(n)\times O(m)$ point.

\section{Scalar--tensor--antisymmetric surface operators}
\label{sec:scalar-tensor-anti}

Section~\ref{sec:SandT} presents the most general 
surface defects reachable via short RG flows triggered by scalar 
and symmetric tensor operators of dimension close to two. Here 
we generalise this to systems with an additional operator in the 
antisymmetric representation of $O(N)$. In this case we do 
not try to find the most general solutions, but we still 
find it instructive to identify lower symmetry defects with 
richer defect conformal manifolds.

The prime example we have in mind is the $O(N)\times O(2)$ model 
in Section~\ref{sec:N2}, but as before, we start with 
a general setting so that this analysis can be applied easily 
to other similar theories.

We extend the defect action \eqref{eq:surface-action} to include a unit normalized antisymmetric operator $Z$ with dimension $\Delta_Z = 2+\gamma_Z$,
\begin{equation}
  S_{\rm def}
  =S_{\rm UV}
  +\int_{\bR^2} d^2\tau\, \left(
  \mu^{-\gamma_S}h^S S
  +\mu^{-\gamma_T}h^T_{ij}T_{ij}
  +\mu^{-\gamma_Z}h^Z_{ij}Z_{ij}
  \right).
\label{eq:surface-action-current}
\end{equation}
One may also combine $h_{ij}=h^S\mathbbm{1}_{ij}+h^{T}_{ij}+h^Z_{ij}$, as a full unconstrained 
$N\times N$ matrix, but we do not use that notation below.

For the structure constants involving $Z$, they are taken to use the same tensor structures as in \eqref{eq:current-three-point-definitions} which involve the conserved current.
Concretely, we have
\begin{equation}
\begin{aligned}
  \langle S(x_1)Z_{ij}(x_2)Z_{kl}(x_3)\rangle
  &=
  \frac{C_{SZZ}P_{ij,kl}}
  {x_{12}^{\Delta_S}x_{13}^{\Delta_S}x_{23}^{2\Delta_Z-\Delta_S}}, 
\\
  \langle T_{ij}(x_1)Z_{kl}(x_2)Z_{mn}(x_3)\rangle
  &=
  \frac{C_{TZZ}\Pi_{ij,ab}P_{kl,ac}P_{mn,cb}}
  {x_{12}^{\Delta_T}x_{13}^{\Delta_T}x_{23}^{2\Delta_Z-\Delta_T}}.
  \label{eq:Z-three-point-definitions}
\end{aligned}
\end{equation}
All other structure constants among the deforming primaries which involve at least one $Z$ insertion vanish.

Using these normalizations, we find the beta functions
\begin{align}
  \beta^S
  &=
  \gamma_S h^S
  +\pi C_{SSS}(h^S)^2
  +\pi C_{STT}|h^T|^2
  +\pi C_{SZZ}|h^Z|^2,
\label{eq:STZ-betaS-functions}
\\
  \beta^T_{ij}
  &=
  \gamma_T h^T_{ij}
  +2\pi C_{STT}h^S h^T_{ij}
  +\pi C_{TTT}\left((h^T)^2_{ij}-\frac{\delta_{ij}}{N}|h^T|^2\right)
  +\pi C_{TZZ}\left((h^Z)^2_{ij}+\frac{\delta_{ij}}{N}|h^Z|^2\right),
\label{eq:STZ-betaT-functions}
\\
  \beta^Z_{ij}
  &=
  \gamma_Z h^Z_{ij}
  +2\pi C_{SZZ}h^S h^Z_{ij}
  -\pi C_{TZZ} (h^T h^Z+h^Z h^T)_{ij},
\label{eq:STZ-betaZ-functions}
\end{align}
where we used matrix multiplication between $h^T$ and $h^Z$ and 
$|h^Z|^2 = h^Z_{ij}h^Z_{ij}$. 

\begin{table}[t]
\centering
\renewcommand{\arraystretch}{1.25}
\begin{tabular}{lll}
\hline
operator & $O(N)$ reps & dimension \\
\hline
$S$ & $\mathbf1$ & $2+\gamma_S$ \\
$T_{ij}$ & $\mathbf T_N$ & $2+\gamma_T$ \\
$Z_{ij}$ & $\mathbf A_N$ & $2+\gamma_Z$ \\
$V_r$ & $\mathbf1$ & $3+\gamma_S$ \\
$U_{ij,r}$ & $\mathbf T_N$ & $3+\gamma_T$ \\
$W_{ij,r}$ & $\mathbf A_N$ & $3+\gamma_Z$ \\
$j_{ij,r}$ & $\mathbf A_N$ & $3-\varepsilon$ \\
\hline
\end{tabular}
\caption{Bulk operators used in the surface operator construction. 
In addition to those in Table~\ref{tab:uv-field-content}, 
we include here the antisymmetric field $Z_{ij}$ and 
its descendant $W_{ij,r}$.
\label{tab:STZ-field-content}}
\end{table}

\subsection{Fixed points}
\label{sec:STZ-fixed-points}

The complete system 
\eqref{eq:STZ-betaS-functions}--\eqref{eq:STZ-betaZ-functions} is complex, 
with many nontrivial saddles. Instead of analysing it in full generality,
we focus on several physically interesting fixed points with residual symmetries.

Obviously, setting $h^Z=0$ is itself a
consistent truncation, reducing to the system analysed in 
Section~\ref{sec:SandT}. We do not repeat those fixed points 
here, but they are presented in the case of the 
$O(N)\times O(2)$ model in Section~\ref{sec:N2-Dn-fixed-points}.
It should be noted that the fixed points with 
$h^Z\neq0$ cannot be stable, as shown below
\eqref{eq:STZ-singlet-Gamma}.

To find new fixed points, let us first look for those with 
$h^{T}=0$. We see that the tensor beta function 
\eqref{eq:STZ-betaT-functions} requires
$(h^Z)^2_{ij}+N^{-1}\delta_{ij}h^Z_{kl}h^Z_{kl}=0$. This is possible
only for even $N$, and up to $O(N)$ rotations it has 
$N/2$ off-diagonal $2\times2$ blocks, all proportional to one 
constant $\eta$
\begin{equation}
  h^{Z,\cC_{N/2}}
  =\eta^{\cC_{N/2}} \bigoplus_{k=1}^{N/2}
  \begin{pmatrix}0&1\\ -1&0\end{pmatrix},
  \qquad
  h^{S,\cC_{N/2}}
  =-\frac{\gamma_Z}{2\pi C_{SZZ}},
\label{eq:STZ-pure-U-h}
\end{equation}
with
\begin{equation}
  h^{Z,\cC_{N/2}}_{ij}h^{Z,\cC_{N/2}}_{ij}
  =\frac{\gamma_Z(2\gamma_SC_{SZZ}-\gamma_ZC_{SSS})}
  {4\pi^2C_{SZZ}^3}.
\label{eq:STZ-pure-U-norm}
\end{equation}
This fixed point breaks $O(N)\to U(N/2)$ and we denote it 
as $\cC_{N/2}$.

Generalising this, it is easy to engineer 
fixed points preserving $U(p)\times O(n)$, with $N=2p+n$ that 
we denote $\cC_p$. They have $p$ antisymmetric $2\times2$ 
blocks where $h^Z_{ij}$ is proportional to
one $\eta$ coupling and the rest of $h^Z_{ij}$ vanish. We 
then take $h^T$ diagonal and respecting this symmetry. Explicitly,
\begin{equation}
  h^{Z,\cC_p}
  =\eta^{\cC_p}
  \left(\bigoplus_{k=1}^{p}\begin{pmatrix}0&1\\ -1&0\end{pmatrix}\right)
  \bigoplus 0_n,
  \qquad
  h^{T,\cC_p}
  =\diag\big(\underbrace{h_{2p}^{\cC_p},\cdots,h_{2p}^{\cC_p}}_{2p},
  \underbrace{h_n^{\cC_p},\cdots,h_n^{\cC_p}}_{n}\big).
\label{eq:STZ-Un-On-ansatz}
\end{equation}
For $\eta^{\cC_p}\neq0$, vanishing of $\beta^Z$ 
\eqref{eq:STZ-betaZ-functions} fixes
\begin{equation}
  h^{S,\cC_p}
  =\frac{C_{TZZ}}{C_{SZZ}}h_{2p}^{\cC_p}
  -\frac{\gamma_Z}{2\pi C_{SZZ}}.
\label{eq:STZ-hS-from-linear-constraint}
\end{equation}
With the shorthand $\varpi=2p-n$, the first $2p$ diagonal 
entries of $\beta^T$ fix
\begin{equation}
  \big(\eta^{\cC_p}\big)^2
  =\frac{N}{n\pi C_{TZZ}}
  \left[\bigl(\gamma_{T}+2\pi C_{STT}h^{S,\cC_p}\bigr)h_{2p}^{\cC_p}
  -\pi C_{TTT}\frac{\varpi}{n}\big(h_{2p}^{\cC_p}\big)^2\right].
\label{eq:STZ-eta2}
\end{equation}
Substituting \eqref{eq:STZ-hS-from-linear-constraint} and
\eqref{eq:STZ-eta2} into $\beta^S=0$ gives one quadratic equation for
$h_{2p}^{\cC_p}$,
\begin{equation}
  A_Z\big(h_{2p}^{\cC_p}\big)^2+B_Zh_{2p}^{\cC_p}+C_Z=0,
\label{eq:STZ-h2p-quadratic}
\end{equation}
where
\begin{equation}
\begin{aligned}
  A_Z
  &=\pi\left[\frac{C_{SSS}C_{TZZ}^2}{C_{SZZ}^2}
  +\frac{2pN}{n}
  \left(3C_{STT}
  -\frac{C_{SZZ}C_{TTT}}{C_{TZZ}}\frac{\varpi}{n}\right)\right],
\\
  B_Z&=\frac{C_{TZZ}}{C_{SZZ}}\gamma_S
  -\frac{C_{SSS}C_{TZZ}}{C_{SZZ}^2}\gamma_Z
  -\frac{2pN}{nC_{TZZ}}
  (\gamma_ZC_{STT}-\gamma_{T}C_{SZZ}),
\\
  C_Z
  &=
  \frac{\gamma_Z}{4\pi C_{SZZ}^2}
  (\gamma_ZC_{SSS}-2\gamma_SC_{SZZ}).
\label{eq:STZ-quadratic-coefficients}
\end{aligned}
\end{equation}
When the discriminant is positive, the quadratic \eqref{eq:STZ-h2p-quadratic} has two roots, so henceforth we distinguish them as 
the fixed points $\cC_{p,+}$ and $\cC_{p,-}$ and denote the 
couplings $h_{2p}^{\cC_{p,\pm}}$, $\eta^{\cC_{p,\pm}}$. 
Both branches have the same residual symmetry $U(p)\times O(n)$ but are generically inequivalent defects since any quantities that depend on the fixed-point couplings can take different values on the two branches.
When the discriminant vanishes, the branches merge.
Tracelessness of $T$ requires
$2ph_{2p}^{\cC_{p,\pm}}+nh_n^{\cC_{p,\pm}}=0$, fixing $h_n^{\cC_{p,\pm}}$.

It is not too hard to generalise the solution to have
$O(N)\to U(p)\times O(n)\times O(m)$ with $N=2p+n+m$ by replacing
the ansatz in \eqref{eq:STZ-Un-On-ansatz} with
\begin{equation}
h^{T}
  =\diag\big(\underbrace{h_{2p},\cdots,h_{2p}}_{2p},
  \underbrace{h_n,\cdots,h_n}_{n},
  \underbrace{h_m,\cdots,h_m}_{m}\big).
\end{equation}
Its main advantage is that it
allows to recover the $\cD_n$ saddle points of
Section~\ref{sec:SandT} by taking $p\to0$.
Still, to our taste, this is one degree of complication beyond
the necessary, so we did not pursue this.

\subsubsection{Defect conformal manifold}
\label{sec:STZ-DCM}
For fixed $p$ and $n=N-2p$, the fixed point $\cC_{p,\pm}$ preserves
$U(p)\times O(n)\subset O(N)$. 
The defect conformal manifold is therefore the homogeneous space
\begin{equation}
  \frac{O(N)}{U(p)\times O(n)}.
\end{equation}
This is an orthogonal flag manifold. Equivalently, it is the space of
orthogonal decompositions
\begin{equation}
  \mathbb R^N=V_{2p}\oplus W_n
\end{equation}
together with an orthogonal complex structure on $V_{2p}$. Thus it may
also be viewed as a fibration over the real Grassmannian 
$\Gr(2p,\bR^N)$ \eqref{eq:grassmannian} from 
Section~\ref{sec:SandT}, with fibre $O(2p)/U(p)$.

The two branches $\cC_{p,\pm}$ have different conformal data. 
The normalisation constants $C^{\cC_{p,\pm}}_{\bt_{\rm bifund}}$ and 
$C^{\cC_{p,\pm}}_{\bt_\wedge}$ take different values in the two branches. 
Since those constants appear in the Zamolodchikov metric of the 
conformal manifold \eqref{eq:Ct-Dn-general}, \eqref{eq:Ct-wedge-definition}, these 
manifolds which are isomorphic as homogeneous spaces 
are not isometric.

Unlike the Grassmannian in Section~\ref{sec:grassmannian}, 
these manifolds are simply connected, so there are no vortex 
configurations. 
But for $\cC_{N/2}$, i.e.~$n=0$, where there is a unique solution to the fixed point equations, the space $O(N)/U(N/2)$ has two 
connected components. On such surface defects there are 
then $\bZ_2$ valued line interfaces. 
Of course one can also introduce line interfaces between inequivalent 
surface defects or between surface defects on a connected component
of the defect conformal manifold, see 
\cite{Diatlyk:2024qpr, Kravchuk:2024qoh}.

\subsection{Stability matrix and near dimension two operators}
\label{sec:STZ-near-two}

We now linearize the beta functions 
\eqref{eq:STZ-betaS-functions}--\eqref{eq:STZ-betaZ-functions} around the
$\cC_{p,\pm}$ fixed points.
The fields of dimension close to two are the UV operators
$S$, $T$, $Z$. Under the preserved $U(p)\times O(n)$ symmetry they decompose
as
\begin{equation}
\begin{aligned}
  &S:\quad&
  \mathbf1&\to(\mathbf1,\mathbf1),
\\
  &T:&
  \mathbf{T}_N&\to(\mathbf1,\mathbf1)
  \oplus(\mathrm{Sym}^2\mathbf p,\mathbf1)
  \oplus(\mathrm{Sym}^2\bar{\mathbf p},\mathbf1)
  \oplus(\mathbf p,\mathbf n)
  \oplus(\bar{\mathbf p},\mathbf n)
  \oplus(\mathbf1,\mathbf T_n)
  \oplus(\mathbf{Adj}_p,\mathbf1),
\\
  &Z:&
  \mathbf{A}_N&\to(\mathbf1,\mathbf1)
  \oplus(\wedge^2\mathbf p,\mathbf1)
  \oplus(\wedge^2\bar{\mathbf p},\mathbf1)
  \oplus(\mathbf p,\mathbf n)
  \oplus(\bar{\mathbf p},\mathbf n)
  \oplus(\mathbf1,\mathbf A_n)
  \oplus(\mathbf{Adj}_p,\mathbf1).
\end{aligned}
\label{eq:STZ-Un-On-operator-decomposition}
\end{equation}
The resulting operators and dimensions are shown in
Table~\ref{tab:STZ-near-two-dimensions}.

\begin{table}[t]
\centering
\renewcommand{\arraystretch}{1.25}
\begin{tabular}{llll}
\hline
operator & $U(p)\times O(n)$ reps & dimension
& WF $O(N)\times O(2)$ dimension \\
\hline
$\cO_{1,2,3}$ & $(\mathbf1,\mathbf1)$ & $2+\gamma_{\cO_i}^{\cC_{p,\pm}}$
& see~\eqref{eq:STZ-N2-singlet-characteristic} \\
$T_n$ & $(\mathbf1,\mathbf T_n)$ & $2+\gamma_{T_n}^{\cC_{p,\pm}}$
& $2+\gamma_{T}+\frac{4\pi}{\sqrt N}h^{S,\cC_{p,\pm},{\rm WF}}
+4\pi h_n^{\cC_{p,\pm},{\rm WF}}$ \\
$T_{\rm sym}$, $\bar T_{\rm sym}$ &
$(\mathrm{Sym}^2\mathbf p,\mathbf1)
\oplus(\mathrm{Sym}^2\bar{\mathbf p},\mathbf1)$
& $2+\gamma_{T_{\rm sym}}^{\cC_{p,\pm}}$
& $2+\gamma_{T}+\frac{4\pi}{\sqrt N}h^{S,\cC_{p,\pm},{\rm WF}}
+4\pi h_{2p}^{\cC_{p,\pm},{\rm WF}}$ \\
$\cO_{{\rm adj},\pm}$ & $(\mathbf{Adj}_p,\mathbf1)$
& $2+\gamma_{{\rm adj},\pm}^{\cC_{p,\pm}}$
& see~\eqref{eq:STZ-N2-adj-gammas} \\
$\bt_{\rm bifund}$, $\bar\bt_{\rm bifund}$ & 
$(\mathbf p,\mathbf n)\oplus(\bar{\mathbf p},\mathbf n)$
& $2$ & $2$ \\
$Z_{\rm bifund}$, $\bar Z_{\rm bifund}$ & $(\mathbf p,\mathbf n)\oplus(\bar{\mathbf p},\mathbf n)$
& $2+\gamma_{Z_{\rm bifund}}^{\cC_{p,\pm}}$
& $2+\gamma_{T}+\frac{4\pi}{\sqrt N}h^{S,\cC_{p,\pm},{\rm WF}}
+4\pi h_n^{\cC_{p,\pm},{\rm WF}}$ \\
$Z_n$ & $(\mathbf1,\mathbf A_n)$ & $2+\gamma_{Z_n}^{\cC_{p,\pm}}$
& $2+4\pi(h_n^{\cC_{p,\pm},{\rm WF}}-h_{2p}^{\cC_{p,\pm},{\rm WF}})$ \\
$\bt_{\wedge}$, $\bar\bt_{\wedge}$ &
$(\wedge^2\mathbf p,\mathbf1)\oplus(\wedge^2\bar{\mathbf p},\mathbf1)$
& $2$ & $2$ \\
\hline
\end{tabular}
\caption{Near-two operators in the closed $S$, $T$, $Z$ sector at the
$\cC_{p,\pm}$ fixed point with preserved $U(p)\times O(n)$ symmetry.}
\label{tab:STZ-near-two-dimensions}
\end{table}

One can recover the $O(N)$ symmetric defect $\cD_N$ from the formulas 
here by setting $p=0$. We do not analyse the $\cD_n$ fixed points 
in general, but do in the case of the chiral $O(N)\times O(2)$ model in 
Section~\ref{sec:N2}.

We denote the nonsinglet fields coming from $T$ by $T_n$, 
$T_{\rm sym}$, $\bt_{\rm bifund}$ and $\cO_{{\rm adj},\pm}$. 
$T_{\rm sym}$, $\bt_{\rm bifund}$ are complex and 
$\bt_{\rm bifund}$ and $\cO_{{\rm adj},\pm}$ mix with the corresponding 
components of $Z$.
The anomalous dimensions of the operators from $T$ 
that do not mix are
\begin{align}
  \gamma_{T_n}^{\cC_{p,\pm}}
  &=
  \gamma_{T}
  +2\pi C_{STT}h^{S,\cC_{p,\pm}}
  +2\pi C_{TTT}h_n^{\cC_{p,\pm}},
\label{eq:STZ-Tn-dimensions}
\\
  \gamma_{T_{\rm sym}}^{\cC_{p,\pm}}
  &=
  \gamma_{T}
  +2\pi C_{STT}h^{S,\cC_{p,\pm}}
  +2\pi C_{TTT}h_{2p}^{\cC_{p,\pm}},
\label{eq:STZ-Tsym-dimensions}
\end{align}

The two fields arising from $Z$ without mixing with $T$ are 
$Z_n$ and the complex $\bt_\wedge$. Their dimensions are
\begin{align}
  \gamma_{Z_n}^{\cC_{p,\pm}}
  &=
  \gamma_Z
  +2\pi C_{SZZ}h^{S,\cC_{p,\pm}}
  -2\pi C_{TZZ}h_n^{\cC_{p,\pm}}
  =
  2\pi C_{TZZ}(h_{2p}^{\cC_{p,\pm}}-h_n^{\cC_{p,\pm}}).
\label{eq:STZ-Zn-dimensions}
\\
  \gamma_{\bt_{\wedge}}^{\cC_{p,\pm}}
  &=
  0.
\label{eq:STZ-twedge-dimensions}
\end{align}
The latter are protected tilts due to $O(2p)\to U(p)$ breaking.

The $U(p)$ adjoint components of $T$ and $Z$, mix.
In an orthonormal basis adapted to the complex structure on the first
$2p$ directions the mixing matrix is
\begin{equation}
  \Gamma_{\rm adj}^{\cC_{p,\pm}}
  =
  \begin{pmatrix}
    \gamma_{T}+2\pi C_{STT}h^{S,\cC_{p,\pm}}+2\pi C_{TTT}h_{2p}^{\cC_{p,\pm}}
    &
    2\pi C_{TZZ}\eta^{\cC_{p,\pm}}
    \\
    2\pi C_{TZZ}\eta^{\cC_{p,\pm}}
    &
    0
  \end{pmatrix}.
\label{eq:STZ-adj-Gamma}
\end{equation}
We denote the eigenvalues 
$\gamma_{{\rm adj},\pm}^{\cC_{p,\pm}}$, and the corresponding dimensions are
$2+\gamma_{{\rm adj},\pm}^{\cC_{p,\pm}}$.

The bifundamental components connecting the $2p$ and $n$ blocks of
$T$ and $Z$ also mix. 
the basis formed by the corresponding components of $T$ and $Z$, the
mixing matrix is
\begin{equation}
  \Gamma_{\rm bifund}^{\cC_{p,\pm}}
  =
  \begin{pmatrix}
    \gamma_{T}+2\pi C_{STT}h^{S,\cC_{p,\pm}}+\pi C_{TTT}(h_{2p}^{\cC_{p,\pm}}+h_n^{\cC_{p,\pm}})
    &
    \pi C_{TZZ}\eta^{\cC_{p,\pm}}
    \\
    \pi C_{TZZ}\eta^{\cC_{p,\pm}}
    &
    \pi C_{TZZ}(h_{2p}^{\cC_{p,\pm}}-h_n^{\cC_{p,\pm}})
  \end{pmatrix}.
\label{eq:STZ-bifund-Gamma}
\end{equation}
The determinant of \eqref{eq:STZ-bifund-Gamma} vanishes after imposing
the fixed point equations. The zero eigenvalue gives the protected tilt
operator $\bt_{\rm bifund}$. The orthogonal
operator has anomalous dimension
\begin{equation}
  \gamma_{Z_{\rm bifund}}^{\cC_{p,\pm}}
  =\gamma_{T}
  +2\pi C_{STT}h^{S,\cC_{p,\pm}}
  +\pi (C_{TTT}+C_{TZZ})h_{2p}^{\cC_{p,\pm}}
  +\pi (C_{TTT}-C_{TZZ})h_n^{\cC_{p,\pm}}.
\label{eq:STZ-B-gamma}
\end{equation}

The singlet sector is spanned by $S$, the component of $T$ parallel to
$h^{T,\cC_{p,\pm}}$, and the component of $Z$ parallel to $h^{Z,\cC_{p,\pm}}$. In the
unit-normalized basis the mixing matrix is
\begin{equation}
  \Gamma_{\rm singlet}^{\cC_{p,\pm}}
  =
  \begin{pmatrix}
    \gamma_S+2\pi C_{SSS}h^{S,\cC_{p,\pm}}
    &
    2\pi C_{STT}\sqrt{\frac{2pN}{n}}\,h_{2p}^{\cC_{p,\pm}}
    &
    2\pi C_{SZZ}\sqrt{2p}\,\eta^{\cC_{p,\pm}}
    \\
    2\pi C_{STT}\sqrt{\frac{2pN}{n}}\,h_{2p}^{\cC_{p,\pm}}
    &
    \gamma_{T}
    +2\pi C_{STT}h^{S,\cC_{p,\pm}}
    -2\pi C_{TTT}\frac{\varpi}{n}h_{2p}^{\cC_{p,\pm}}
    &
    -2\pi C_{TZZ}\eta^{\cC_{p,\pm}}\sqrt{\frac{n}{N}}
    \\
    2\pi C_{SZZ}\sqrt{2p}\,\eta^{\cC_{p,\pm}}
    &
    -2\pi C_{TZZ}\eta^{\cC_{p,\pm}}\sqrt{\frac{n}{N}}
    &
    0
  \end{pmatrix}.
\label{eq:STZ-singlet-Gamma}
\end{equation}
The eigenvalues of \eqref{eq:STZ-singlet-Gamma} give the
dimensions $2+\gamma_{\cO_i}^{\cC_{p,\pm}}$ of the three singlet operators.

For the $\cC_{N/2}$ fixed point with $n=0$ and $h^{T,\cC_{N/2}}=0$, 
$T_n$, $\bt_{\rm bifund}$, $Z_{\rm bifund}$ and $Z_n$ are absent. The $U(N/2)$ singlet
component of $T$ decouples from $S$ and $Z$, with anomalous dimension
\begin{equation}
  \gamma_{T}^{\cC_{N/2}}
  =
  \gamma_{T}
  +2\pi C_{STT}h^{S,\cC_{N/2}}.
\label{eq:STZ-pure-U-T-dimension}
\end{equation}
The remaining two singlets are governed by
\begin{equation}
  \Gamma_{\rm singlet}^{\cC_{N/2}}
  =
  \begin{pmatrix}
    \gamma_S+2\pi C_{SSS}h^{S,\cC_{N/2}}
    &
    2\pi C_{SZZ}\sqrt{N}\,\eta^{\cC_{N/2}}
    \\
    2\pi C_{SZZ}\sqrt{N}\,\eta^{\cC_{N/2}}
    &
    0
  \end{pmatrix}.
\label{eq:STZ-pure-U-singlet-Gamma}
\end{equation}

\subsubsection{Reality and stability analysis}
\label{sec:STZ-real-and-stable}

In Section~\ref{sec:real-and-stable} it is shown, 
that the symmetric fixed point
$\cD_N$ is stable in part of the parameter range~\cite{Pannell:2024hbu}, and that when it is, the other
$\cD_n$ are real. Adding the antisymmetric coupling $h^Z$, one must check whether
$\cD_N$ remains stable against $Z$ deformations and whether the old 
$\cD_n$ and new $\cC_{p,\pm}$ fixed points are real. 
The gradient structure carries over with 
$\beta^I=\partial A/\partial h^I$ and $A$ as in
\eqref{eq:A-function} but now summing over $I\in\{S,T_{ij},Z_{ij}\}$, generalising
\cite{Pannell:2024hbu} which treats only the symmetric coupling.

At $\cD_N$, using \eqref{eq:STZ-betaZ-functions} and 
\eqref{eq:DN-h-couplings}, the $Z$ perturbation has
\begin{equation}
  \gamma_Z^{\cD_N}
  =\gamma_Z-2\frac{C_{SZZ}}{C_{SSS}}\gamma_S,
\end{equation}
providing an extra condition for stability.

This is related also to reality of the $U(N/2)$ 
saddle, $\cC_{N/2}$ where $h^T=0$ and
\begin{equation}
|h^{Z,\cC_{N/2}}|^2
  =\frac{\gamma_Z(2\gamma_S C_{SZZ}-\gamma_Z C_{SSS})}
  {4\pi^2 C_{SZZ}^3}
  =-\frac{C_{SSS}}{4\pi^2 C_{SZZ}^3}
  \gamma_Z\gamma_Z^{\cD_N}.
\end{equation}
Thus, for positive $C_{SSS}$ and $C_{SZZ}$, the pure
$\cC_{N/2}$ saddle is real iff
\begin{equation}
  \gamma_Z\gamma_Z^{\cD_N}\le0.
\end{equation}
In particular, if $\gamma_Z<0$, stability of $\cD_N$ in the $Z$ direction,
$\gamma_Z^{\cD_N}>0$, implies reality of $\cC_{N/2}$.

The $U(p)\times O(n)$ fixed points are real when
\eqref{eq:STZ-h2p-quadratic} has a real root and the corresponding
$(\eta^{\cC_{p,\pm}})^2$ in \eqref{eq:STZ-eta2} is nonnegative. For
$\cC_{N/2}$, reality requires the right hand side of
\eqref{eq:STZ-pure-U-norm} to be nonnegative.

Whenever $\eta^{\cC_{p,\pm}}\neq0$, however, these real fixed points are
unstable. For the $U(p)\times O(n)$ fixed point the
singlet matrix \eqref{eq:STZ-singlet-Gamma} has a zero in the bottom
right entry, while its last row is nonzero. Therefore it cannot be
positive definite. Equivalently, the component of $Z$ parallel to
$h^{Z,\cC_{p,\pm}}$
mixes with the other singlets but has zero diagonal anomalous dimension,
so at least one singlet eigenvalue $\gamma_{\cO_i}^{\cC_{p,\pm}}$ is negative.

For $p>1$ there is a second, independent obstruction. The adjoint block
\eqref{eq:STZ-adj-Gamma} has determinant
\begin{equation}
  \det\Gamma_{\rm adj}^{\cC_{p,\pm}}
  =
  -4\pi^2 C_{TZZ}^2\big(\eta^{\cC_{p,\pm}}\big)^2,
\end{equation}
and hence has one negative eigenvalue.

The same argument applies to the $\cC_{N/2}$ fixed point. The singlet
matrix \eqref{eq:STZ-pure-U-singlet-Gamma} has determinant
\begin{equation}
  \det\Gamma_{\rm singlet}^{\cC_{N/2}}
  =
  -4\pi^2 C_{SZZ}^2N\big(\eta^{\cC_{N/2}}\big)^2,
\end{equation}
so one of its two eigenvalues is negative whenever the fixed point is
real and $\eta^{\cC_{N/2}}\neq0$.

\subsection{Operators of dimension near three}
\label{sec:STZ-near-three}

We now consider the transverse vectors 
of dimension near three in the $S$, $T$, $Z$ system
at a $\cC_{p,\pm}$ fixed point. 
In the UV, in addition to the descendants $V_r$ and $U_{ij,r}$, we now
also require the descendant of $Z$, denoted $W_{ij,r}$ with the 
usual normalisation \eqref{eq:uv-descendant-normalization}
\begin{equation}
  W_{ij,r}=\frac{1}{\sqrt{2\Delta_Z}}\partial_rZ_{ij}
  =
  \frac12\partial_rZ_{ij}+O(\varepsilon).
\label{eq:W-normalization}
\end{equation}

The descendant structure constants follow from the parent
ones by the relations
\eqref{eq:descendant-structure-constants} and their obvious extension to
$Z$, i.e.\ $C_{WWS}=\tfrac12 C_{SZZ}$,
$C_{WWT}=C_{UWZ}=\tfrac12 C_{TZZ}$, and
$C_{VUT}=\tfrac12 C_{STT}$.

Mirroring the derivation of $C_{TjU}=\sqrt{(d-2)C_j}$ 
\eqref{eq:ward-CjU-sqrt}, $C_{ZjW}$ is determined by using 
the action of the $O(N)$ generators $E_{kl}$
\eqref{eq:adjoint-generator-convention} on the antisymmetric $Z_{mn}$
\begin{equation}
\begin{aligned}
\partial_\mu
\left\langle j_{kl}^{\mu}(x)Z_{ij}(x_1)Z_{mn}(x_2)\right\rangle
&=
-\delta^{(d)}(x-x_1)
\big((E_{kl})_i{}^a\delta_j^b-(E_{kl})_j{}^a\delta_i^b\big)
\left\langle Z_{ab}(x_1)Z_{mn}(x_2)\right\rangle
\\
&\quad
-\delta^{(d)}(x-x_2)
\big((E_{kl})_m{}^a\delta_n^b-(E_{kl})_n{}^a\delta_m^b\big)
\left\langle Z_{ij}(x_1)Z_{ab}(x_2)\right\rangle .
\end{aligned}
\end{equation}
Using the descendant
normalisation $W_{ij,r}=(1/\sqrt{2\Delta_Z})\partial_rZ_{ij}$ and matching
the residue against the three-point function
$\langle Z(x_1)j_{kl,r}(x_2)W_{mn,s}(x_3)\rangle$ in the same way as in
\eqref{eq:current-three-point-definitions}, this fixes
\begin{equation}
  \frac{C_{ZjW}^2}{2C_j}=\frac{d-2}{\Delta_Z}.
\label{eq:ward-CjW-ratio-abstract}
\end{equation}
At leading order, and
with the sign matching the orientation in
\eqref{eq:adjoint-generator-convention},
\begin{equation}
  C_{ZjW}=\sqrt{(d-2)C_j}+O(\varepsilon).
\label{eq:ward-CjW-sqrt}
\end{equation}

Under $O(N)\to U(p)\times O(n)$ with $N=2p+n$ the four operators 
$V_r$, $U_{ij,r}$, $W_{ij,r}$ \eqref{eq:uv-descendant-normalization} 
together with the current $j_{ij,r}$ which has $\Delta_j=d-1$ 
and is close to 3 for $d\simeq4$  
decompose in the same pattern as their dimension-two operators in
\eqref{eq:STZ-Un-On-operator-decomposition}
\begin{equation}
\begin{aligned}
  &V_r:&
  \mathbf1&\to(\mathbf1,\mathbf1),
\\
  &U_{ij,r}:&
  \mathbf{T}_N&\to(\mathbf1,\mathbf1)
  \oplus(\mathrm{Sym}^2\mathbf p,\mathbf1)
  \oplus(\mathrm{Sym}^2\bar{\mathbf p},\mathbf1)
  \oplus(\mathbf p,\mathbf n)
  \oplus(\bar{\mathbf p},\mathbf n)
  \oplus(\mathbf1,\mathbf T_n)
  \oplus(\mathbf{Adj}_p,\mathbf1),
\\
  &W_{ij,r}:&
  \mathbf{A}_N&\to(\mathbf1,\mathbf1)
  \oplus(\wedge^2\mathbf p,\mathbf1)
  \oplus(\wedge^2\bar{\mathbf p},\mathbf1)
  \oplus(\mathbf p,\mathbf n)
  \oplus(\bar{\mathbf p},\mathbf n)
  \oplus(\mathbf1,\mathbf A_n)
  \oplus(\mathbf{Adj}_p,\mathbf1),
\\
  &j_{ij,r}:&
  \mathbf{A}_N&\to(\mathbf1,\mathbf1)
  \oplus(\wedge^2\mathbf p,\mathbf1)
  \oplus(\wedge^2\bar{\mathbf p},\mathbf1)
  \oplus(\mathbf p,\mathbf n)
  \oplus(\bar{\mathbf p},\mathbf n)
  \oplus(\mathbf1,\mathbf A_n)
  \oplus(\mathbf{Adj}_p,\mathbf1).
\end{aligned}
\label{eq:STZ-near-three-rep-decomposition}
\end{equation}
The resulting operators are listed in
Table~\ref{tab:STZ-near-three-dimensions}.

\begin{table}[t]
\centering
\renewcommand{\arraystretch}{1.25}
\begin{tabular}{lll}
\hline
operator & $U(p)\times O(n)$ reps & dimension \\
\hline
$\bD_r$ & $(\mathbf1,\mathbf1)$ & $3$ \\
$\cO_{\pm,r}$ & $(\mathbf1,\mathbf1)$
& $3+\gamma_{\cO_{\pm,r}}^{\cC_{p,\pm}}$ \\
$U_{{\rm sym},r}$, $\bar U_{{\rm sym},r}$ &
$(\mathrm{Sym}^2\mathbf p,\mathbf1)\oplus(\mathrm{Sym}^2\bar{\mathbf p},\mathbf1)$
& $3+\gamma_{U_{\rm sym}}^{\cC_{p,\pm}}$ \\
$U_{n,r}$ & $(\mathbf1,\mathbf T_n)$
& $3+\gamma_{U_n}^{\cC_{p,\pm}}$ \\
$\cO_{{\rm adj},\pm,r}$ & $(\mathbf{Adj}_p,\mathbf1)$
& $3+\gamma_{{\rm adj},\pm,r}^{\cC_{p,\pm}}$ \\
$u_{i,\hat\imath\check\jmath,r}$, $i=1,2,3$ &
$(\mathbf p,\mathbf n)\oplus(\bar{\mathbf p},\mathbf n)$
& $3+\gamma_{u_i}^{\cC_{p,\pm}}$ \\
$W_{n,\check\imath\check\jmath,r}$ & $(\mathbf1,\mathbf A_n)$
& $3+\gamma_{W_n}^{\cC_{p,\pm}}$ \\
$\cO_{\wedge,\pm,r}$, $\bar\cO_{\wedge,\pm,r}$ &
$(\wedge^2\mathbf p,\mathbf1)\oplus(\wedge^2\bar{\mathbf p},\mathbf1)$
& $3+\gamma_{\wedge,\pm}^{\cC_{p,\pm}}$ \\
$j_{U(1),r}$ & $(\mathbf1,\mathbf1)$ & $3-\varepsilon$ \\
$j_{{\rm adj},r}$ & $(\mathbf{Adj}_p,\mathbf1)$ & $3-\varepsilon$ \\
$j_{n,\check\imath\check\jmath,r}$ & $(\mathbf1,\mathbf A_n)$
& $3-\varepsilon$ \\
\hline
\end{tabular}
\caption{Transverse vector operators of dimension near three in the 
$\cC_{p,\pm}$ fixed point preserving $U(p)\times O(n)$ symmetry.}
\label{tab:STZ-near-three-dimensions}
\end{table}

The fields that do not mix are:
\begin{itemize}
\item
The components of the current in the $O(n)$ adjoint, $U(p)$ adjoint 
and the $U(1)\subset U(p)$ singlet. They remain as conserved currents.
\item
From $U$, the descendant of $T_{ij}$ there is 
$U_n$ in $(\mathbf1,\mathbf T_n)$ with
\begin{equation}
  \gamma_{U_n}^{\cC_{p,\pm}}
  =
  \gamma_{T}
  +\pi C_{STT}h^{S,\cC_{p,\pm}}
  +\pi C_{TTT}h_n^{\cC_{p,\pm}}.
\label{eq:STZ-Un-vector-dimension}
\end{equation}
Likewise complex symmetric $U_{\rm sym}$ 
in $(\mathrm{Sym}^2\mathbf p,\mathbf1)\oplus
(\mathrm{Sym}^2\bar{\mathbf p},\mathbf1)$ with
\begin{equation}
  \gamma_{U_{\rm sym}}^{\cC_{p,\pm}}
  =
  \gamma_{T}
  +\pi C_{STT}h^{S,\cC_{p,\pm}}
  +\pi C_{TTT}h_{2p}^{\cC_{p,\pm}}.
\label{eq:STZ-Usym-vector-dimension}
\end{equation}

\item 
The component of $W$ in $(\mathbf1,\mathbf A_n)$ cannot 
mix with the conserved current. Instead it has
\begin{equation}
  \gamma_{W_n}^{\cC_{p,\pm}}
  =
  \gamma_Z
  +\pi C_{SZZ}h^{S,\cC_{p,\pm}}
  -\pi C_{TZZ}h_n^{\cC_{p,\pm}}.
\label{eq:STZ-Wn-vector-dimension}
\end{equation}

\end{itemize}

The components of $U$ and $W$ in the $(\mathbf{Adj}_p,\mathbf1)$ 
representation mix. In the orthonormal basis
$(U_{{\rm adj},r}^{\cC_{p,\pm}},\,W_{{\rm adj},r}^{\cC_{p,\pm}})$ the mixing matrix
is
\begin{equation}
  \Gamma_{{\rm adj},r}^{\cC_{p,\pm}}
  =
  \begin{pmatrix}
    \gamma_T+\pi C_{STT}h^{S,\cC_{p,\pm}}+\pi C_{TTT}h_{2p}^{\cC_{p,\pm}}
    &
    \pi C_{TZZ}\eta^{\cC_{p,\pm}}
    \\[.6em]
    \pi C_{TZZ}\eta^{\cC_{p,\pm}}
    &
    \gamma_Z+\pi C_{SZZ}h^{S,\cC_{p,\pm}}-\pi C_{TZZ}h_{2p}^{\cC_{p,\pm}}
  \end{pmatrix},
\label{eq:STZ-adj-vector-Gamma}
\end{equation}

$j$ and $W$ both have components in the 
$(\wedge^2\mathbf p,\mathbf1)\oplus
(\wedge^2\bar{\mathbf p},\mathbf1)$ representation that mix. 
In the basis
$(W_{\wedge,r}^{\cC_{p,\pm}},\,j_{\wedge,r}^{\cC_{p,\pm}}/\sqrt{C_j})$, 
the mixing matrix is
\begin{equation}
  \Gamma_{\wedge,r}^{\cC_{p,\pm}}
  =
  \begin{pmatrix}
    \displaystyle
    \gamma_Z+\pi C_{SZZ}h^{S,\cC_{p,\pm}}-\pi C_{TZZ}h_{2p}^{\cC_{p,\pm}}
    &
    \displaystyle
    2\pi\frac{C_{ZjW}}{\sqrt{C_j}}\,\eta^{\cC_{p,\pm}}
    \\[.6em]
    \displaystyle
    2\pi\frac{C_{ZjW}}{\sqrt{C_j}}\,\eta^{\cC_{p,\pm}}
    &
    -\varepsilon
\end{pmatrix}.
\label{eq:STZ-Wwedge-vector-Gamma}
\end{equation}
Using $\gamma_Z+2\pi C_{SZZ}h^{S,\cC_{p,\pm}}-2\pi C_{TZZ}h_{2p}^{\cC_{p,\pm}}=0$
\eqref{eq:STZ-hS-from-linear-constraint} 
and the relation \eqref{eq:ward-CjW-ratio-abstract}, this 
simplifies to
\begin{equation}
  \Gamma_{\wedge,r}^{\cC_{p,\pm}}
  =
  \begin{pmatrix}
    \displaystyle
    \gamma_Z/2
    &
    \displaystyle
    2\pi\sqrt{d-2}\,\eta^{\cC_{p,\pm}}
    \\[.6em]
    \displaystyle
    2\pi\sqrt{d-2}\,\eta^{\cC_{p,\pm}}
    &
    -\varepsilon
\end{pmatrix}.
\label{eq:STZ-Wwedge-vector-Gamma-simplified}
\end{equation}

All three of $U$, $W$ and $j$ have components in the bifundamental 
$(\mathbf p,\mathbf n)\oplus(\bar{\mathbf p},\mathbf n)$ 
that mix. In the basis
$(U_{b,r}^{\cC_{p,\pm}},\,W_{b,r}^{\cC_{p,\pm}},\,j_{b,r}^{\cC_{p,\pm}}/\sqrt{C_j})$, 
and where the starred entries are completed by symmetrization 
and the fixed point superscripts removed for brevity,
\begin{equation}
  \Gamma_{b,r}^{\cC_{p,\pm}}
  =
  \begin{pmatrix}
    \displaystyle
    \gamma_T+\pi C_{STT}h^{S}
    +\tfrac{\pi C_{TTT}}{2}(h_{2p}+h_n)
    & \star
    & \star
    \\[.6em]
    \tfrac{\pi C_{TZZ}}{2}\eta
    & \gamma_Z+\pi C_{SZZ}h^{S}
    -\tfrac{\pi C_{TZZ}}{2}(h_{2p}+h_n)
    & \star
    \\[.6em]
    \pi\tfrac{C_{TjU}}{\sqrt{C_j}}
      (h_{2p}-h_n)
    & \pi\tfrac{C_{ZjW}}{\sqrt{C_j}}\eta
    & -\varepsilon
  \end{pmatrix},
\label{eq:STZ-bifund-vector-Gamma}
\end{equation}
giving three anomalous dimensions $\gamma_{u_i}^{\cC_{p,\pm}}$, $i=1,2,3$.

The singlet components of $V$, $U$ and $W$ all mix. 
In the unit-normalised basis
\\
$(V_r^{\cC_{p,\pm}},\;
h^{T,\cC_{p,\pm}}_{ij}U_{ij,r}/|h^{T,\cC_{p,\pm}}|,\;
h^{Z,\cC_{p,\pm}}_{ij}W_{ij,r}/|h^{Z,\cC_{p,\pm}}|)$
the matrix (again with superscripts suppressed) is
\begin{equation}
  \Gamma_{\rm sing,r}^{\cC_{p,\pm}}
  =
  \begin{pmatrix}
    \displaystyle
    \gamma_S+\pi C_{SSS}h^{S}
    & \pi C_{STT}\sqrt{\tfrac{2pN}{n}}h_{2p}
    & \pi C_{SZZ}\sqrt{2p}\,\eta
    \\[.6em]
    \pi C_{STT}\sqrt{\tfrac{2pN}{n}}h_{2p}
    & \gamma_T+\pi C_{STT}h^{S}
    -\pi C_{TTT}\tfrac{\varpi}{n}h_{2p}
    & -\pi C_{TZZ}\eta\sqrt{\tfrac{n}{N}}
    \\[.6em]
    \pi C_{SZZ}\sqrt{2p}\,\eta
    & -\pi C_{TZZ}\eta\sqrt{\tfrac{n}{N}}
    & \gamma_Z+\pi C_{SZZ}h^{S}
    -\pi C_{TZZ}h_{2p}
  \end{pmatrix}.
\label{eq:STZ-singlet-vector-Gamma}
\end{equation}
The determinant of this matrix vanishes by using 
$\beta^S=\beta^T=\beta^Z=0$ 
\eqref{eq:STZ-betaS-functions}--\eqref{eq:STZ-betaZ-functions}. 
The zero eigenvector is the protected displacement 
\begin{equation}
  \big(
    h^{S,\cC_{p,\pm}},
    \sqrt{\tfrac{2pN}{n}}h_{2p}^{\cC_{p,\pm}},
    \sqrt{2p}\,\eta^{\cC_{p,\pm}}
  \big)
\end{equation}
with
\begin{equation}
  \Delta_{\bD}^{\cC_{p,\pm}}=3.
\label{eq:STZ-displacement-dimension}
\end{equation}
The two non-zero eigenvalues are denoted $\gamma_{\cO_{\pm,r}}^{\cC_{p,\pm}}$.

Finally, let us look at the special case of the $\cC_{N/2}$ fixed 
point with $n=0$. In this case $h^{T,\cC_{N/2}}=0$, 
and the $(\mathbf p,\mathbf n)$,
$(\mathbf1,\mathbf A_n)$ and $(\mathbf1,\mathbf T_n)$ blocks all
disappear.

$U_{\rm sym}$ in $(\mathrm{Sym}^2\mathbf p,\mathbf1)\oplus
(\mathrm{Sym}^2\bar{\mathbf p},\mathbf1)$ decouples, since there is no
$W$ analogue in this representation, and has
\begin{equation}
  \gamma_{U_{\rm sym}}^{\cC_{N/2}}
  =\gamma_T+\pi C_{STT}h^{S,\cC_{N/2}}.
\end{equation}
The mixing of the $U$ and $W$ components in 
$(\mathbf{Adj}_p,\mathbf1)$ \eqref{eq:STZ-adj-vector-Gamma} 
persists, but the matrix simplifies to
\begin{equation}
  \Gamma_{{\rm adj},r}^{\cC_{N/2}}
  =
  \begin{pmatrix}
    \gamma_T+\pi C_{STT}h^{S,\cC_{N/2}}
    & \pi C_{TZZ}\eta^{\cC_{N/2}}
    \\[.4em]
    \pi C_{TZZ}\eta^{\cC_{N/2}}
    & \gamma_Z+\pi C_{SZZ}h^{S,\cC_{N/2}}
  \end{pmatrix},
\label{eq:STZ-adj-vector-Gamma-N2}
\end{equation}
The $SU(p)$ and $U(1)$ currents remain decoupled and conserved.

The broken-$O(N)/U(N/2)$ components in 
$(\wedge^2\mathbf p,\mathbf1)\oplus(\wedge^2\bar{\mathbf p},\mathbf1)$ 
are again as in 
\eqref{eq:STZ-Wwedge-vector-Gamma} now simplified with 
$h_{2p}^{\cC_{N/2}}=0$ on the diagonal. 

Among the singlets, the component of $U$ no longer mixes with 
those of $V$ or $W$, since there is no $h^T$
to source the mixing. It decouples at
\begin{equation}
  \gamma_{U_{\rm sing}}^{\cC_{N/2}}
  =\gamma_T+\pi C_{STT}h^{S,\cC_{N/2}}.
\label{eq:STZ-Using-N2-decoupled}
\end{equation}
The remaining mixing in
\eqref{eq:STZ-singlet-vector-Gamma} reduces to the $2\times2$ block in
the $(V_r,h^{Z,\cC_{p,\pm}}_{ij}W_{ij,r}/|h^{Z,\cC_{p,\pm}}|)$ basis,
\begin{equation}
  \Gamma_{\rm sing,r}^{\cC_{N/2}}
  =
  \begin{pmatrix}
    \gamma_S+\pi C_{SSS}h^{S,\cC_{N/2}}
    & \pi C_{SZZ}\sqrt{N}\,\eta^{\cC_{N/2}}
    \\[.4em]
    \pi C_{SZZ}\sqrt{N}\,\eta^{\cC_{N/2}}
    & -\pi C_{SZZ}h^{S,\cC_{N/2}}
  \end{pmatrix},
\label{eq:STZ-singlet-vector-Gamma-N2}
\end{equation}
whose determinant vanishes by $\beta^S=0$ and whose zero
eigenvector is the protected displacement.

\section{Protected operators and their flows}
\label{sec:protected-flow}
At fixed points of defect RG flows, 
tilt and displacement operators have protected integer 
dimensions, and their normalizations $C_\bt$ and $C_\bD$ 
are fixed by their relation 
to the broken translation and global symmetry currents. In this section
we first define these operators rigorously by varying the action in the 
symmetry-breaking directions. 
Next, we determine the fixed-point 
values of these normalizations from Ward identities and finally 
we extend the treatment of displacements and tilts along the RG flows. 
We treat the $\cD_n$ defects first, then the full system 
including the antisymmetric coupling and in 
Section~\ref{sec:DTflows} we analyse their flows between 
fixed points.

\subsection{Tilt and displacement normalisations for 
scalar--tensor defects}
\label{sec:tD-in-ST}

The formal definition of the tilt and displacement operators 
is by coupling the action~\eqref{eq:surface-action} 
to extra sources $u\in O(N)$ and $v\in\bR^d$, such that the 
deformed defect is
\begin{equation}
  S_{\rm def}
  =S_{\rm UV}
  +\mu^{-\gamma_S}h^S 
  \int_{\bR^2} d^2\tau\,S(\tau+v(\tau))
  +\mu^{-\gamma_T}
  \int_{\bR^2} d^2\tau\,(u(\tau)h^Tu(\tau)^\top)_{ij}
  T_{ij}(\tau+v(\tau)).
\label{eq:deformed-action}
\end{equation}
For constant $u\notin O(n)\times O(m)$ and for $v$ perpendicular 
to $\bR^2$, this is a different defect, but equivalent by a 
global broken symmetry. For nonconstant 
parameters (ignoring equivalences), this is no longer a uniform 
flat defect.%
\footnote{We do not consider $v$ with components parallel to  
$\bR^2$, so as not to worry about writing reparametrisation 
invariant actions.}
$u$ is defined modulo right $O(n)\times O(m)$ action, so parametrises 
the Grassmannian $O(N)/(O(n)\times O(m))$, the defect conformal 
manifold. We are concerned with infinitesimal perturbations where 
$u=\mathbbm{1}+w$ with $w\in \sof(N)/(\sof(n)\oplus\sof(m))$.

Denoting $Z[u,v]$ (or alternatively $Z[w,v]$) as the defect partition 
function with the sources, the displacement and tilt are 
defined as
\begin{align}
\big\langle\bt_{\hat\imath_1\check\jmath_1}(\tau_1)\cdots
\bt_{\hat\imath_k\check\jmath_k}(\tau_k)\big\rangle
&=\frac{\delta}{\delta u^{\hat\imath_1\check\jmath_1}(\tau_1)}
\cdots\frac{\delta}{\delta u^{\hat\imath_k\check\jmath_k}(\tau_k)}
\log Z[u,v]\Big\vert_{u=\mathbbm1,v=0},
\\
\big\langle\bD_{r_1}(\tau_1)\cdots\bD_{r_k}(\tau_k)\big\rangle
&= \frac{\delta}{\delta v^{r_1}(\tau_1)}
\cdots\frac{\delta}{\delta v^{r_k}(\tau_k)}
\log Z[u,v]\Big\vert_{u=\mathbbm1,v=0}.
\end{align}
At separated points and for infinitesimal $v$ and $w$ we can 
also use
\begin{align}
Z_\text{lin}[v, w]
=
\left\langle \exp\int d^2\tau\,
\left(  v_r(\tau)\bD^r(\tau) 
+  w_{\hat\imath\check\jmath}(\tau)\bt^{\hat\imath\check\jmath}(\tau) \right)
\right\rangle_\text{DCFT}. 
\label{eq:dCFT_partition_with_sources}
\end{align}

The tilt and displacement are expressed in terms of the operators 
$S$, $T_{ij}$ and their normal derivatives $V_r$, $U_{ij,r}$. 
Prior to setting the fixed-point values, at a scale $\mu$, the sources 
$v$, $w$ couple to the operators
\begin{align}
  \bt_{\hat\imath\check\jmath}
  =-\frac{\delta S_{\rm def}}
  {\delta w_{\hat\imath\check\jmath}}\Big\vert_{w=v=0}
  &=
  2\mu^{-\gamma_T}\big(h_n-h_m\big)T_{\hat\imath\check\jmath}.
  \label{eq:t-explicit-from-variation}
\\
  \bD_r=-\frac{\delta S_{\rm def}}{\delta v^r}\Big\vert_{w=v=0}
  &=
  -\mu^{-\gamma_S}h^S\partial_rS
  -\mu^{-\gamma_T}h^T_{ij}\partial_rT_{ij}
\nonumber  \\
  &=
  -\sqrt{2\Delta_S}\,\mu^{-\gamma_S}h^S V_r
  -\sqrt{2\Delta_T}\,\mu^{-\gamma_T}h^T_{ij}U_{ij,r}.
  \label{eq:D-explicit-from-variation}
\end{align}
The expression for $\bt$ assumes the diagonal coupling $h_{ij}^T$ 
as in \eqref{eq:hn-hm-ansatz}, the sign is due to $\hat\imath$ 
taking values in the first $n$ rows/columns of $h^T$ and the factor of two is due to the sum over $ij$ being unconstrained.

By setting the fixed point values for the couplings determined 
in Section~\ref{sec:fixed-points}, these become the
protected operators of dimension two 
\eqref{eq:tensor-gammas-generic} and three 
\eqref{eq:displacement-dimension-generic}. Their two-point functions at separated points are then 
\eqref{eq:CD-Ct}
\begin{equation}
  \left\langle
    \bt_{\hat\imath\check\jmath}(\tau)\bt_{\hat k\check l}(0)
  \right\rangle
  =
  \frac{C_{\bt}\delta_{\hat\imath\hat k}\delta_{\check\jmath\check l}}{|\tau|^4}
\qquad
  \left\langle
    \bD_r(\tau)\bD_s(0)
  \right\rangle
  =
  \frac{C_{\bD}\delta_{rs}}{|\tau|^6}.
\label{eq:CD-Ct-def}
\end{equation}
The values of $C_\bt$ and $C_\bD$ follow by substituting the
fixed-point couplings in \eqref{eq:D-explicit-from-variation}, 
\eqref{eq:t-explicit-from-variation} and
using the normalizations \eqref{eq:uv-two-point-normalization},
\eqref{eq:uv-descendant-normalization}. 
Noting the factor of $1/2$ in 
$\Pi_{ij,kl}$ \eqref{eq:st-and-adj-projectors}, this gives
\begin{equation}
  \big\langle\bt_{\hat\imath\check\jmath}(\tau)
  \bt_{\hat k\check l}(0)\big\rangle
  =
  2\big(h_n^{\cD_n}-h_m^{\cD_n}\big)^2
  \frac{\delta_{\hat\imath\hat k}\delta_{\check\jmath\check l}}{|\tau|^4}.
\qquad\Rightarrow\qquad
  C_{\bt}^{\cD_n}
  =
  2\big(h_n^{\cD_n}-h_m^{\cD_n}\big)^2.
  \label{eq:Ct-Dn-general}
\end{equation}
Equivalently, using the tracelessness condition
$n h_n^{\cD_n}+m h_m^{\cD_n}=0$,
\begin{equation}
  C_{\bt}^{\cD_n}
  =
  \frac{2N}{nm}\,
  h^{T,\cD_n}_{ij}h^{T,\cD_n}_{ij}.
  \label{eq:Ct-Dn-tensor-norm}
\end{equation}
For a fixed point with $n\neq m$ and $C_{TTT}\neq0$, this 
evaluates to \eqref{eq:hn-hm-generic}
\begin{equation}
  C_{\bt}^{\cD_n}
  =
  \frac{2N^2}{\pi^2\nu^2C_{TTT}^2}
  \left[
    \gamma_T+2\pi C_{STT}h^{S,\cD_n}
  \right]^2
  \label{eq:Ct-Dn-generic}
\end{equation}
with $h^{S,\cD_n}$ in \eqref{eq:hS-roots-sigma-gen-generic}. 
When $n=m=N/2$, use \eqref{eq:Ct-Dn-general} with $h_{N/2}^{\cD_{N/2}}$ from \eqref{eq:hT-half}.

As both $\gamma_T$ and $h^{S,\cD_n}$ are of order $\varepsilon$, 
$C_\bt$ (and likewise $C_\bD$ below) are of order $\varepsilon^2$. 
The derivation above is precise, except for \eqref{eq:Ct-Dn-generic} 
which is at first order in conformal perturbation theory.

One can evaluate $C_\bt$ also in the case when $C_{TTT}=0$, 
using the formulas around \eqref{eq:hT-O2}. But as noted, the 
possibility of there being a large nonsymmetric conformal manifold is 
unlikely, so we do not elaborate on it.

The same analysis determines $C_\bD$ from 
\eqref{eq:D-explicit-from-variation}
\begin{equation}
  C_{\bD}^{\cD_n}
  =
  2\Delta_S\big(h^{S,\cD_n}\big)^2
  +
  2\Delta_T h^{T,\cD_n}_{ij}h^{T,\cD_n}_{ij}.
  \label{eq:CD-endpoint-general}
\end{equation}
For a fixed point with $n\neq m$, use
\eqref{eq:hTnorm-generic} in \eqref{eq:CD-endpoint-general}
\begin{equation}
\begin{aligned}
  C_{\bD}^{\cD_n}
  &=
  2\Delta_S\big(h^{S,\cD_n}\big)^2
  +
  2\Delta_T
  \frac{nmN}{\nu^2\pi^2C_{TTT}^2}
  \left[
    \gamma_T+2\pi C_{STT}h^{S,\cD_n}
  \right]^2 ,
  \label{eq:CD-Dn-general}
\end{aligned}
\end{equation}
with $h^{S,\cD_n}$ in \eqref{eq:hS-roots-sigma-gen-generic}. 
For $n=m=N/2$, replace the tensor norm in
\eqref{eq:CD-endpoint-general} by \eqref{eq:hT-O2}.

For the $O(N)$ symmetric endpoint $\cD_N$ \eqref{eq:DN-h-couplings},
the tensor coupling vanishes and the displacement normalization follows
from \eqref{eq:CD-endpoint-general}
\begin{equation}
  C_{\bD}^{\cD_N}
  =
  2\Delta_S
  \left(
    \frac{\gamma_S}{\pi C_{SSS}}
  \right)^2.
  \label{eq:CD-DN-general}
\end{equation}
The tilt normalization at the same endpoint follows from the Ward
identity \eqref{eq:Ct-Dn-general}. The $O(N)$ symmetric point is a
degenerate case in which $h_n^{\cD_N}=h_m^{\cD_N}=0$. Therefore
\begin{equation}
  C_{\bt}^{\cD_N}=0.
  \label{eq:Ct-DN}
\end{equation}

\subsection{Tilt and displacement normalisations for 
scalar--tensor--antisymmetric defects}
\label{sec:tD-in-STZ}

The same analysis applies to the scalar--tensor--antisymmetric 
fixed points $\cC_{p,\pm}$ of Section~\ref{sec:scalar-tensor-anti}.
The only change is that the defect action
\eqref{eq:surface-action-current} carries the additional coupling
$h^Z_{ij}Z_{ij}$, whose fixed-point value is given in
\eqref{eq:STZ-Un-On-ansatz}. Coupling the defect to the broken rotation
$u\in O(N)$ and the vector $v$ as in \eqref{eq:deformed-action}
gives the deformed action
\begin{equation}
\begin{aligned}
  S_{\rm def}
  =S_{\rm UV}
  &+\mu^{-\gamma_S}h^S\int_{\bR^2} d^2\tau\,S(\tau+v)
  +\mu^{-\gamma_T}\int_{\bR^2} d^2\tau\,\big(u\,h^Tu^\top\big)_{ij}T_{ij}(\tau+v)
\\
  &+\mu^{-\gamma_Z}\int_{\bR^2} d^2\tau\,\big(u\,h^Zu^\top\big)_{ij}Z_{ij}(\tau+v).
\end{aligned}
\label{eq:STZ-deformed-action}
\end{equation}
The tilt is the $w$-variation of this action at $u=\mathbbm1+w$. Writing
$w=w^AE_A$, where $E_A$ is a basis of broken generators, and noting that the
broken generators act on both $h^T$ and $h^Z$, the analog of
\eqref{eq:t-explicit-from-variation} is
\begin{equation}
  \bt_A
  =-\frac{\delta S_{\rm def}}{\delta w^A}\Big\vert_{w=v=0}
  =-\mu^{-\gamma_T}\big[E_A,h^T\big]_{ij}T_{ij}
  -\mu^{-\gamma_Z}\big[E_A,h^Z\big]_{ij}Z_{ij},
\label{eq:STZ-tilt-commutator}
\end{equation}
with $E_A$ tangent to the conformal manifold $O(N)/(U(p)\times O(n))$ of
Section~\ref{sec:STZ-DCM}. 
Following the decomposition of the fields 
in \eqref{eq:STZ-Un-On-operator-decomposition}, the
broken generators decompose under $U(p)\times O(n)$ into 
the representations
\begin{equation}
  \sof(N)/(\mathfrak u(p)\oplus\sof(n))
  =
  (\mathbf p,\mathbf n)\oplus(\bar{\mathbf p},\mathbf n)
  \oplus(\wedge^2\mathbf p,\mathbf1)\oplus(\wedge^2\bar{\mathbf p},\mathbf1).
\label{eq:STZ-tangent-decomposition}
\end{equation}
The generators in $(\mathbf p,\mathbf n)\oplus(\bar{\mathbf p},\mathbf n)$
mix the $2p$ blocks with the $n$ blocks of both $h^T$ and $h^Z$,
so both commutators in \eqref{eq:STZ-tilt-commutator} contribute. 
The generators in $(\wedge^2\mathbf p,\mathbf1)\oplus
(\wedge^2\bar{\mathbf p},\mathbf1)$ act within the $2p$ blocks, where
$h^T\propto\mathbbm1_{2p}$, so $[E_A,h^T]=0$ and only $h^Z$ varies.
We study the two cases in turn.

The bifundamental tilts carry indices $(a,\hat\imath)$ and
$(\bar a,\hat\imath)$, with $a$ a $U(p)$ fundamental index and
$\hat\imath$ an $O(n)$ index. Then
$[E_{a\hat\imath},h^T]\propto h_{2p}^{\cC_{p,\pm}}-h_n^{\cC_{p,\pm}}$ and
$[E_{a\hat\imath},h^Z]\propto\eta^{\cC_{p,\pm}}$, so \eqref{eq:STZ-tilt-commutator} gives
\begin{equation}
  \bt_{a\hat\imath}
  =
  2\big(h_{2p}^{\cC_{p,\pm}}-h_n^{\cC_{p,\pm}}\big)T_{a\hat\imath}
  -2\eta^{\cC_{p,\pm}}Z_{a\hat\imath},
  \qquad
  \bar\bt_{\bar a\hat\imath}
  =
  2\big(h_{2p}^{\cC_{p,\pm}}-h_n^{\cC_{p,\pm}}\big)T_{\bar a\hat\imath}
  -2\eta^{\cC_{p,\pm}}Z_{\bar a\hat\imath}.
  \label{eq:Cp-bifund-tilt-source}
\end{equation}
This is the Ward-identity derivation of the zero eigenvalues in the
dimension-two mixing problem 
\eqref{eq:STZ-bifund-Gamma}. In the same 
convention as \eqref{eq:Ct-Dn-general} we get
\begin{equation}
  C_{\bt_{\rm bifund}}^{\cC_{p,\pm}}
  =
  2\left[
    \big(h_{2p}^{\cC_{p,\pm}}-h_n^{\cC_{p,\pm}}\big)^2
    +\big(\eta^{\cC_{p,\pm}}\big)^2
  \right].
  \label{eq:Ct-Cp-STZ-bifund}
\end{equation}

The wedge tilts $\bt_\wedge$ \eqref{eq:STZ-twedge-dimensions} in the 
$(\wedge^2\mathbf p,\mathbf1)\oplus(\wedge^2\bar{\mathbf p},\mathbf1)$ 
representation carry antisymmetric indices $a,b$ and $\bar a,\bar b$. 
Now only $h^Z$ varies, so \eqref{eq:STZ-tilt-commutator} gives
\begin{equation}
  \bt_{ab}
  =
  2\eta^{\cC_{p,\pm}}Z_{ab},
  \qquad
  \bar\bt_{\bar a\bar b}
  =
  2\eta^{\cC_{p,\pm}}Z_{\bar a\bar b}.
  \label{eq:Cp-wedge-tilt-source}
\end{equation}
The complex
components inherit their normalization from the real projector
\eqref{eq:st-and-adj-projectors},
\begin{equation}
  \big\langle
    Z_{ab}(\tau)Z_{\bar c\bar d}(0)
  \big\rangle
  =
  \frac{
    \frac12\left(
      \delta_{a\bar c}\delta_{b\bar d}
      -\delta_{a\bar d}\delta_{b\bar c}
    \right)}
  {|\tau|^4}.
\end{equation}
Defining $C_{\bt_\wedge}^{\cC_{p,\pm}}$ via
\begin{equation}
  \big\langle\bt_{ab}(\tau)\bar\bt_{\bar c\bar d}(0)
  \big\rangle
  =
  \frac{C_{\bt_\wedge}^{\cC_{p,\pm}}
  \left(\delta_{a\bar c}\delta_{b\bar d}-\delta_{a\bar d}\delta_{b\bar c}\right)}
  {|\tau|^4},
  \label{eq:Ct-wedge-definition}
\end{equation}
we get
\begin{equation}
  C_{\bt_\wedge}^{\cC_{p,\pm}}
  =
  2\big(\eta^{\cC_{p,\pm}}\big)^2.
  \label{eq:Ct-Cp-STZ-wedge}
\end{equation}

The displacement is the $v$-variation of
\eqref{eq:STZ-deformed-action}. Using the descendant normalizations
\eqref{eq:uv-descendant-normalization} and the definition of $W_{ij,r}$, 
the derivative of $Z$ \eqref{eq:W-normalization},
\begin{equation}
  \bD_r
  =
  -\sqrt{2\Delta_S}\,h^{S,\cC_{p,\pm}}V_r
  -\sqrt{2\Delta_T}\,h^{T,\cC_{p,\pm}}_{ij}U_{ij,r}
  -\sqrt{2\Delta_Z}\,h^{Z,\cC_{p,\pm}}_{ij}W_{ij,r}.
\label{eq:STZ-D-source}
\end{equation}
The corresponding protected operator is the zero-eigenvalue vector of 
the vector mixing matrix \eqref{eq:STZ-singlet-vector-Gamma}. Its
components agree with the vector displayed below
\eqref{eq:STZ-singlet-vector-Gamma}.
Therefore
\begin{equation}
  C_{\bD}^{\cC_{p,\pm}}
  =
  2\Delta_S\big(h^{S,\cC_{p,\pm}}\big)^2
  +2\Delta_T h^{T,\cC_{p,\pm}}_{ij}h^{T,\cC_{p,\pm}}_{ij}
  +2\Delta_Z h^{Z,\cC_{p,\pm}}_{ij}h^{Z,\cC_{p,\pm}}_{ij}.
  \label{eq:CD-Cp-STZ-general}
\end{equation}
For the representative $\cC_{p,\pm}$ in
\eqref{eq:STZ-Un-On-ansatz}, with
$n=N-2p>0$, tracelessness gives
$h_n^{\cC_{p,\pm}}=-(2p/n)h_{2p}^{\cC_{p,\pm}}$ and hence
\begin{equation}
  C_{\bD}^{\cC_{p,\pm}}
  =
  2\Delta_S\big(h^{S,\cC_{p,\pm}}\big)^2
  +2\Delta_T\frac{2pN}{n}\big(h_{2p}^{\cC_{p,\pm}}\big)^2
  +4p\Delta_Z\big(\eta^{\cC_{p,\pm}}\big)^2.
  \label{eq:CD-Cp-STZ}
\end{equation}
For the pure $U(N/2)$ endpoint $\cC_{N/2}$ one instead uses
$h^{T,\cC_{N/2}}=0$ in \eqref{eq:STZ-pure-U-h} and the norm in
\eqref{eq:STZ-pure-U-norm}. For generic $\cC_{p,\pm}$, one may also eliminate
$h^{S,\cC_{p,\pm}}$ and $\eta^{\cC_{p,\pm}}$ using
\eqref{eq:STZ-hS-from-linear-constraint} and \eqref{eq:STZ-eta2}, with
$h_{2p}^{\cC_{p,\pm}}$ fixed by \eqref{eq:STZ-h2p-quadratic}.

\subsection{Two-point functions along flows}
\label{sec:DTflows}

Having determined the tilt and displacement operators at the conformal fixed points, we now study the two-point functions along the RG flows.
To be concrete, 
we look at the scalar--tensor system and restrict to a two-dimensional 
subspace in the space of coupling with $O(n)\times O(m)$ symmetry 
along the entire flow. This is the same subspace as in 
Section~\ref{sec:deformations-of-the-defect} and it includes 
$\cD_0$, $\cD_N$ and two points on the $\cD_n$ defect conformal 
manifold.

This space consists of the running scalar coupling $\bar h^S(s)$ 
with $s=\mu|\tau|$ and the running tensor coupling
\begin{equation}
  \bar h^T_{ij}(s)
  =\bar h_n(s)
  \diag\big(
  \underbrace{1,\cdots,1}_{n},
  \underbrace{-\tfrac{n}{m},\cdots,-\tfrac{n}{m}}_{m}\big).
  \label{eq:running-hn-hm-ansatz}
\end{equation}
In the rest of this subsection, all occurrences of the barred couplings 
$\bar h^S$, $\bar h^T_{ij}$ and $\bar h_n$ should be 
viewed as the running coupling, which is a function of $s$. Those 
without a bar are still the renormalised couplings at a scale $\mu$.

The running of the couplings $\bar h^S(s)$ and $\bar h_n(s)$ is 
controlled by the beta functions 
\begin{equation}
\begin{aligned}
  -s\frac{d\bar h^S(s)}{ds}
  &=\beta^S\equiv
  \gamma_S \bar h^S
  +\pi C_{SSS}(\bar h^S)^2
  +\pi C_{STT}\frac{nN}{m}\bar h_n^2,
  \\
  -s\frac{d\bar h_n(s)}{ds}
  &=\beta_n
  \equiv
  \gamma_T \bar h_n
  +2\pi C_{STT}\bar h^S \bar h_n
  +\pi C_{TTT}\frac{m-n}{m}\bar h_n^2.
  \end{aligned}
  \label{eq:distance-CS-hS-hn-hm}
\end{equation}
These carry an extra minus sign relative to the beta functions
\eqref{eq:singlet-h-beta-functions}, \eqref{eq:tensor-h-beta-functions},
which govern the running with the energy scale, 
$\beta=\mu\,dh/d\mu$. In \eqref{eq:distance-CS-hS-hn-hm} the
couplings are functions of the dimensionless distance
$s=\mu|\tau|$, and the scale probed at separation $|\tau|$ is
$1/|\tau|\propto1/s$. Increasing $s$ therefore lowers the energy,
$d\ln(\text{scale})=-d\ln s$, so that $s\,d\bar h/ds=-\beta$. 
Equivalently, the
flow runs from the UV ($s\to0$) to the IR ($s\to\infty$). 
Note that all couplings on the
right-hand side are evaluated at the same scale $s$.

In addition to the singlet and this particular component of the tensor, 
we follow the fields in the bifundamental representation of 
$O(n)\times O(m)$. In the absence of fields beyond $S$ and 
$T_{ij}$, this sector is
characterized by a single running anomalous dimension
$\gamma_{T_{\hat\imath\check\jmath}}(s)$, which
vanishes at the $\cD_n$ fixed-point \eqref{eq:tensor-gammas-generic}. 
Its value along the flow is also determined by the tensor beta functions
\eqref{eq:tensor-h-beta-functions}
\begin{equation}
  \gamma_{T_{\hat\imath\check\jmath}}(s)
  =
  \gamma_T+2\pi C_{STT}\bar h^S(s)
  +\pi C_{TTT}\frac{m-n}{m}\bar h_n(s).
  \label{eq:running-tilt-gamma}
\end{equation}

The tilt in \eqref{eq:t-explicit-from-variation} has the 
prefactor $\bar h_n-\bar h_m=\frac{N}{m}\bar h_n$, so its full evolution is
governed by \eqref{eq:running-tilt-gamma} together with the evolution 
of $\bar h_n$, which is
$-\beta_n$ \eqref{eq:distance-CS-hS-hn-hm}. These two exactly 
cancel each other. 
A similar argument applies to the displacement 
operator and the RG-improved two-point functions take the form
\begin{equation}
\begin{aligned}
  \big\langle
    \bt_{\hat\imath\check\jmath}(\tau)\bt_{\hat k\check l}(0)
  \big\rangle_{\rm RG}
  &=
  \frac{C_{\bt}(s,h^S,h_n)
  \delta_{\hat\imath\hat k}\delta_{\check\jmath\check l}}{|\tau|^4},
\\
  \big\langle\bD_r(\tau)\bD_s(0)\big\rangle_{\rm RG}
  &=
  \frac{C_{\bD}(s,h^S,h_n)\delta_{rs}}{|\tau|^6}.
\end{aligned}
\label{eq:protected-two-point-running-form}
\end{equation}
The overall power is the same as it is at the fixed points but there 
is further scale dependence in the 
normalisations $C_\bt$ and $C_\bD$ in the numerator.
Their values can be read off directly from 
squaring the expression \eqref{eq:t-explicit-from-variation}, 
\eqref{eq:D-explicit-from-variation} at the running 
couplings and using the fixed-point normalizations 
\eqref{eq:Ct-Dn-general}, \eqref{eq:CD-endpoint-general}, giving
\begin{equation}
\begin{aligned}
  \left\langle\bD_r(\tau)\bD_s(0)\right\rangle
  &=
  \frac{\delta_{rs}}{|\tau|^6}
  \left[
    2\Delta_S\big(\bar h^S(\mu|\tau|)\big)^2
    +2\Delta_T\frac{nN}{m}\bar h_n(\mu|\tau|)^2
  \right],
  \\
  \left\langle
    \bt_{\hat\imath\check\jmath}(\tau)
    \bt_{\hat k\check l}(0)
  \right\rangle
  &=
  \frac{
  2\frac{N^2}{m^2}\bar h_n(\mu|\tau|)^2
  \delta_{\hat\imath\hat k}\delta_{\check\jmath\check l}}
  {|\tau|^4},
  \label{eq:protected-two-point-CS-solution}
\end{aligned}
\end{equation}
At the order we consider, the normalization $C_{\bt}$ does not 
depend explicitly on the singlet coupling $\bar h^S$, as seen in
\eqref{eq:t-explicit-from-variation}. However, $\bar h^S$ still
affects the evolution of $C_{\bt}$ implicitly through the running
of $\bar h_n$ in \eqref{eq:distance-CS-hS-hn-hm}. 
$C_\bt$ and $C_\bD$ arise from two-point functions so 
satisfy the Callan--Symanzik equations
\begin{align}
&
\left[
s\frac{\partial}{\partial s}
+\beta^S\frac{\partial}{\partial h^S}
+\beta_n\frac{\partial}{\partial h_n}
\right]C_{\bD}(s,h^S,h_n)=0,
\label{eq:D-protected-two-point-CS}
\\[.5em]
&
\left[
s\frac{\partial}{\partial s}
+\beta^S\frac{\partial}{\partial h^S}
+\beta_n\frac{\partial}{\partial h_n}
\right]C_{\bt}(s,h^S,h_n)=0.
\label{eq:t-protected-two-point-CS}
\end{align}

The expressions for $C_\bt$ and $C_\bD$ 
above are valid up to $O(h^3,\gamma h^2)$ corrections. 
We now focus on the three specific flows 
considered in this paper.

First, let us examine the flow along the $O(N)$-symmetric 
trajectory, where the tensor coupling vanishes and 
there are no bifundamental fields to follow. 
The flow is governed by a single running coupling $\bar h^S(s)$ satisfying
\begin{equation}
  s\frac{d\bar h^S}{ds}
  =
  -\gamma_S \bar h^S
  -\pi C_{SSS}\big(\bar h^S\big)^2,
  \qquad
  \bar h^S(0)=0,
  \qquad
  \lim_{s\to\infty}\bar h^S(s)=h^{S,\cD_N}=-\frac{\gamma_S}{\pi C_{SSS}},
  \label{eq:D0-DN-running-coupling}
\end{equation}
with the explicit endpoint value from \eqref{eq:DN-h-couplings}. This 
equation is solved by
\begin{equation}
  \bar h^S(s)
  =
  \frac{h^S s^{-\gamma_S}}
  {1+\frac{h^S}{h^{S,\cD_N}}
 (s^{-\gamma_S}-1)}.
  \label{eq:D0-DN-running-solution}
\end{equation}
Here $h^S$ is the renormalized coupling at the scale $\mu$, entering as 
an integration constant, as a reference value $\bar h^S(1)=h^S$ at 
$s=\mu|\tau|=1$. 
The RG-improved displacement two-point function along this flow
\eqref{eq:protected-two-point-CS-solution} simplifies to
\begin{equation}
\begin{aligned}
  \big\langle\bD_r(\tau)\bD_s(0)\big\rangle_{\cD_0\to\cD_N}
  &=
  \frac{2\Delta_S\big(\bar h^S(\mu|\tau|)\big)^2
  \delta_{rs}}{|\tau|^6}
  +O(h^3,\gamma h^2).
  \label{eq:D0-DN-protected-two-point}
\end{aligned}
\end{equation}
It is easy to check that this solves the Callan-Symanzik equation 
\eqref{eq:D-protected-two-point-CS}. Related calculations in conformal perturbation 
theory of two-point functions of protected and unprotected 
operators along short RG flows using Callan--Symanzik equation are presented in \cite{Cappelli:1989yu,Karateev:2024skm}. 

The symmetry-breaking $\cD_0\to\cD_n$ flow is governed by
 the coupled evolution of $\bar h^S(s)$ and $\bar h_n(s)$, interpolating between the boundary conditions
\eqref{eq:hS-roots-sigma-gen-generic}, \eqref{eq:hn-hm-generic}
\begin{equation}
\begin{aligned}
  \bar h^S(0)&=0,
  &\qquad
  \lim_{s\to\infty}\bar h^S(s)&=h^{S,\cD_n},
  \\
  \bar h_n(0)&=0,
  &
  \lim_{s\to\infty}\bar h_n(s)&=h_n^{\cD_n}.
  \label{eq:D0-Dn-running-boundaries}
\end{aligned}
\end{equation}
$\bar h^S(s)$ and $\bar h_n(s)$ evolve as a coupled
pair in \eqref{eq:distance-CS-hS-hn-hm} and 
unlike the $O(N)$-symmetric case this system has no closed-form
solution. The profiles along the separatrix
\eqref{eq:D0-Dn-running-boundaries} may be obtained numerically and
substituted into \eqref{eq:protected-two-point-CS-solution}.

For the symmetry-restoring $\cD_n\to\cD_N$ flow, using the framework 
in Section~\ref{sec:deformations-of-the-defect}, the $O(N)$-symmetric 
endpoint is identified by $g_+^*$, $g_-^*$, so the flow is in between
\begin{equation}
  g_\pm(0)=0,
  \qquad
  \lim_{s\to\infty}g_\pm(s)=g_\pm^* .
  \label{eq:Dn-DN-Opm-rho-definition-running}
\end{equation}
The system may also be described by $g_+(s)$ and $\rho(s)=g_-(s)/g_+(s)$ 
\eqref{eq:dn-rho-definition}. The latter is a root of the cubic 
\eqref{eq:dn-rho-fixed-point-cubic} at all the fixed points. It satisfies
\begin{equation}
\lim_{s\to\infty}\rho(s)=\frac{g_-^*}{g_+^*}.
\label{eq:Dn-DN-rho-boundary}
\end{equation}
Its value at $s=0$ is not well defined, because $g_+(0)=g_-(0)=0$.

We write the two beta functions \eqref{eq:dn-two-singlet-beta-functions} for 
$g_\pm$ as flow equations for $g_+(s)$ and $\rho(s)$ as
\begin{equation}
\begin{aligned}
  s\frac{dg_+(s)}{ds}
  &=
  -\gamma_+^{\cD_n}g_+
  -\pi g_+^2
  \left(C_{+++}^{\cD_n}
    +2C_{++-}^{\cD_n}\rho 
    +C_{+--}^{\cD_n}\rho^2\right),
  \\
  s\frac{d\rho(s)}{ds}
  &=
  \big(\gamma_+^{\cD_n}-\gamma_-^{\cD_n}\big)\rho
  -\pi g_+
  \left[C_{++-}^{\cD_n}
  -\big(C_{+++}^{\cD_n}-2C_{+--}^{\cD_n}\big)\rho
  -\big(2C_{++-}^{\cD_n}-C_{---}^{\cD_n}\big)\rho^2
  -C_{+--}^{\cD_n}\rho^3
  \right].
\label{eq:Dn-DN-Opm-rho-flow}
\end{aligned}
\end{equation}
This coupled system is again not integrable. The cubic nonlinearity in
$\rho$ admits no closed-form solution, so $g_+(s)$ and $\rho(s)$ 
may be solved numerically between the endpoints
\eqref{eq:Dn-DN-Opm-rho-definition-running},
\eqref{eq:Dn-DN-rho-boundary}.

Finally, let us briefly mention the sum rules of~\cite{Baume:2024poj}, 
which apply to any integer dimension operator including 
tilts and displacements. 
For surface operators, with the normalizations in 
\eqref{eq:CD-Ct-def}, they are
\begin{equation}
\begin{aligned}
  C_{\bt}^{\rm UV}-C_{\bt}^{\rm IR}
  &=
  \frac{1}{16\pi}\int d^2\tau\,\tau^2
  \left(\tau^2\square-16\right)
  \big\langle
    \bt_{\hat\imath\check\jmath}(\tau)
    \bt_{\hat\imath\check\jmath}(0)
  \big\rangle_{\rm RG}.
\\
  C_{\bD}^{\rm UV}-C_{\bD}^{\rm IR}
  &=
  \frac{1}{24\pi}\int d^2\tau\,(\tau^2)^2
  \left(\tau^2\square-36\right)
  \big\langle\bD_r(\tau)\bD_r(0)\big\rangle_{\rm RG}.
  \label{eq:surface-BMP-CD-Ct}
\end{aligned}
\end{equation}
Given an explicit flow, like \eqref{eq:D0-DN-running-solution}, 
\eqref{eq:D0-DN-protected-two-point}, 
the integrand can be computed explicitly. It is not clear what 
the utility of this is, since the proof of the identity is by 
rewriting \eqref{eq:surface-BMP-CD-Ct} 
in terms of $s$, where they become total derivatives,
\begin{equation}
\begin{aligned}
  \frac{1}{8}\int_0^\infty\frac{ds}{s}
  \left(s\frac{d}{ds}\right)
  \left(s\frac{d}{ds}-8\right)
  C_{\bt}(s,h^S,h_n)
  &=
  \left(\frac{s}{8}\frac{d}{ds}-1\right)C_{\bt}(s,h^S,h_n)\bigg|_0^\infty,
\\
  \frac{1}{12}\int_0^\infty\frac{ds}{s}
  \left(s\frac{d}{ds}\right)
  \left(s\frac{d}{ds}-12\right)
  C_{\bD}(s,h^S,h_n)
  &=
  \left(\frac{s}{12}\frac{d}{ds}-1\right)C_{\bD}(s,h^S,h_n)\bigg|_0^\infty,
  \label{eq:BMP-radial-reduction}
\end{aligned}
\end{equation}
immediately reproducing the endpoint values 
\eqref{eq:Ct-Dn-tensor-norm}, \eqref{eq:CD-Dn-general}. It should 
be noted that the formulas hold even if the protected operator 
is absent in one of the limits, with the replacement $C_\bt=0$ 
or $C_\bD=0$. 
This is relevant for all the flows originating in the bulk (a.k.a 
the trivial defect), or in the $O(N)$ restoring flow.

\section{Examples}
\label{sec:examples}

In this section we specialise the calculations in the previous 
sections to concrete 
models of perturbative CFTs: The usual WF $O(N)$ model, its long-range 
variant, the chiral $O(N)\times O(2)$ model, and $O(N)$ 
tricritical model in $d=3-\varepsilon$. We heavily rely on the 
constructions in the previous sections, plug in the values for these 
theories when it leads to simplifications; where the CFT data is 
anyhow complicated, one may as well retain the generic expressions.

We also discuss what happens to other fields that exist in some of 
these models and are absent in the general construction. 
We do not plug in the data for operators of dimension close to
three, as those are rather cumbersome and unintuitive and 
anyhow very easy to reproduce from
Sections~\ref{sec:dimension-three-operators}
and~\ref{sec:STZ-near-three}.

\subsection{Critical \texorpdfstring{$O(N)$}{O(N)} 
Wilson--Fisher theory in \texorpdfstring{$d=4-\varepsilon$}{4-epsilon}}
\label{sec:WFON}

We start with the usual Wilson--Fisher (WF) $O(N)$ model 
in $d=4-\varepsilon$, for which the defects were already 
studied in the $\varepsilon$ expansion 
in~\cite{Trepanier:2023tvb, 
Raviv-Moshe:2023yvq, Giombi:2023dqs, Diatlyk:2024ngd}.

The CFT data can be found for example in \cite{Henriksson:2022rnm} 
and the dimensions of $S$ and $T_{ij}$ differ from two by
\begin{equation}
  \gamma_S^{\rm WF}
  =
  -\frac{6}{N+8}\varepsilon+O(\varepsilon^2),
  \qquad
  \gamma_T^{\rm WF}
  =
  -\frac{N+6}{N+8}\varepsilon+O(\varepsilon^2),
  \qquad
  \Delta_j^{\rm WF}
  =3-\varepsilon.
\label{eq:wf-uv-gammas}
\end{equation}
It is easy to check that, at leading order, the structure constants 
for unit normalised operators are
\begin{equation}
  C_{SSS}^{\rm WF}
  =
  C_{STT}^{\rm WF}
  =
  2\sqrt{\frac2N}+O(\varepsilon^2),
  \qquad
  C_{TTT}^{\rm WF}
  =
  2\sqrt2+O(\varepsilon^2).
\label{eq:wf-scalar-constants}
\end{equation}

To reproduce the CFT data of operators of dimension close to three 
based on Section~\ref{sec:dimension-three-operators}, one requires 
information about the current $j$. As mentioned, we do not 
write explicitly this data, yet we provide here the 
required extra information. 
In the free and critical WF theory the central charges 
are \cite{Petkou:1995vu}
\begin{equation}
  C_j^{\rm free}=4(d-2)\kappa_d^2=\frac{1}{2\pi^4}+O(\varepsilon)\,,
\qquad
  C_j^{\rm WF}
  =C_j^{\rm free}
  \left[1-\frac{3(N+2)}{4(N+8)^2}\varepsilon^2+O(\varepsilon^3)\right].
  \label{eq:wf-current-central-charge}
\end{equation}
Using \eqref{eq:ward-CjU-sqrt}, we find
\begin{equation}
  C_{TjU}^{\rm WF}=\frac{1}{\pi^2}+O(\varepsilon).
\label{eq:wf-CTjU}
\end{equation}
All this CFT data can be easily found in the literature, see 
e.g.~\cite{Henriksson:2022rnm}. In fact these are known to 
far higher order in $\varepsilon$.

In the current analysis we just need the information above, 
but for completeness we recall that the 
basic fields are $\phi_i$ and at leading order, 
the unit-normalised operators of dimension near two are
\begin{equation}
  S=\frac{1}{\sqrt{2N}\,\kappa_d}\,\phi_i\phi_i,
  \qquad
  T_{ij}=\frac{1}{\sqrt2\,\kappa_d}
  \left(\phi_i\phi_j-\frac{\delta_{ij}}{N}\phi_k\phi_k\right).
\label{eq:wf-unit-bilinears}
\end{equation}
Here we use the canonical propagators
\begin{equation}
  \langle\phi_i(x)\phi_j(0)\rangle
  =\frac{\kappa_d\delta_{ij}}{x^{d-2}},
\qquad
\kappa_d
  =\frac{\Gamma(d/2-1)}{4\pi^{d/2}},
  \qquad
  d=4-\varepsilon.
\label{eq:wf-field-propagator}
\end{equation}
Likewise, the canonical $O(N)$ Noether current is
\begin{equation}
  j_{ij,\mu}
  =\phi_i\partial_\mu\phi_j-\phi_j\partial_\mu\phi_i,
\label{eq:wf-noether-current}
\end{equation}
and Wick contraction gives
\begin{equation}
  \big\langle j_{ij,r}(x)j_{kl,s}(0)\big\rangle_{\rm free}
  =4(d-2)\kappa_d^2\frac{P_{ij,kl}\delta_{rs}}{x^{2d-2}}\,,
\label{eq:wf-current-two-point-free}
\end{equation}
reproducing the free-field expression in 
\eqref{eq:wf-current-central-charge}.

Of course, at leading order in $\varepsilon$, we may simply set 
$\kappa_4=1/4\pi^2$. At higher order in $\varepsilon$, if the 
theory is not free, one anyhow has to renormalise the operators.

\subsubsection{Fixed points}
Using the results of Section~\ref{sec:fixed-points} and plugging in 
the CFT data above, the $O(N)$ symmetric fixed 
point is
\begin{equation}
  \cD_N^{\rm WF}:
  \qquad
  h^{S,\cD_N,{\rm WF}}=\frac{3\sqrt{2N}}{2\pi(N+8)}\varepsilon,
  \qquad
  h^{T,\cD_N,{\rm WF}}_{ij}=0,
  \label{eq:DN-WF-h-couplings}
\end{equation}
When the tensor ${\cal I}_{ij,kl,pq}$ of \eqref{eq:st-cubic-invariant} is
nonzero and $n\neq m$, the two roots \eqref{eq:hS-roots-sigma-gen-generic} are labeled $\cD_n$ and $\cD_m$. 
The discriminant \eqref{eq:discriminant} is
\begin{equation}
  \sigma_n=\sqrt{9-nm}.
  \label{eq:WF-fixed-point-shorthands}
\end{equation}
Then (recall $\nu=n-m$)
\begin{equation}
  \cD_n^{\rm WF}:\qquad
  h^{S,\cD_n,{\rm WF}}
  =\frac{\sqrt{2N}\varepsilon}{4\pi N(N+8)}
  [3N+2nm+\nu\sigma_n],
  \label{eq:Dn-WF-hS}
\end{equation}
and the tensor eigenvalues in \eqref{eq:hn-hm-ansatz} are
\begin{equation}
  h_n^{\cD_n,{\rm WF}}
  =
  -\frac{\sqrt2\,m\varepsilon}{4\pi N(N+8)}
  (\nu-2\sigma_n),
  \qquad
  h_m^{\cD_n,{\rm WF}}
  =
  \frac{\sqrt2\,n\varepsilon}{4\pi N(N+8)}
  (\nu-2\sigma_n).
  \label{eq:Dn-WF-hn-hm}
\end{equation}

The WF fixed points are real when
\begin{equation}
  \sigma_n^2=9-nm\ge0.
  \label{eq:WF-reality-condition}
\end{equation}
The cases of $\sigma_n=0$, 
which are $N=6$ with $n=m=3$ and $N=10$ with $(n,m)=(1,9), (9,1)$, 
were treated in \cite{deSabbata:2024xwn}. 
In all these cases the degeneracy is lifted at the next order and 
the fixed points remain real. The values of the couplings $h$ and 
the dimensions get 
corrections at order $\varepsilon^{3/2}$, see the appendix of 
\cite{deSabbata:2024xwn}.

When $n=m=N/2$, the fixed point is obtained by substituting
$\nu=0$ into \eqref{eq:Dn-WF-hS} and \eqref{eq:Dn-WF-hn-hm} 
and the reality
condition \eqref{eq:WF-reality-condition} becomes $N\le6$. For $N=2$ the
tensor ${\cal I}_{ij,kl,pq}$ \eqref{eq:st-cubic-invariant} vanishes, so
the same couplings are obtained from the $C_{TTT}=0$ equations
\eqref{eq:hS-O2} and \eqref{eq:hT-O2}.

The fixed points above are the same as found 
in~\cite{Trepanier:2023tvb, Raviv-Moshe:2023yvq, 
Giombi:2023dqs, Diatlyk:2024ngd}.

\subsubsection{Operators of dimension near two}
\label{sec:WF-dim2}

To find the spectrum, we substitute the UV data \eqref{eq:wf-uv-gammas},
\eqref{eq:wf-scalar-constants}, \eqref{eq:Dn-WF-hS}, and
\eqref{eq:Dn-WF-hn-hm} into \eqref{eq:tensor-gammas-generic} to 
find the dimensions in Table~\ref{tab:Dn-near-two-dimensions}. 
They agree with those derived in~\cite{Trepanier:2023tvb, 
Raviv-Moshe:2023yvq, Giombi:2023dqs}.

For the two singlets, substituting \eqref{eq:wf-uv-gammas},
\eqref{eq:wf-scalar-constants}, \eqref{eq:Dn-WF-hS}, and
\eqref{eq:Dn-WF-hn-hm} into \eqref{eq:mixed-singlet-Gamma-generic} gives
\begin{equation}
  \Gamma_{\rm singlet}^{\cD_n,{\rm WF}}
  =\frac{\varepsilon}{N(N+8)}
  \begin{pmatrix}
    2(2nm+\nu\sigma_n)
    & -2\sqrt{nm}(\nu-2\sigma_n)
    \\
    -2\sqrt{nm}(\nu-2\sigma_n)
    & \nu(\nu-2\sigma_n)
  \end{pmatrix}.
  \label{eq:WF-mixed-singlet-Gamma}
\end{equation}
Its two eigenvalues are
\begin{equation}
  \gamma_\pm^{\cD_n,{\rm WF}}
  =
  \frac{\varepsilon}{2(N+8)}
  \left[N\pm\sqrt{N^2+16\sigma_n^2-8\nu\sigma_n}\right].
  \label{eq:WF-mixed-singlets-gamma}
\end{equation}

When $n=m=N/2$, the dimensions are obtained by setting 
the WF values in \eqref{eq:n-half-tensor-gammas-generic} and 
$\eqref{eq:n-half-mixed-singlets-gamma-generic}$. In fact, 
this is the same as setting 
$\nu=0$ and $\sigma_n^2=9-N^2/4$ in Table~\ref{tab:Dn-near-two-dimensions} 
and \eqref{eq:WF-mixed-singlets-gamma}. 
If $C_{TTT}=0$, one should use
\eqref{eq:hS-O2} and \eqref{eq:hT-O2} and diagonalize
\eqref{eq:Gamma-definition} directly.

For the WF theory, the condition for the stability of the 
symmetric defect $\cD_N$, $\gamma_TC_{SSS}>2\gamma_SC_{STT}$ is 
simply $\gamma_T>2\gamma_S$, or using \eqref{eq:wf-uv-gammas} 
$N<6$ \cite{Pannell:2024hbu}. 
For $N=6$ one has $\gamma_{T}^{\cD_6,{\rm WF}}=0$, 
but using the 
results in the appendix of \cite{deSabbata:2024xwn}, this 
is lifted in the next order, becoming irrelevant, so also 
$N=6$ is stable. 
For $N>6$ some symmetry-breaking defects are complex 
and for $N>10$ all of them are.

The spectrum of operators of dimension close to three can be 
easily gotten from Section~\ref{sec:dimension-three-operators}.

\subsubsection{Displacements and tilts}
\label{sec:WF-displacements-and-tilts}

At the $O(N)$ symmetric $\cD_N$ fixed point, there 
are no tilt operators, or equivalently, 
$C_\bt^{\cD_N,\rm WF}$ vanishes. The displacement 
normalisation at this fixed point is given in 
\eqref{eq:CD-DN-general}. When evaluated with the WF 
data in \eqref{eq:wf-uv-gammas} and \eqref{eq:wf-scalar-constants}, 
we find
\begin{equation}
  C_{\bD}^{\cD_N,{\rm WF}}
  =
  \frac{18N}{\pi^2(N+8)^2}\varepsilon^2
  +O(\varepsilon^3).
  \label{eq:CD-WF-evaluated}
\end{equation}

For the symmetry-breaking fixed point $\cD_n^{\rm WF}$, the tilt normalisation is given by \eqref{eq:Ct-Dn-general}.
Using \eqref{eq:Dn-WF-hn-hm},
\begin{equation}
  h_n^{\cD_n,{\rm WF}}-h_m^{\cD_n,{\rm WF}}
  =
  -\frac{(\nu-2\sigma_n)}{2\sqrt{2}\pi(N+8)}\varepsilon\,,
  \label{eq:Dn-WF-h-difference}
\end{equation}
the tilt normalisation is
\begin{equation}
  C_{\bt}^{\cD_n,{\rm WF}}
  =
  \frac{(\nu-2\sigma_n)^2}{4\pi^2(N+8)^2}\varepsilon^2
  +O(\varepsilon^3).
  \label{eq:Ct-Dn-WF-evaluated}
\end{equation}
For the displacement normalisation, we need the tensor norm, which follows from \eqref{eq:Dn-WF-hn-hm} and is
\begin{equation}
  h^{T,\cD_n,{\rm WF}}_{ij}h^{T,\cD_n,{\rm WF}}_{ij}
  =
  \frac{nm(\nu-2\sigma_n)^2}{8\pi^2N(N+8)^2}\varepsilon^2.
  \label{eq:hTnorm-Dn-WF}
\end{equation}
Since the fixed-point couplings are of order $\varepsilon$, the
dimensions $\Delta_S$ and $\Delta_T$ in the general formula for the displacement normalisation \eqref{eq:CD-endpoint-general} may be replaced by their leading value
$2$ at this order. Combining this with the value of $h^{S,\cD_n,\rm WF}$ in \eqref{eq:Dn-WF-hS} gives
\begin{equation}
  C_{\bD}^{\cD_n,{\rm WF}}
  =
  \frac{3\varepsilon^2}{\pi^2(N+8)^2}
  \left(3N+2nm+\nu\sigma_n\right)
  +O(\varepsilon^3).
  \label{eq:CD-Dn-WF-evaluated}
\end{equation}

\subsection{Long range Wilson--Fisher theory in \texorpdfstring{$d=4-\varepsilon$}{4-epsilon}}
\label{sec:nonlocal4-ep}

A generalisation of the usual WF $O(N)$ model is the long range WF 
CFT~\cite{Fisher:1972zz} with nonlocal action
\begin{equation}
  S_{\rm LR}
  =\frac{\mathcal N_\delta}{2}
  \int d^d x\, d^d y\,
  \frac{\phi_a(x)\phi_a(y)}{|x-y|^{d+2-\delta}}
  +\frac{\lambda_0}{4!}
  \int d^d x\,\big(\phi_a\phi_a\big)^2,
\qquad
\mathcal N_\delta
  =\frac{2^{2-\delta}
  \Gamma\big(\frac{d+2-\delta}{2}\big)}
  {\pi^{d/2}\Gamma\big(-1+\frac{\delta}{2}\big)}.
\label{eq:NL-action}
\end{equation}
We take the nonlocality parameter as $\delta=\alpha\varepsilon$ 
and an interacting fixed point exists for $0<\alpha<1/2$.

In this case (see e.g.~\cite{Benedetti:2020rrq})
\begin{equation}
\gamma_S^\text{LR}
=-\frac{6}{N+8}\varepsilon
-\frac{N-4}{N+8}\delta
+O(\varepsilon^2),
\qquad
\gamma_T^\text{LR}
=-\frac{N+6}{N+8}\varepsilon
+\frac{N+4}{N+8}\delta
+O(\varepsilon^2).
\label{eq:LR-WF-gammas}
\end{equation}
For $N\geq2$, both  $S$ and $T$ are relevant deformations 
of the trivial defect for every value of $\alpha$ for which the 
interacting bulk fixed point exists.

The structure constants at leading order are as in the regular 
WF theory \eqref{eq:wf-scalar-constants} with all the resulting 
simplifications as in that case.

This model is being actively studied also for its connection 
to massive field theories in $AdS$. 
Defects in the long-range WF theory were previously studied 
in~\cite{Bianchi:2024eqm}. Using the mapping to their notation 
$\hat\varepsilon^\text{BCdS}$ and $\kappa^\text{BCdS}$ 
\begin{equation}
  \hat\varepsilon^\text{BCdS}=\varepsilon,
  \qquad
  \kappa^\text{BCdS}=1-2\alpha,
\end{equation}
our general analysis reproduces their results for the $O(N)$ 
invariant fixed point coupling $h^{S,\cD_N,{\rm LR}}$ 
\eqref{eq:DN-LR-h-couplings} and the dimension
$\gamma_S^{\cD_N,{\rm LR}}$ in 
\eqref{eq:DN-LR-stability-dimensions}. Our results for the 
other dimensions as well as the symmetry breaking defects are new.

One could worry whether usual conformal 
perturbation theory is valid for a nonlocal theory 
and whether there are 
extra nonlocal counterterms that render the usual treatment invalid. 
We attempt to circumvent this by not allowing for extra nonlocal 
terms on the defect, and our results agree with those 
of~\cite{Bianchi:2024eqm}, but this question deserves further study.
We should also mention that despite the theory being nonlocal, its 
defects are expected to have displacement and tilt operators, 
as recently proved in~\cite{Bianchi:2026sax, Qiao:2026ijh}.

One variance from the local theory is the lack of a local conserved 
current $j_{ij,r}$. It is replaced with the antisymmetric 
vector bilinear, with free-field realisation
\begin{equation}
  j^{\rm LR}_{ij,\mu}
  =\frac{1}{\cN_{j^{\rm LR}}}
  (\phi_i\partial_\mu\phi_j-\phi_j\partial_\mu\phi_i).
\label{eq:JNL}
\end{equation}
It is not conserved. At leading order in the long-range theory
\begin{equation}
  \gamma_{j^{\rm LR}}
  =-\varepsilon+\delta=(\alpha-1)\varepsilon 
  +O(\varepsilon^2).
\end{equation}
Writing
\begin{equation}
  \langle \phi_i(x)\phi_j(0)\rangle
  =\frac{\kappa_{d,\delta}\delta_{ij}}{x^{d-2+\delta}},
\qquad
  \kappa_{d,\delta}
  =\frac{\Gamma\big(\frac{d-2+\delta}{2}\big)}
  {4^{1-\delta/2}\pi^{d/2}\Gamma\big(1-\frac{\delta}{2}\big)}
\end{equation}
Wick contraction gives
\begin{equation}
  \big\langle j^{\rm LR}_{ij,r}(x)j^{\rm LR}_{kl,s}(0)\big\rangle
  =
  \frac{4(d-2+\delta)\kappa_{d,\delta}^2}{\cN_{j^{\rm LR}}^2}
  \frac{P_{ij,kl}\delta_{rs}}{x^{2d-2+2\delta}},
\end{equation}
with $P_{ij,kl}$ the usual projector 
\eqref{eq:st-and-adj-projectors}. 
Insisting now on unit normalization fixes
\begin{equation}
  \cN_{j^{\rm LR}}^2
  =4(d-2+\delta)\kappa_{d,\delta}^2 .
\end{equation}
The same Wick contraction gives the replacement for the Ward-identity
coefficient,
\begin{equation}
  C_{Tj^{\rm LR}U}^{\rm LR}
  =\sqrt{d-2+\delta}
  +O(\varepsilon,\delta)
  =\sqrt{2}
  +O(\varepsilon,\delta),
\end{equation}

\subsubsection{Fixed points}

As far as the CFT data is concerned, at the relevant order, 
the operators in the long-range WF theory have dimension shifted 
by the parameter $\delta$ \eqref{eq:NL-action}, but the structure constants 
are as in the local theory \eqref{eq:wf-scalar-constants}, 
with all the resulting simplifications. 

The $O(N)$ symmetric fixed point is
\begin{equation}
  \cD_N^{\rm LR}:
  \qquad
  h^{S,\cD_N,{\rm LR}}
  =
  \frac{\sqrt{2N}}{4\pi(N+8)}
  (6+(N-4)\alpha)\varepsilon,
  \qquad
  h^{T,\cD_N,{\rm LR}}_{ij}=0 .
  \label{eq:DN-LR-h-couplings}
\end{equation}
As for the other fixed points, with $n\neq m$, the discriminant is
\begin{equation}
  \sigma_n^2
  =
  \frac{9\nu^2}{N^2}
  -
  \frac{36nm}{N^2}
  \frac{
  (N+6-(N+4)\alpha)
  (N-6-(3N-4)\alpha)}
  {(6+(N-4)\alpha)^2}.
  \label{eq:LR-discriminant}
\end{equation}
The corresponding fixed point couplings are
\begin{align}
  h^{S,\cD_n,{\rm LR}}
  &=
  \frac{\varepsilon}{4\sqrt2\,\pi N^{3/2}(N+8)}
  \left[4nm(N+6-(N+4)\alpha)
    +\left(\nu^2+\frac{N\nu}{3}\sigma_n\right)
    (6+(N-4)\alpha)\right],
\nonumber\\
  h_n^{\cD_n,{\rm LR}}
  &=
  \frac{m}{\nu}
  \frac{\gamma_T^{\rm LR}
  +2\pi C_{STT}^{\rm WF}h^{S,\cD_n,{\rm LR}}}
  {\pi C_{TTT}^{\rm WF}},
  \qquad
  h_m^{\cD_n,{\rm LR}}
  =
  -\frac{n}{m}h_n^{\cD_n,{\rm LR}} .
  \label{eq:Dn-LR-h-couplings}
\end{align}
For $n=m=N/2$, the couplings are obtained from
\eqref{eq:n-half-hS} and \eqref{eq:hT-half} with
\eqref{eq:wf-scalar-constants} and the dimensions in
\eqref{eq:LR-WF-gammas}.

\subsubsection{Operators of dimension near two}
\label{sec:near2-LR}

At the $O(N)$ symmetric long-range WF fixed point,
\eqref{eq:DN-stability-dimensions} gives
\begin{equation}
  \gamma_S^{\cD_N,{\rm LR}}
  =
  \frac{6\varepsilon+(N-4)\delta}{N+8},
  \qquad
  \gamma_T^{\cD_N,{\rm LR}}
  =
  \frac{(6-N)\varepsilon+(3N-4)\delta}{N+8}.
  \label{eq:DN-LR-stability-dimensions}
\end{equation}
For $n\neq m$, the dimensions of the tensors 
$T_n$ and $T_m$ are obtained by inserting
\eqref{eq:Dn-LR-h-couplings} into
\eqref{eq:tensor-gammas-generic}, namely
\begin{equation}
\begin{aligned}
  \gamma_{T_n}^{\cD_n,{\rm LR}}
  &=
  \frac{N}{\nu}
  \left(
  \gamma_T^{\rm LR}
  +2\pi C_{STT}^{\rm WF}h^{S,\cD_n,{\rm LR}}\right),
\\
  \gamma_{T_m}^{\cD_n,{\rm LR}}
  &=
  -\frac{N}{\nu}
  \left(
  \gamma_T^{\rm LR}
  +2\pi C_{STT}^{\rm WF}h^{S,\cD_n,{\rm LR}}\right),
  \qquad
  \gamma_\bt^{\cD_n,{\rm LR}}=0 .
  \label{eq:tensor-gammas-LR}
\end{aligned}
\end{equation}
The anomalous dimensions of the two singlet operators 
are the eigenvalues of
\eqref{eq:mixed-singlet-Gamma-generic}, with the WF structure constants
and the long-range fixed point couplings. Explicitly, the matrix is
\begin{equation}
  \Gamma_{\rm singlet}^{\cD_n,{\rm LR}}
  =
  \begin{pmatrix}
    \displaystyle
    \gamma_S^{\rm LR}
    +4\pi\sqrt{\frac2N}\,h^{S,\cD_n,{\rm LR}}
    &
    \displaystyle
    4\pi\sqrt{\frac2N}\,
    \big|h^{T,\cD_n,{\rm LR}}\big|
    \\[.5em]
    \displaystyle
    4\pi\sqrt{\frac2N}\,
    \big|h^{T,\cD_n,{\rm LR}}\big|
    &
    \displaystyle
    -\gamma_T^{\rm LR}
    -4\pi\sqrt{\frac2N}\,h^{S,\cD_n,{\rm LR}}
  \end{pmatrix}.
  \label{eq:LR-mixed-singlet-Gamma}
\end{equation}
For $n=m=N/2$, the same substitution should be made in
\eqref{eq:n-half-tensor-gammas-generic} and
\eqref{eq:n-half-mixed-singlets-gamma-generic}.

Within the range $0<\alpha<1/2$ where the interacting bulk fixed
point exists, the symmetric defect $\cD_N^{\rm LR}$ is stable when
\begin{equation}
 \max\left(\frac{N-6}{3N-4},0\right)<\alpha<\frac{1}{2}.
 \label{eqn:alpha-stability}
\end{equation}
In this range, the other fixed points are real by
\eqref{eq:discriminant-rewritten}. In particular,
$\alpha\in[1/3,1/2)$ lies in this range for all $N\geq2$.

For $N\geq6$, at the lower limit of the interval \eqref{eqn:alpha-stability} 
$\cD_N^{\rm LR}$ becomes marginal, 
i.e.\ $\gamma_T^{\cD_N,{\rm LR}}=0$ \eqref{eq:DN-LR-stability-dimensions}. 
For even $N$ it also collides with the $\cD_{N/2}^{\rm LR}$ fixed point. 

All symmetry-breaking fixed points are saddles by the analysis 
of~\cite{Pannell:2024hbu} and 
Section~\ref{sec:real-and-stable}. For $n\neq m$, the 
$\sigma_n$ \eqref{eq:LR-discriminant} vanishes 
when
\begin{equation}
 \alpha
 =
 \frac{2\left((N+8)\sqrt{nm}-(4nm+3N-12)\right)}
 {(N-4)^2-16nm}.
\label{eq:LR-collision-values}
\end{equation}
There are no solutions in the allowed interval for $N\leq6$ and
otherwise the solutions are (up to $n\leftrightarrow m$)
\begin{equation}
\begin{aligned}
7\leq N\leq10:
\qquad
n=2,\cdots,\left\lfloor\frac{N-1}{2}\right\rfloor,
\\
N\geq11:
\qquad
n=1,\cdots,\left\lfloor\frac{N-1}{2}\right\rfloor.
\end{aligned}
\end{equation}
Note that the numerator and denominator of \eqref{eq:LR-collision-values} 
vanish together when $(N-4)^2=16nm$. In such cases the 
quadratic for $\alpha$ obtained by setting
$\sigma_n^2=0$ in \eqref{eq:LR-discriminant}
degenerates and has a single root
\begin{equation}
 \alpha=\frac{nm-9}{4nm+3N-12},
\label{eq:LR-collision-degenerate}
\end{equation}
the smallest such case being $N=68$, $n=4$ or $n=64$, 
with $\alpha=13/64$.

All the above mentioned degeneracies are based on the values at the 
leading-order solution and must be
resolved by higher-order perturbation theory, as explained below
\eqref{eq:discriminant}.

The spectrum of transverse vectors can be easily found
from Section~\ref{sec:dimension-three-operators}.

\subsubsection{Displacements and tilts}
\label{sec:LR-displacements-and-tilts}

As already stated, despite the theory being nonlocal, its 
defects should have protected displacement and tilt 
operators~\cite{Bianchi:2026sax, Qiao:2026ijh}.

$C_\bD$ and $C_\bt$ are again fixed by plugging 
\eqref{eq:DN-LR-h-couplings}, \eqref{eq:Dn-LR-h-couplings} into 
\eqref{eq:CD-endpoint-general}, \eqref{eq:Ct-Dn-general}.
In fact, for $O(N)$ symmetric defect, 
this exactly mirrors the one-loop mechanism described 
in~\cite{Qiao:2026ijh}, where in the notation there, 
for a defect generated by weakly relevant scalar operators 
$\cO_A$, the protected displacement eigenvector is proportional to 
$\lambda_*^A\partial_i\cO_A$, or in our notation 
$\bD_r=-\sqrt{2\Delta_S}\, h^{S,\cD_N,{\rm LR}}V_r$
\eqref{eq:D-explicit-from-variation}, leading to
\begin{equation}
  C_{\bD}^{\cD_N,{\rm LR}}
  =
  2\Delta_S\big(h^{S,\cD_N,{\rm LR}}\big)^2
  =
  4\big(h^{S,\cD_N,{\rm LR}}\big)^2
  +O(\varepsilon^3).
  \label{eq:CD-DN-LR-general-evaluated}
\end{equation}
The same result can be inferred from the fixed-point coupling of
\cite{Bianchi:2024eqm}, after translating the coupling to the 
unnormalised $\phi_a\phi_a$ there to $h^S$.

The tilt normalisation $C_\bt^{\cD_N,\rm LR}$ vanishes at the symmetric fixed point, and the displacement normalisation is
\begin{equation}
  C_{\bD}^{\cD_N,{\rm LR}}
  =
  \frac{N[6+(N-4)\alpha]^2}{2\pi^2(N+8)^2}\varepsilon^2
  +O(\varepsilon^3).
\label{eq:Ct-CD-DN-LR-evaluated}
\end{equation}
For the symmetry-breaking fixed point $\cD_n^{\rm LR}$ with $n\neq m$, 
the tilt normalisation obtained from \eqref{eq:Dn-LR-h-couplings} and \eqref{eq:Ct-Dn-general} is
\begin{equation}
  C_{\bt}^{\cD_n,{\rm LR}}
  =
  \frac{\varepsilon^2}{4\pi^2(N+8)^2}
  \left[\nu(2\alpha-1)
  +\frac{\sigma_n}{3}\big(6+(N-4)\alpha\big)
  \right]^2
  +O(\varepsilon^3).
  \label{eq:Ct-Dn-LR-evaluated}
\end{equation}
Using the discriminant relation \eqref{eq:LR-discriminant}, the 
displacement normalisation can be written as
\begin{equation}
\begin{aligned}
  C_{\bD}^{\cD_n,{\rm LR}}
  & =
  \frac{\varepsilon^2}{12\pi^2(N+8)^2}
  \big(6+(N-4)\alpha\big)
  \Big[3N\big(N+6-(N+4)\alpha\big)
  +3(2\alpha-1)\nu^2
  \\
  &\hspace{8em}
  +\nu\sigma_n\big(6+(N-4)\alpha\big)\Big]
  +O(\varepsilon^3).
\end{aligned}
\label{eq:CD-Dn-LR-evaluated}
\end{equation}
This expression is analytic in $\nu$, so the $n=m=N/2$ result is 
simply its $\nu=0$ limit.

\subsection{The chiral 
\texorpdfstring{$O(N)\times O(2)$}{O(N) x O(2)} model}
\label{sec:N2}

There are several variants of the WF theory with symmetry beyond 
$O(N)$ including the $MN$ model, with $M\times N$ fundamental fields, 
the biconical $O(N)\times O(M)$ model with $N+M$ fundamental fields 
and the chiral $O(N)\times O(M)$ model with $N\times M$ fields. See 
\cite{Osborn:2017ucf, Henriksson:2020fqi, Henriksson:2021lwn} 
for some recent works on them.

To avoid making the discussion too general, we choose to 
focus on a simple representative, the chiral $O(N)\times O(2)$ model. 
The theory has a real bifundamental field $\phi_{ia}$, 
where $i=1,\cdots,N$ and $a=1,2$. In the $\varepsilon$ 
expansion, one takes the action
\begin{equation}
  \mathcal L
  =
  \frac12 \partial_\mu\phi_{ia}\partial^\mu\phi_{ia}
  +\frac{\lambda}{4!}(\phi_{ia}\phi_{ia})^2
  +\frac{\zeta}{4!}
  \left(
    \phi_{ia}\phi_{ib}\phi_{ja}\phi_{jb}
    -\phi_{ia}\phi_{ia}\phi_{jb}\phi_{jb}
  \right) .
\label{eq:N2-model-lagrangian}
\end{equation}
The two quartic couplings allow, besides the Gaussian and $O(2N)$
Wilson--Fisher fixed points, two fully interacting fixed points usually
called chiral and antichiral~\cite{Kawamura:1988zz}.

Defining
\begin{equation}
  R=N^2-24N+48,
  \qquad
  D=2(N+4)(N-3).
\label{eq:N2-finite-shorthands}
\end{equation}
When $R>0$, i.e.~$N\geq22$, the model has two fixed points, known as chiral 
and antichiral, with couplings
\begin{equation}
\begin{aligned}
  \lambda_{*\pm}
  &=
  \frac{3\big((3N-8)\pm \sqrt{R}\big)}
  {2\big(D\pm 6\sqrt{R}\big)}\,\varepsilon,\\
  \zeta_{*\pm}
  &=
  \frac{6(N-2)}
  {D\pm 6\sqrt{R}}\,\varepsilon .
\end{aligned}
\label{eq:N2-finite-fixed-couplings}
\end{equation}
These fixed points and their critical exponents have been studied
perturbatively to high order in the $\varepsilon$
expansion~\cite{Calabrese:2004nt,Kompaniets:2019xez}.

The stability of these fixed points in the bulk is governed by the
eigenvalues of $\partial\beta_a/\partial g_b$ in the $(\lambda,\zeta)$
plane. At the two fixed points
\eqref{eq:N2-finite-fixed-couplings} these are~\cite{Osborn:2017ucf}
\begin{equation}
  \varepsilon,
  \qquad
  \pm\frac{2(N-2)\sqrt R}{D\pm6\sqrt R}\,\varepsilon.
\label{eq:N2-bulk-stability-eigenvalues}
\end{equation}
For $N\geq22$ one
has $R>0$ and $D\pm6\sqrt R>0$, so the chiral point ($+$) is stable, 
while at the antichiral point ($-$), one eigenvalue is negative, so it 
is a saddle. 

The bilinear operators in the $\phi_{ia}\times\phi_{jb}$ OPE decompose
under $O(N)\times O(2)$ as
\begin{equation}
  S=(\mathbf1_N,\mathbf1_2),\qquad
  T=(\mathbf T_N,\mathbf 1_2),\qquad
  T_2=(\mathbf 1_N,\mathbf T_2),\qquad
  Y=(\mathbf T_N,\mathbf T_2),\qquad
  Z=(\mathbf A_N,\mathbf A_2),
\label{eq:N2-bilinear-representations}
\end{equation}
Here $\mathbf 1, \mathbf T, \mathbf A$ 
denote respectively the singlet, symmetric traceless tensor
and antisymmetric tensor of the corresponding orthogonal group. The
symmetric traceless representation $\mathbf T_2$ 
of $O(2)$ is two-dimensional and
the antisymmetric representation $\mathbf A_2$ is one-dimensional. Thus
the $Z$ sector is an $O(N)$ antisymmetric tensor with a pseudoscalar
$O(2)$ factor. Using the projectors in
\eqref{eq:st-and-adj-projectors}, with a superscript indicating the
orthogonal factor on which they act, the unit-normalized bilinears are
\begin{equation}
\begin{gathered}
S
=
\frac{1}{2\sqrt N\,\kappa_d}\phi_{ia}\phi_{ia},
\qquad
T_{ij}
=
\frac{1}{2\kappa_d}
\left(
  \phi_{ia}\phi_{ja}
  -\frac{\delta_{ij}}{N}\phi_{ka}\phi_{ka}
\right),
\qquad
Z_{ij}
=
\frac{1}{2\kappa_d}
P^{(N)}_{ij,kl}\epsilon_{ab}\phi_{ka}\phi_{lb}.
\\
T_{2,ab}
=
\frac{1}{\sqrt{2N}\,\kappa_d}
\left(
  \phi_{ia}\phi_{ib}
  -\frac{\delta_{ab}}{2}\phi_{ic}\phi_{ic}
\right),
\qquad
Y_{ij,ab}
=
\frac{1}{\sqrt2\,\kappa_d}
\Pi^{(N)}_{ij,kl}\Pi^{(2)}_{ab,cd}\phi_{kc}\phi_{ld},
\\
\end{gathered}
\label{eq:N2-dimension-two-fields}
\end{equation}
with $\epsilon_{12}=1$. These normalizations give the two-point functions
in \eqref{eq:uv-two-point-normalization}, with $\Pi^{(2)}_{ab,cd}$ for
$T_2$, a product of projectors for $Y$, and $P^{(N)}_{ij,kl}$ for $Z_{ij}$.

Their one-loop anomalous dimensions are
\begin{equation}
\begin{aligned}
  \gamma_{S,\pm}^{\rm N2}
  &=
  -\varepsilon+\frac{2(N+1)}{3}\lambda_{*\pm}
  -\frac{N-1}{3}\zeta_{*\pm}+O(\varepsilon^2),\\
  \gamma_{T,\pm}^{\rm N2}
  &=
  -\varepsilon+\frac{2}{3}\lambda_{*\pm}
  +\frac{1}{3}\zeta_{*\pm}+O(\varepsilon^2),\\
  \gamma_{T_2,\pm}^{\rm N2}
  &=
  -\varepsilon+\frac{2}{3}\lambda_{*\pm}
  +\frac{N-1}{3}\zeta_{*\pm}+O(\varepsilon^2),\\
  \gamma_{Y,\pm}^{\rm N2}
  &=
  -\varepsilon+\frac{2}{3}\lambda_{*\pm}
  -\frac{1}{3}\zeta_{*\pm}+O(\varepsilon^2),\\
  \gamma_{Z,\pm}^{\rm N2}
  &=
  -\varepsilon+\frac{2}{3}\lambda_{*\pm}
  -\zeta_{*\pm}+O(\varepsilon^2).
\end{aligned}
\label{eq:N2-finite-gammas}
\end{equation}
Substituting \eqref{eq:N2-finite-fixed-couplings} one finds 
that all three of $\gamma_S$, $\gamma_T$ and $\gamma_Z$ 
are negative for $N\geq22$ at both the chiral and antichiral 
fixed points.

In three dimensions one can study the model in the large-$N$ 
limit~\cite{Pelissetto:2001fi,Gracey:2002ze}. In that case, the chiral fixed 
point has two Hubbard--Stratonovich fields of dimension close to two in 
the same representations as $S$ and $T_2$ with 
anomalous dimensions
\begin{equation}
  \gamma_{S,+}^{\rm N2}
  =-\frac{16}{\pi^2N}+O(N^{-2}),
  \qquad
  \gamma_{T_2,+}^{\rm N2}
  =-\frac{8}{\pi^2N}+O(N^{-2}).
\label{eq:N2-HS-chiral-gammas}
\end{equation}
At the antichiral fixed point only the analogue of $T_2$ appears 
and
\begin{equation}
  \gamma_{T_2,-}^{\rm N2}
  =
  -\frac{8}{3\pi^2N}+O(N^{-2}).
\label{eq:N2-HS-antichiral-gamma}
\end{equation}
These should match with the chiral and antichiral 
fixed points in the $\varepsilon$ expansion.

Formulas in the rest of this section do not distinguish the chiral 
or antichiral theories and the $\pm$ subscripts are dropped 
from $\gamma^{\rm N2}$. The superscript N2 is meant to indicate 
the specialisation of all quantities from 
section~\ref{sec:scalar-tensor-anti} to the chiral/antichiral 
$O(N)\times O(2)$ model.

As for the structure constants, we use the tensor normalizations of
\eqref{eq:uv-two-point-normalization}--\eqref{eq:scalar-three-point-definitions} and their obvious generalisations, including the $Z$ structures in \eqref{eq:Z-three-point-definitions}. 
Specifically, 
the constant $C_{TT_2Y}^{\rm N2}$ multiplies
$\Pi^{(N)}\Pi^{(2)}$, $C_{TYY}^{\rm N2}$ multiplies
${\cal I}^{(N)}\Pi^{(2)}$, $C_{TZZ}^{\rm N2}$ multiplies $\Pi^{(N)}P^{(N)}P^{(N)}$, and $C_{YYZ}^{\rm N2}$ multiplies a tensor of 
the form $\Pi^{(N)}\Pi^{(N)}P^{(N)}\Pi^{(2)}\Pi^{(2)}$. 

With these conventions, the nonzero leading structure constants are
\begin{equation}
\begin{gathered}
  C_{SSS}^{\rm N2}
  =C_{STT}^{\rm N2}
  =C_{ST_2T_2}^{\rm N2}
  =C_{SYY}^{\rm N2}
  =C_{SZZ}^{\rm N2}
  =C_{TT_2Y}^{\rm N2}
  =\frac{2}{\sqrt N}+O(\varepsilon),
\\
  C_{TTT}^{\rm N2}
  =C_{TYY}^{\rm N2}=2+O(\varepsilon),
  \qquad
  C_{TZZ}^{\rm N2}=-2+O(\varepsilon),
  \qquad
  C_{YYZ}^{\rm N2}=2\sqrt2+O(\varepsilon).
\end{gathered}
\label{eq:N2-nonzero-structure-constants}
\end{equation}
The minus sign in $C_{TZZ}^{\rm N2}$ follows
from using real antisymmetric matrices for the two antisymmetric factors
in $Z$. The sign of $C_{YYZ}^{\rm N2}$ depends on the orientation chosen for the
$O(2)$ antisymmetric tensor. All omitted structure 
constants vanish at this order. For
low-rank groups this list can shorten further because some
representations or invariant tensors are absent.

The beta functions for a surface deformation is as in 
\eqref{eq:STZ-betaS-functions}--\eqref{eq:STZ-betaZ-functions} 
with extra source terms from $C_{ST_2T_2}^{\rm N2}$, $C_{SYY}^{\rm N2}$, 
$C_{TT_2Y}^{\rm N2}$ and $C_{TYY}^{\rm N2}$. The reduction to the system in 
Section~\ref{sec:scalar-tensor-anti} is valid if 
we can set $h^Y_{ij,ab}=h^{T_2}_{ab}=0$.

For that we need to examine the beta functions for $T_2$ and $Y$ 
themselves
\begin{align}
  \beta^{T_2}_{ab}
  &=
  \left(\gamma_{T_2}^{\rm N2}+\frac{4\pi}{\sqrt N}h^S\right)h^{T_2}_{ab}
  +\frac{4\pi}{\sqrt N}h^T_{ij}h^Y_{ij,ab},
\label{eq:STZ-N2-betaT2-functions}
\\
  \beta^Y_{ij,ab}
  &=
  \left(\gamma_Y^{\rm N2}+\frac{4\pi}{\sqrt N}h^S\right)h^Y_{ij,ab}
  +\frac{4\pi}{\sqrt N}h^T_{ij}h^{T_2}_{ab}
  +2\pi\Pi^{(N)}_{ij,kl}
  \big(h^T_{ko}h^Y_{ol,ab}+h^Y_{ko,ab}h^T_{ol}\big)
\nonumber\\
  &\quad
  +4\sqrt2\pi\,
  \Pi^{(N)}_{ij,ro}P^{(N)}_{mn,rs}\Pi^{(N)}_{kl,so}
  \Pi^{(2)}_{ab,eg}\Pi^{(2)}_{cd,fg}
  h^Y_{kl,cd}h^Z_{mn}\epsilon_{ef}.
\label{eq:STZ-N2-betaY-functions}
\end{align}
We see that those are automatically zero when 
$h^Y_{ij,ab}=h^{T_2}_{ab}=0$. 
This is simply because representation theory excludes 
$C_{T_2AB}$ and $C_{YAB}$ with $A,B\in\{S,T,Z\}$. 
In any case, the reduction 
\eqref{eq:STZ-betaS-functions}--\eqref{eq:STZ-betaZ-functions}
is a consistent truncation.

\subsubsection{\texorpdfstring{$\cD_n$}{Dn} fixed points}
\label{sec:N2-Dn-fixed-points}

A further consistent reduction of this system is to also set 
$h^Z=0$, giving back the $S$ and $T$ surface defects of 
Section~\ref{sec:SandT}. Since both $S$ and $T_{ij}$ are 
$O(2)$ singlets, all resulting fixed points retain the 
$O(2)$ symmetry of the bulk fixed points and operators 
retain their UV $O(2)$ quantum numbers. Due to this, 
we mostly omit the $O(2)$ structures in this section, 
leaving it implicit. We 
describe the surface defect in the chiral and antichiral 
bulk theories by the values of
$\gamma_S^{\rm N2},\gamma_T^{\rm N2}$ from \eqref{eq:N2-finite-gammas}. 
As they are complicated, we do not
substitute them explicitly.

Using $C_{SSS}^{\rm N2}$, $C_{STT}^{\rm N2}$, and $C_{TTT}^{\rm N2}$ from 
\eqref{eq:N2-nonzero-structure-constants}
the $O(N)$ symmetric fixed point has
\begin{equation}
  h^{S,\cD_N,{\rm N2}}
  =-\frac{\sqrt N}{2\pi}\gamma_S^{\rm N2},
  \qquad
  h^{T,\cD_N,{\rm N2}}_{ij}=0 .
\label{eq:DN-chiral-h-couplings}
\end{equation}
For $n\neq m$, the discriminant becomes (recall that $\nu=n-m$)
\begin{equation}
  \sigma_n^2
  =\frac{9\nu^2}{N^2}
  +\frac{36nm}{N^2}
  \frac{\gamma_T^{\rm N2}(2\gamma_S^{\rm N2}-\gamma_T^{\rm N2})}{(\gamma_S^{\rm N2})^2}.
\label{eq:chiral-ST-discriminant}
\end{equation}
The values of the couplings at the fixed points are
\begin{equation}
\begin{aligned}
  h^{S,\cD_n,{\rm N2}}
  &=
  -\frac{1}{4\pi N^{3/2}}
  \left(\nu^2\gamma_S^{\rm N2}+4nm\gamma_T^{\rm N2}
  +\frac{N\nu}{3}\gamma_S^{\rm N2}\sigma_n\right),
\\
  h_n^{\cD_n,{\rm N2}}
  &=\frac{m}{2\pi\nu}
  \left(\gamma_T^{\rm N2}+\frac{4\pi}{\sqrt N}h^{S,\cD_n,{\rm N2}}\right),
  \qquad
  h_m^{\cD_n,{\rm N2}}
  =-\frac{n}{m}h_n^{\cD_n,{\rm N2}} .
\label{eq:Dn-chiral-h-couplings}
\end{aligned}
\end{equation}
For $n=m=N/2$, one uses
\begin{equation}
  h^{S,\cD_{N/2},{\rm N2}}
  =-\frac{\sqrt N}{4\pi}\gamma_T^{\rm N2},
  \qquad
  h_n^{\cD_{N/2},{\rm N2}}
  =\frac{1}{4\pi}
  \sqrt{\gamma_T^{\rm N2}(2\gamma_S^{\rm N2}-\gamma_T^{\rm N2})},
  \qquad
  h_m^{\cD_{N/2},{\rm N2}}=-h_n^{\cD_{N/2},{\rm N2}} .
\label{eq:Dhalf-chiral-h-couplings}
\end{equation}

\subsubsection{Near dimension two operators at \texorpdfstring{$\cD_n$}{Dn} fixed points}

The anomalous dimensions of $S$ and $T$ at the $O(N)$ 
symmetric fixed point follow from \eqref{eq:DN-stability-dimensions} and 
\eqref{eq:DN-chiral-h-couplings},
\begin{equation}
  \gamma_S^{\cD_N,{\rm N2}}=-\gamma_S^{\rm N2},
  \qquad
  \gamma_T^{\cD_N,{\rm N2}}=\gamma_T^{\rm N2}-2\gamma_S^{\rm N2} .
  \label{eq:DN-chiral-stability-dimensions}
\end{equation}

For $n\neq m$, the fields $T_n$, $T_m$, the tilt, and the two singlets
are obtained from \eqref{eq:tensor-gammas-generic} and
\eqref{eq:mixed-singlet-Gamma-generic}, with
\eqref{eq:N2-nonzero-structure-constants} and
\eqref{eq:Dn-chiral-h-couplings}.
Thus
\begin{equation}
\begin{aligned}
  \gamma_{T_n}^{\cD_n,{\rm N2}}
  &=
  \frac{N}{\nu}
  \left(\gamma_T^{\rm N2}+\frac{4\pi}{\sqrt N}
  h^{S,\cD_n,{\rm N2}}\right),
\\
  \gamma_{T_m}^{\cD_n,{\rm N2}}
  &=
  -\frac{N}{\nu}
  \left(\gamma_T^{\rm N2}+\frac{4\pi}{\sqrt N}
  h^{S,\cD_n,{\rm N2}}\right),
  \qquad
  \gamma_\bt^{\cD_n,{\rm N2}}=0 .
  \label{eq:tensor-gammas-chiral}
\end{aligned}
\end{equation}
The singlet mixing matrix is as in \eqref{eq:mixed-singlet-Gamma-generic}
with the structure constant from \eqref{eq:N2-nonzero-structure-constants}.

The remaining bilinears in \eqref{eq:N2-bilinear-representations} split
under $O(N)\times O(2)\to O(n)\times O(m)\times O(2)$ as
\begin{equation}
\begin{aligned}
  &T_{2,ab}:&
  (\mathbf 1_N,\mathbf T_2)&\to(\mathbf 1,\mathbf 1,\mathbf T_2),
\\
  &Z_{ij}:&
  (\mathbf A_N,\mathbf A_2)&\to
  \big((\mathbf A_n,\mathbf 1)
  \oplus(\mathbf 1,\mathbf A_m)
  \oplus(\mathbf n,\mathbf m)\big)\otimes\mathbf A_2,
\\
  &Y_{ij,ab}:&
  (\mathbf T_N,\mathbf T_2)&\to
  \big((\mathbf 1,\mathbf 1)
  \oplus(\mathbf T_n,\mathbf 1)
  \oplus(\mathbf 1,\mathbf T_m)
  \oplus(\mathbf n,\mathbf m)\big)\otimes\mathbf T_2.
\end{aligned}
\label{eq:ST-N2-ZYT2-splitting}
\end{equation}
We analyse these components to complete the stability analysis.

The $T_2$ field mixes with
the $Y$ component parallel to $h^{T,\cD_n,{\rm N2}}_{ij}$. In the basis
\begin{equation}
  \left(
    T_{2,ab},
    \frac{h^{T,\cD_n,{\rm N2}}_{ij}Y_{ij,ab}}
    {\big|h^{T,\cD_n,{\rm N2}}\big|}
  \right),
\end{equation}
the mixing matrix is
\begin{equation}
  \Gamma_{T_2Y}^{\cD_n,{\rm N2}}
  =
  \begin{pmatrix}
    \displaystyle
    \gamma_{T_2}^{\rm N2}+\frac{4\pi}{\sqrt N}h^{S,\cD_n,{\rm N2}}
    &
    \displaystyle
    \frac{4\pi}{\sqrt N}\big|h^{T,\cD_n,{\rm N2}}\big|
    \\[.5em]
    \displaystyle
    \frac{4\pi}{\sqrt N}\big|h^{T,\cD_n,{\rm N2}}\big|
    &
    \displaystyle
    \gamma_Y^{\rm N2}-2\gamma_T^{\rm N2}
  -\frac{4\pi}{\sqrt N}h^{S,\cD_n,{\rm N2}}
    \end{pmatrix}.
  \label{eq:chiral-T2Y-Gamma}
\end{equation}
The other components of $Y$ are in the traceless symmetric representations 
of $O(n)$ and $O(m)$ and the bifundamental. And all are $O(2)$ doublets. 
Their anomalous dimensions are
\begin{equation}
\begin{aligned}
  \gamma_{Y_n}^{\cD_n,{\rm N2}}
  &=\gamma_Y^{\rm N2}+\frac{4\pi}{\sqrt N}h^{S,\cD_n,{\rm N2}}
  +4\pi h_n^{\cD_n,{\rm N2}},
\\
  \gamma_{Y_m}^{\cD_n,{\rm N2}}
  &=\gamma_Y^{\rm N2}+\frac{4\pi}{\sqrt N}h^{S,\cD_n,{\rm N2}}
  +4\pi h_m^{\cD_n,{\rm N2}},
\\
  \gamma_{Y_{\rm bifund}}^{\cD_n,{\rm N2}}
  &=
  \gamma_Y^{\rm N2}+\frac{4\pi}{\sqrt N}h^{S,\cD_n,{\rm N2}}
  +2\pi\big(h_n^{\cD_n,{\rm N2}}+h_m^{\cD_n,{\rm N2}}\big).
\label{eq:chiral-Y-split-gammas}
\end{aligned}
\end{equation}
$Z$ decomposes into two antisymmetrics and a bifundamental and 
all are $O(2)$ pseudoscalars. With the sign $C_{TZZ}^{\rm N2}=-2$, they have
\begin{equation}
\begin{aligned}
  \gamma_{Z_n}^{\cD_n,{\rm N2}}
  &=\gamma_Z^{\rm N2}+\frac{4\pi}{\sqrt N}h^{S,\cD_n,{\rm N2}}
  -4\pi h_n^{\cD_n,{\rm N2}},
\\
  \gamma_{Z_m}^{\cD_n,{\rm N2}}
  &=\gamma_Z^{\rm N2}+\frac{4\pi}{\sqrt N}h^{S,\cD_n,{\rm N2}}
  -4\pi h_m^{\cD_n,{\rm N2}},
\\
  \gamma_{Z_{\rm bifund}}^{\cD_n,{\rm N2}}
  &=\gamma_Z^{\rm N2}+\frac{4\pi}{\sqrt N}h^{S,\cD_n,{\rm N2}}
  -2\pi\big(h_n^{\cD_n,{\rm N2}}+h_m^{\cD_n,{\rm N2}}\big).
\label{eq:chiral-Z-split-gammas}
\end{aligned}
\end{equation}
For $n=m=N/2$, one uses \eqref{eq:Dhalf-chiral-h-couplings} and 
finds that the expressions above remain valid.

The $O(N)$ symmetric defect $\cD_N^{\rm N2}$ 
is stable to all of the above perturbations when
\begin{equation}
  -\gamma_S^{\rm N2}>0,\qquad
  \gamma_T^{\rm N2},\,\gamma_Z^{\rm N2},\,
   \gamma_{T_2}^{\rm N2},\,\gamma_Y^{\rm N2}>2\gamma_S^{\rm N2}.
\label{eq:N2-DN-full-stability}
\end{equation}
Substituting the one-loop data \eqref{eq:N2-finite-gammas},
the symmetric defect in the antichiral bulk theory satisfies these
inequalities for all integer $N\geq22$, although the antichiral bulk
fixed point is itself a saddle by \eqref{eq:N2-bulk-stability-eigenvalues}.
In the same bulk theory,
\eqref{eq:chiral-ST-discriminant} is nonnegative for all $n$, so all
the symmetry-breaking fixed points are real. They are saddles by the
analysis of Section~\ref{sec:real-and-stable}.

In the chiral bulk theory, which is stable in the bulk for $N\geq22$,
the symmetric fixed point is real but unstable for all integer $N\geq22$. 
For $N=22$ all symmetry-breaking fixed points are real. 
For $23\leq N\le36$ only some branches near $n=1$ or $n=N-1$
remain real, and for $N\ge37$ none of the $n\neq m$ branches are real.
For $2<N<22$, both bulk fixed points are complex.

Given the complexity of the anomalous dimensions
\eqref{eq:N2-finite-gammas}, we found no integer solutions 
with degenerate fixed points, i.e.\ $\sigma_n=0$ 
\eqref{eq:chiral-ST-discriminant}.

Near dimension three transverse vectors are easy to get from the 
expressions in Section~\ref{sec:dimension-three-operators}. 
The one new ingredient is the need to include also 
the operators arising from the descendants of $Z$, $T_2$ and $Y$. 
These do not mix with $V_r$, $U_{ij,r}$ 
but the singlet component of $Y$ does mix with $T_2$.

\subsubsection{\texorpdfstring{$\cC_p$}{Cp} fixed points in the 
\texorpdfstring{$O(N)\times O(2)$}{O(N) x O(2)} model}

We now turn to the fixed points involving nonzero 
$h^Z$, following Section~\ref{sec:scalar-tensor-anti}. 
The main simplification 
compared to the general formulas there is that the particular values 
of the structure constants
\eqref{eq:N2-nonzero-structure-constants}
\begin{equation}
  C_{SSS}^{\rm N2}=C_{STT}^{\rm N2}=C_{SZZ}^{\rm N2}=\frac{2}{\sqrt N},
  \qquad
  C_{TTT}^{\rm N2}=2,
  \qquad
  C_{TZZ}^{\rm N2}=-2.
\label{eq:STZ-N2-constants}
\end{equation}
The anomalous dimensions $\gamma_S^{\rm N2},\gamma_{T}^{\rm N2},\gamma_Z^{\rm N2}$ are those in
\eqref{eq:N2-finite-gammas}, evaluated at either the chiral or the
antichiral fixed point. We do not plug in their values, as 
they are rather complicated.

One particular point to notice is that compared to the general formalism 
in Section~\ref{sec:scalar-tensor-anti}, the $O(N)\times O(2)$ model has 
an extra $O(2)$ symmetry. Since $Z$ is in the antisymmetric representation 
of $O(2)$ (see the $\epsilon_{ab}$ in \eqref{eq:N2-dimension-two-fields}), 
turning on $h^Z$ breaks the extra $O(2)$ to $SO(2)$. The fields arising 
from $S$, $T$ and $Z$ are now singlets and those from $T_2$ and $Y$ are 
$SO(2)$ doublets. 

The $U(N/2)$ symmetric $\cC_{N/2}^{\rm N2}$ fixed point becomes
\begin{equation}
  h^{S,\cC_{N/2},{\rm N2}}=-\frac{\sqrt N}{4\pi}\gamma_Z^{\rm N2},
  \qquad
  h^{Z,\cC_{N/2},{\rm N2}}_{ij}h^{Z,\cC_{N/2},{\rm N2}}_{ij}
  =
  \frac{N}{16\pi^2}\gamma_Z^{\rm N2}(2\gamma_S^{\rm N2}-\gamma_Z^{\rm N2}).
\label{eq:STZ-N2-pure-U}
\end{equation}
For the $U(p)\times O(n)$ fixed point, we write
\begin{equation}
  \varpi=2p-n,
\end{equation}
and express all the data in terms of $h_{2p}^{\cC_{p,\pm},{\rm N2}}$ as
\begin{align}
  h^{S,\cC_{p,\pm},{\rm N2}}
  &=
  -\frac{\sqrt N}{4\pi}\big(\gamma_Z^{\rm N2}
  +4\pi h_{2p}^{\cC_{p,\pm},{\rm N2}}\big).
\label{eq:STZ-N2-hS}
\\
  \big(\eta^{\cC_{p,\pm},{\rm N2}}\big)^2
  &=
  \frac{N}{2\pi n}h_{2p}^{\cC_{p,\pm},{\rm N2}}
  \left(\gamma_Z^{\rm N2}-\gamma_{T}^{\rm N2}
  +\frac{2\pi N}{n}h_{2p}^{\cC_{p,\pm},{\rm N2}}\right).
\label{eq:STZ-N2-eta2}
\end{align}
Then the discriminant of equation \eqref{eq:STZ-h2p-quadratic} becomes
\begin{equation}
  \sigma_{2p}^2
  =
  \frac{1}{n^2}
  \left[
    \big(n\gamma_S^{\rm N2}+2p\gamma_T^{\rm N2}\big)^2
    +4pN\gamma_Z^{\rm N2}(\gamma_S^{\rm N2}-\gamma_T^{\rm N2})
  \right].
\label{eq:STZ-N2-discriminant}
\end{equation}
When $\sigma_{2p}^2\ge0$, the two roots are
\begin{equation}
\begin{aligned}
  h_{2p}^{\cC_{p,\pm},{\rm N2}}
  =
  \frac{n}{4\pi N}
  \left[
  \frac{n\gamma_S^{\rm N2}+2p\gamma_T^{\rm N2}}{N}
  -\gamma_Z^{\rm N2}
  \pm\frac{n}{N}\sigma_{2p}
  \right].
\end{aligned}
\label{eq:STZ-N2-h2p-roots}
\end{equation}

Let us finally spell out which of these saddles are real at
finite $N$ restricting to $N\geq22$, where the
bulk fixed points exist and connect to the
large-$N$ branches in \eqref{eq:N2-HS-chiral-gammas}, 
\eqref{eq:N2-HS-antichiral-gamma}. In the antichiral bulk fixed 
point, all $U(p)\times O(n)$ symmetric saddles $\cC_{p,\pm}$
are real, irrespective of the choice of sign in 
\eqref{eq:STZ-N2-h2p-roots}. The same is true for the $\cC_{N/2}$ 
fixed point.
For the chiral bulk fixed point $\cC_{N/2}^{\rm N2}$ is not 
real in this range, since
$\gamma_Z^{\rm N2}(2\gamma_S^{\rm N2}-\gamma_Z^{\rm N2})<0$. 
The $\cC_{p,\pm}$ saddles are real only for
$22\leq N\le33$. More explicitly, for $N=22$ only the $+$ root is real
for $1\leq p\le5$; for $N=23$, the $p=1$ saddle has only the $+$ root,
while $p=2,3$ have both roots; for $N=24,25$, the saddles with $p=1,2$
have both roots; and for $26\leq N\le33$, only the $p=1$ saddles have
both roots. For $N\ge34$ none of the chiral mixed saddles are real.
All real saddles listed here remain unstable since it is shown in 
Section~\ref{sec:STZ-real-and-stable} that only $\cD_N$ may be 
stable.

\subsubsection{Near dimension two operators at \texorpdfstring{$\cC_p$}{Cp} fixed points}

Here we use the results from Section~\ref{sec:STZ-near-two}, specialised 
to this model. The fields split at the fixed point according to 
\eqref{eq:STZ-Un-On-operator-decomposition}. Then using 
\eqref{eq:STZ-N2-constants}, \eqref{eq:STZ-N2-hS} and the tracelessness of $h^T$ we find
\begin{equation}
\begin{aligned}
  \gamma_{T_n}^{\cC_{p,\pm},{\rm N2}}
  &=
  \frac{n}{N}\left(\gamma_T^{\rm N2}-\gamma_S^{\rm N2}\mp\sigma_{2p}\right),
\\
  \gamma_{Z_n}^{\cC_{p,\pm},{\rm N2}}
  &=
  \gamma_Z^{\rm N2}-\frac{n\gamma_S^{\rm N2}+2p\gamma_T^{\rm N2}}{N}
  \mp\frac{n}{N}\sigma_{2p}.
\\
  \gamma_{T_{\rm sym}}^{\cC_{p,\pm},{\rm N2}}
  &=
  \gamma_T^{\rm N2}-\gamma_Z^{\rm N2}.
\label{eq:STZ-N2-Tn-Zn-sym-WF}
\end{aligned}
\end{equation}
The dimensions of the adjoint fields are
\begin{equation}
  \gamma_{\cO_{\rm adj},\pm}^{\cC_{p,\pm},{\rm N2}}
  =
  \frac12\left[\gamma_T^{\rm N2}-\gamma_Z^{\rm N2}
  \pm\sqrt{(\gamma_T^{\rm N2}-\gamma_Z^{\rm N2})^2
  +64\pi^2\big(\eta^{\cC_{p,\pm},{\rm N2}}\big)^2}\right],
\label{eq:STZ-N2-adj-gammas}
\end{equation}
with $\eta^{\cC_{p,\pm},{\rm N2}}$ from \eqref{eq:STZ-N2-eta2}, \eqref{eq:STZ-N2-h2p-roots}.
The bifundamental dimensions are 0 for the tilt and
\begin{equation}
  \gamma_{Z_{\rm bifund}}^{\cC_{p,\pm},{\rm N2}}
  =\frac{n}{N}\left(\gamma_T^{\rm N2}-\gamma_S^{\rm N2}\mp\sigma_{2p}\right).
\label{eq:STZ-N2-B-gamma}
\end{equation}
The singlets are the roots of the characteristic polynomial 
\begin{equation}
\begin{aligned}
  0
  &=
  -\gamma_\cO^3
  +\frac{2p\gamma_S^{\rm N2}+n\gamma_T^{\rm N2}-N\gamma_Z^{\rm N2}
  \mp n\sigma_{2p}}{N}\gamma_\cO^2
  +16\pi^2\big(\eta^{\cC_{p,\pm},{\rm N2}}\big)^2\gamma_\cO
\\
  &\quad
  +\frac{2pn(\gamma_S^{\rm N2}-\gamma_T^{\rm N2})^2
  \pm n\sigma_{2p}
  (2p\gamma_S^{\rm N2}+n\gamma_T^{\rm N2}-N\gamma_Z^{\rm N2})}
  {N^2}\gamma_\cO
  \pm\frac{16\pi^2n}{N}\,
  \sigma_{2p}\big(\eta^{\cC_{p,\pm},{\rm N2}}\big)^2 .
\end{aligned}
\label{eq:STZ-N2-singlet-characteristic}
\end{equation}

At the $\cC_{N/2}^{\rm N2}$ fixed point, the decoupled $T$ singlet has
\begin{equation}
  \gamma_{T}^{\cC_{N/2},{\rm N2}}
  =
  \gamma_{T}^{\rm N2}-\gamma_Z^{\rm N2},
\label{eq:STZ-N2-pure-U-T-gamma}
\end{equation}
and the remaining two singlets are mixed as in 
\eqref{eq:STZ-pure-U-singlet-Gamma}. Using
\eqref{eq:STZ-N2-pure-U}, its two eigenvalues are
\begin{equation}
  \gamma_{\pm}^{\cC_{N/2},{\rm N2}}
  =
  \frac12\left[\gamma_S^{\rm N2}-\gamma_Z^{\rm N2}
  \pm\sqrt{(\gamma_S^{\rm N2})^2+6\gamma_S^{\rm N2}\gamma_Z^{\rm N2}-3(\gamma_Z^{\rm N2})^2}\right].
\label{eq:STZ-N2-pure-U-singlet-eigenvalues}
\end{equation}

The $O(N)\times O(2)$ model also contains the bilinears $T_2$ and $Y$.
They do not mix linearly with the $S$, $T$, $Z$ fields around the fixed points
above. This is due to the fact that for the nonvanishing structure 
constants
\eqref{eq:N2-nonzero-structure-constants}, with $h^{T_2}=h^Y=0$, a
single insertion of $T_2$ or $Y$ can only produce $T_2$ or $Y$ again. So 
these operators do not mix, they just split and their different components 
get appropriate shifts.

To avoid cluttering notations, we do not assign the fields new names and 
simply use the original ones $T_2$ and $Y$ with the appropriate indices. 
When the two of them mix, we use the collective name $T_2Y_\pm$.

Under the symmetry breaking $O(N)\times O(2)\to U(p)\times O(n)\times SO(2)$, 
they decompose as (c.f.\ \eqref{eq:STZ-Un-On-operator-decomposition})
\begin{equation}
\begin{aligned}
  &T_2:&
  (\mathbf1_N,\mathbf T_2)&\to(\mathbf1,\mathbf1)\otimes\mathbf T_2,
\\
  &Y:&
  (\mathbf T_N,\mathbf T_2)&\to
  \big((\mathbf1,\mathbf1)
  \oplus(\mathrm{Sym}^2\mathbf p,\mathbf1)
  \oplus(\mathrm{Sym}^2\bar{\mathbf p},\mathbf1)
  \oplus(\mathbf p,\mathbf n)
  \oplus(\bar{\mathbf p},\mathbf n)
  \oplus(\mathbf1,\mathbf T_n)
  \oplus(\mathbf{Adj}_p,\mathbf1)\big)\otimes\mathbf T_2.
\end{aligned}
\label{eq:STZ-N2-T2Y-decomposition}
\end{equation}
Both are $SO(2)$ doublets, but we omit the indices below for simplicity.

The $O(N)$ singlet component of $Y$ mixes with $T_2$. In the basis
\begin{equation}
  \left(
    T_{2,ab},\,
    \frac{h^{T,\cC_{p,\pm},{\rm N2}}_{ij}Y_{ij,ab}}{|h^{T,\cC_{p,\pm},{\rm N2}}|}
  \right),
\end{equation}
the anomalous dimensions are the eigenvalues of
\begin{equation}
  \Gamma_{T_2Y}^{\cC_{p,\pm},{\rm N2}}
  =
  \begin{pmatrix}
    \displaystyle
    \gamma_{T_2}^{\rm N2}+\frac{4\pi}{\sqrt N}h^{S,\cC_{p,\pm},{\rm N2}}
    &
    \displaystyle
    \frac{4\pi}{\sqrt N}\big|h^{T,\cC_{p,\pm},{\rm N2}}\big|
    \\[.6em]
    \displaystyle
    \frac{4\pi}{\sqrt N}\big|h^{T,\cC_{p,\pm},{\rm N2}}\big|
    &
    \displaystyle
    \gamma_Y^{\rm N2}+\frac{4\pi}{\sqrt N}h^{S,\cC_{p,\pm},{\rm N2}}
    -4\pi\,\frac{\varpi}{n}h_{2p}^{\cC_{p,\pm},{\rm N2}}
  \end{pmatrix},
  \label{eq:STZ-N2-T2Y-singlet-Gamma}
\end{equation}
and recall from \eqref{eq:STZ-Un-On-ansatz} and tracelessness that
\begin{equation}
  \big|h^{T,\cC_{p,\pm},{\rm N2}}\big|^2
  =
  \frac{2pN}{n}\big(h_{2p}^{\cC_{p,\pm},{\rm N2}}\big)^2.
\end{equation}

The remaining components of $Y$ do not mix and have the anomalous dimensions
\begin{equation}
\begin{aligned}
\gamma_{Y_n}^{\cC_{p,\pm},{\rm N2}}
  &=
  \gamma_Y^{\rm N2}+\frac{4\pi}{\sqrt N}h^{S,\cC_{p,\pm},{\rm N2}}
  +4\pi h_n^{\cC_{p,\pm},{\rm N2}} .
\\
\gamma_{Y_{\rm sym},\pm}^{\cC_{p,\pm},{\rm N2}}
  &=
  \gamma_Y^{\rm N2}+\frac{4\pi}{\sqrt N}h^{S,\cC_{p,\pm},{\rm N2}}
  +4\pi h_{2p}^{\cC_{p,\pm},{\rm N2}}
  \pm4\sqrt2\,\pi\,\eta^{\cC_{p,\pm},{\rm N2}} .
\\
\gamma_{Y_{\rm bifund},\pm}^{\cC_{p,\pm},{\rm N2}}
  &=
  \gamma_Y^{\rm N2}+\frac{4\pi}{\sqrt N}h^{S,\cC_{p,\pm},{\rm N2}}
  -2\pi\,\frac{\varpi}{n}h_{2p}^{\cC_{p,\pm},{\rm N2}}
  \pm2\sqrt2\,\pi\,\eta^{\cC_{p,\pm},{\rm N2}} .
\\
  \gamma_{Y_{\rm adj}}^{\cC_{p,\pm},{\rm N2}}
  &=
  \gamma_Y^{\rm N2}+\frac{4\pi}{\sqrt N}h^{S,\cC_{p,\pm},{\rm N2}}
  +4\pi h_{2p}^{\cC_{p,\pm},{\rm N2}} .
\label{eq:STZ-N2-Ycomp-gamma}
\end{aligned}
\end{equation}
The $\pm$ subscripts on the symmetric and bifundamental fields correspond 
to the real and imaginary parts of these complex fields that get 
different contributions from the coupling to $Z$.

For the $U(N/2)$ symmetric $\cC_{N/2}^{\rm N2}$ saddle, one simply sets 
$h^{T,\cC_{N/2},{\rm N2}}=0$ and $h^{S,\cC_{N/2},{\rm N2}}$ is in 
\eqref{eq:STZ-N2-pure-U}. There are no bifundamental fields since $n=0$ 
and no mixing among the singlet components.
The anomalous dimensions are
\begin{equation}
  \gamma_{T_2}^{\cC_{N/2},{\rm N2}}
  =\gamma_{T_2}^{\rm N2}-\gamma_Z^{\rm N2},
\quad
\gamma_{Y_{\rm sing}}^{\cC_{N/2},{\rm N2}}
  =
  \gamma_{Y_{\rm adj}}^{\cC_{N/2},{\rm N2}}
  =\gamma_Y^{\rm N2}-\gamma_Z^{\rm N2},
\quad
\gamma_{Y_{\rm sym},\pm}^{\cC_{N/2},{\rm N2}}
  =\gamma_Y^{\rm N2}-\gamma_Z^{\rm N2}
  \pm4\sqrt2\,\pi\,\eta^{\cC_{N/2},{\rm N2}} .
\end{equation}

Stability in the full space requires all anomalous dimensions
in this $T_2,Y$ sector, in addition to the $S$, $T$, $Z$ eigenvalues above, to
be nonnegative.

Transverse vector operators of dimension near three at 
$\cC_{p,\pm}$ are gotten from Section~\ref{sec:STZ-near-three}. 
As with dimension near two operators, one should also 
consider the operators arising from the descendants of 
$T_2$ and $Y$. They decompose as in 
\eqref{eq:STZ-N2-T2Y-decomposition}, do not mix with 
$V_r$, $U_{ij,r}$ and $W_{ij,r}$, mix among themselves 
in the singlet sector and the symmetric and bifundamental 
components of $Y$ get different corrections for their real 
and imaginary parts. All very analogously to the 
discussion above.

\subsubsection{Displacements and tilts}
\label{sec:N2-displacements-and-tilts}

The displacement and tilt normalisations follow from the formulas in
Section~\ref{sec:tD-in-STZ} with the structure constants
\eqref{eq:STZ-N2-constants} and fixed-point couplings 
in \eqref{eq:STZ-N2-pure-U} and \eqref{eq:STZ-N2-hS}--\eqref{eq:STZ-N2-h2p-roots}.
As elsewhere in the $O(N)\times O(2)$ analysis we keep the
bulk dimensions $\gamma_S^{\rm N2}$, $\gamma_T^{\rm N2}$, 
$\gamma_Z^{\rm N2}$ unevaluated, 
and since the fixed-point couplings are of order
$\varepsilon$ we replace $\Delta_S$, $\Delta_T$, $\Delta_Z$ by their leading
value $2$.

At the $O(N)$ symmetric fixed point \eqref{eq:DN-chiral-h-couplings} the
tensor and antisymmetric couplings vanish, so
$C_{\bt}^{\cD_N,{\rm N2}}=0$ and, from \eqref{eq:CD-DN-general},
\begin{equation}
  C_{\bD}^{\cD_N,{\rm N2}}
  =2\Delta_S\big(h^{S,\cD_N,{\rm N2}}\big)^2
  =\frac{N}{\pi^2}\big(\gamma_S^{\rm N2}\big)^2
  +O(\varepsilon^3).
\label{eq:CD-DN-N2-evaluated}
\end{equation}
The expressions at the symmetry breaking points $\cD_n$ do not 
simplify. See \eqref{eq:Ct-Dn-general}, \eqref{eq:CD-endpoint-general}.

Using $C_{TTT}^{\rm N2}=2$ and $C_{STT}^{\rm N2}=2/\sqrt N$ in
\eqref{eq:Ct-Dn-generic}, the tilt normalisation can be written as
\begin{equation}
  C_{\bt}^{\cD_n,{\rm N2}}
  =\frac{N^2}{2\pi^2\nu^2}
  \left[\gamma_T^{\rm N2}+\frac{4\pi}{\sqrt N}h^{S,\cD_n,{\rm N2}}\right]^2 .
\label{eq:Ct-Dn-N2-evaluated}
\end{equation}

For the $\cC_{p,\pm}$ 
fixed points with $h^Z\neq0$ the conformal manifold is
$O(N)/(U(p)\times O(n))$, whose tangent space
\eqref{eq:STZ-tangent-decomposition} splits into the bifundamental and
wedge directions, giving two distinct tilt families.

At the $U(N/2)$ symmetric fixed point $\cC_{N/2}^{\rm N2}$ 
\eqref{eq:STZ-N2-pure-U}, 
the bifundamental directions are absent and the tensor coupling
vanishes, so only the wedge tilt survives. Equations
 \eqref{eq:Ct-Cp-STZ-wedge}, \eqref{eq:STZ-N2-pure-U} give
\begin{equation}
  C_{\bt_\wedge}^{\cC_{N/2},{\rm N2}}
  =\frac{1}{8\pi^2}\gamma_Z^{\rm N2}\big(2\gamma_S^{\rm N2}-\gamma_Z^{\rm N2}\big),
\label{eq:Ct-CNhalf-N2-evaluated}
\end{equation}
while \eqref{eq:CD-Cp-STZ-general} with $h^T=0$ collapses to
\begin{equation}
  C_{\bD}^{\cC_{N/2},{\rm N2}}
  =\frac{N}{2\pi^2}\gamma_S^{\rm N2}\gamma_Z^{\rm N2}
  +O(\varepsilon^3).
\label{eq:CD-CNhalf-N2-evaluated}
\end{equation}

At the generic $U(p)\times O(n)$ fixed point $\cC_{p,\pm}^{\rm N2}$ both
tilt families are present. There are no particular simplifications, so 
see \eqref{eq:Ct-Cp-STZ-bifund}, \eqref{eq:Ct-Cp-STZ-wedge}, 
and \eqref{eq:CD-Cp-STZ}.

\subsection{Tricritical Wilson--Fisher theory in \texorpdfstring{$d=3-\varepsilon$}{3-epsilon}}
\label{sec:3-ep}

Lastly we look at the tricritical 
$O(N)$ WF theory. This theory has a sextic potential, a 
perturbative bulk fixed point in $d=3-\varepsilon$ dimensions 
and three weakly relevant operators.
They are quartics of the fundamental field which decompose into $S$, $T_{ij}$ and the rank four symmetric traceless operator $T_4$.
They have anomalous dimensions~\cite{Osborn:2017ucf}
\begin{equation}
  \gamma_S^{\rm tri}
  =
-\frac{2(N+14)}{3N+22}\varepsilon+O(\varepsilon^2),
\qquad
  \gamma_T^{\rm tri}
  =
-\frac{4(N+7)}{3N+22}\varepsilon+O(\varepsilon^2),
\qquad
    \gamma_{T_4}^{\rm tri}=
-\frac{6(N+6)}{3N+22}\varepsilon+O(\varepsilon^2).
\label{eq:tri-uv-gammas}
\end{equation}
The appearance of a third weakly relevant operator takes us outside the two-coupling framework of Section \ref{sec:SandT}. 
The current is just as in \eqref{eq:wf-noether-current} 
but with $\Delta_j^{\rm tri}=2-\varepsilon$.
In contrast with the earlier examples, the current is therefore a fourth weakly relevant operator on the two-dimensional surface and could be included in the perturbation. 
We choose to not turn on the current and instead study perturbations involving only $S$, $T_{ij}$ and $T_{4,ijkl}$. 
As we discussed after \eqref{eq:tri-CTjU}, this is a consistent truncation.

In terms of the basic fields, the tree-level normalised operators are
\begin{equation}
  S=
  \frac{(\phi_k\phi_k)^2}
  {\sqrt{8N(N+2)}\,\kappa_3^2},
\qquad
  T_{ij} =
  \frac{
  \phi_i\phi_j\phi_k\phi_k
  -\tr}
  {2\sqrt{N+4}\,\kappa_3^2},
\qquad
    T_{4,ijkl} = \frac{\phi_i \phi_j \phi_k \phi_l -\tr}{2\sqrt{6}\kappa_3^2},
\qquad
  \kappa_3 =\frac1{4\pi}+O(\varepsilon),
\label{eq:tri-unit-quartics}
\end{equation}
and the current is kept with the canonical normalisation.

To avoid a glut of indices, we introduce the auxiliary complex null vectors $u^i$ with $u^2=0$, which means that contractions with $u$'s automatically discard traces. We write
\begin{equation}
    T(u) = u^i u^j T_{ij} = \frac{(u\cdot\phi)^2(\phi_k\phi_k)}{2\sqrt{N+4}\kappa_3^2},
\qquad
    T_4(u) = u^i u^j u^k u^l T_{4,ijkl} = \frac{(u\cdot\phi)^4}{2\sqrt{6}\kappa_3^2}.
\end{equation}
In particular,
\begin{equation}
    u_1^i u_1^j u_2^k u_2^l \Pi_{ij,kl} = (u_1\cdot u_2)^2,
\end{equation}
so that the two-point function reads
\begin{equation}
    \langle T(u_1,\tau) T(u_2,0) \rangle = \frac{(u_1\cdot u_2)^2}{\left\lvert \tau \right\rvert^{2\Delta_T}},
\end{equation}
and similarly for $T_4$ with $(u_1\cdot u_2)^4$ in the numerator.

There are six new structure constants which involve the rank four operator. 
Two of them are zero, $C_{SST_4}=0$ and $C_{STT_4}=0$, by $O(N)$ representation theory since the rank four representation doesn't appear in $\mathbf{1}\otimes\mathbf{1}$ or in $\mathbf{1}\otimes\mathbf{T}_N$. We define the four which are generically non-zero by
\begin{equation}
\begin{aligned}
    \left\langle S(0) T_4(u_1,1) T_4(u_2,\infty) \right\rangle & = (u_1\cdot u_2)^4 C_{S T_4 T_4},\\
\left\langle T(u_1,0) T(u_2,1) T_4(u_3,\infty) \right\rangle & = (u_1\cdot u_3)^2(u_2\cdot u_3)^2 C_{TTT_4},\\
\left\langle T(u_1,0) T_4(u_2,1) T_4(u_3,\infty) \right\rangle & = (u_1\cdot u_2)(u_1\cdot u_3)(u_2\cdot u_3)^3 C_{TT_4T_4},\\
\left\langle T_4(u_1,0)T_4(u_2,1)T_4(u_3,\infty) \right\rangle & = (u_1 \cdot u_2)^2(u_1 \cdot u_3)^2(u_2 \cdot u_3)^2 C_{T_4T_4T_4},
\end{aligned}
\end{equation}
where $\mathcal{O}(u,\infty) = \lim_{|\tau|\to\infty}|\tau|^{2\Delta_{\mathcal{O}}}\mathcal{O}(u,\tau)$.

The structure constants to leading order are 
\begin{equation}
\begin{aligned}
    C_{SSS}^{\rm tri}
    & =
    \frac{2\sqrt2\,(N+8)}{\sqrt{N(N+2)}}+O(\varepsilon),
\quad
    C_{STT}^{\rm tri}
     =
    \frac{\sqrt2\,(N+16)}{\sqrt{N(N+2)}}+O(\varepsilon),
\quad
    C_{TTT}^{\rm tri}
    =
    \frac{4(5N+32)}{(N+4)^{3/2}}+O(\varepsilon),
\\
    C_{ST_4T_4}& =\frac{12\sqrt{2}}{\sqrt{N(N+2)}}+O(\varepsilon),
\quad
    C_{TTT_4} =\frac{\sqrt{6}(N+16)}{N+4}+O(\varepsilon),
\quad
    C_{TT_4T_4} = \frac{24}{\sqrt{N+4}}+O(\varepsilon),
\\
    C_{T_4T_4T_4} &= 6\sqrt{6}+O(\varepsilon).
\label{eq:tri-scalar-constants}
\end{aligned}
\end{equation}
The current two-point function is still given by
\eqref{eq:wf-current-two-point-free}, now evaluated at $d=3$,
\begin{equation}
  C_j^{\rm tri}
  =
  4\kappa_3^2+O(\varepsilon)
  =
  \frac1{4\pi^2}+O(\varepsilon).
\label{eq:tri-current-central-charge}
\end{equation}
From the Ward identity \eqref{eq:ward-CjU-sqrt} the 
structure constant with the current in the tricritical theory is
\begin{equation}
  C_{TjU}^{\rm tri}
  =
  \sqrt{C_j^{\rm tri}}+O(\varepsilon)
  =
  \frac1{2\pi}+O(\varepsilon).
\label{eq:tri-CTjU}
\end{equation}

Since the current is weakly relevant on the surface, 
we could include it as a perturbation in addition to 
$S$, $T_{ij}$ and $T_{4,ijkl}$.
We do not include this term, and this is a consistent truncation because the beta function for the current $\beta^j$ contains no term independent of the coupling for the current $h^j$.

Indeed, the only candidate sources are the quadratic terms $C_{j\mathcal{O}_1\mathcal{O}_2}$ with $\mathcal{O}_1, \mathcal{O}_2 \in \{S,T,T_4\}$.
For the pairs $SS$, $ST$, $ST_4$ and $TT_4$, $O(N)$ symmetry means that there is no invariant tensor for each of these three point functions, since the antisymmetric representation $\mathbf A_N$ doesn't appear in any of $\mathbf 1\otimes \mathbf 1$, $\mathbf1\otimes\mathbf T_N$, $\mathbf1\otimes\mathbf T_{4,N}$ or $\mathbf T\otimes\mathbf T_{4,N}$,
where $\mathbf T_{4,N}$ denotes the $O(N)$ traceless symmetric representation of rank 4.

The antisymmetric representation does appear in both $\mathbf T_N\otimes\mathbf T_N$
and $\mathbf T_{4,N}\otimes\mathbf T_{4,N}$. 
While the structure constants $C_{jTT}$ and $C_{jT_4T_4}$ are generally non-zero,
the corresponding terms in the current beta function are proportional to 
\begin{equation}
    h^T_{ik}h^T_{jk} - h^T_{jk}h^T_{ik}, 
    \qquad
    h^{T_4}_{iklp} h^{T_4}_{jklp} - h^{T_4}_{jklp} h^{T_4}_{iklp},
\end{equation}
respectively, both of which vanish identically. 

The remaining current data is then not needed. 
Every term containing $C_{Sjj}$, $C_{Tjj}$ or $C_{T_4jj}$ carries at least one power of $h^j$, appearing as $C_{\mathcal{O}jj}h^{\mathcal{O}}h^j$ in $\beta^j$ or as $C_{\mathcal{O}jj}(h^j)^2$ in $\beta^{\mathcal{O}}$.
These terms all drop out when $h^j=0$.

Setting $h^j=0$ is therefore consistent. 

One particular reason not to turn on $h^j$ is that this direction is not under perturbative control.
The structure constants $C_{Sjj}$, $C_{Tjj}$ and $C_{T_4jj}$ vanish at leading order,
since the only allowed Wick contractions 
are between all the fields in the two $j$s and the quartics $S$, 
$T_{ij}$ and $T_{4,ijkl}$. Given that $j$ is antisymmetric while $S$, $T_{ij}$ and $T_{4,ijkl}$ 
are symmetric, this vanishes. 
The framework 
in Sections~\ref{sec:SandT} and~\ref{sec:scalar-tensor-anti}
solves the fixed point equations with $h\sim\varepsilon$ and $O(1)$ structure constants.
With $C_{\mathcal{O}jj}=O(\varepsilon)$, the terms $C_{\mathcal{O}jj}h^{\mathcal{O}}h^j$ in $\beta^j$ are $O(\varepsilon h^2)$, the same order as the terms already dropped. 
Therefore, a zero at $h^j\sim\varepsilon$ cannot be located at this order.
A fixed point with $h^j\neq0$ would require $h^j=O(1)$ and the beta functions would no longer be describing short RG flows.
Therefore, we do not turn on $j$ deformations.

The perturbation from the trivial surface defect is
\begin{equation}
\label{eq:tricritical-perturbation}
  S_{\rm def}
  =S_{\rm UV}
  +\mu^{-\gamma_S}h^S 
  \int_{\bR^2} d^2\tau\,S
  +\mu^{-\gamma_T}h^T_{ij}
  \int_{\bR^2} d^2\tau\,T_{ij}
  +\mu^{-\gamma_{T_4}} h^{T_4}_{ijkl}
  \int_{\bR^2} d^2\tau\, T_{4,ijkl}.
\end{equation}
The beta functions are modified from \eqref{eq:singlet-h-beta-functions}, \eqref{eq:tensor-h-beta-functions} due to the rank four operator.
They follow from the same first-order conformal perturbation theory procedure as in Section \ref{sec:SandT} with the operator set enlarged to include $T_4$,
\begin{align}
  \beta^S
  & =
  \gamma_S h^S
  +\pi C_{SSS}(h^S)^2
  +\pi C_{STT}h^T_{ij}h^T_{ij}
  +\pi C_{ST_4T_4}
  h^{T_4}_{ijkl}h^{T_4}_{ijkl}
  +O(h^3,\varepsilon h^2),
\label{eq:tri-beta-S}\\
  \beta^T(u)
  & =
  \gamma_T h^T(u)
  +2\pi C_{STT}h^S h^T(u)
  +\pi C_{TTT}u^i u^j h^T_{ik}h^T_{kj}\nonumber
  \\
  &\quad +2\pi C_{TTT_4}
  u^i u^j h^{T_4}_{ijkl}h^T_{kl}
  +\pi C_{TT_4T_4}
  u^i u^j h^{T_4}_{iklp}h^{T_4}_{jklp}
  +O(h^3,\varepsilon h^2),
\label{eq:tri-beta-T}\\
  \beta^{T_4}(u)
  & =
  \gamma_{T_4}h^{T_4}(u)
  +2\pi C_{ST_4T_4}h^S h^{T_4}(u)
  +\pi C_{TTT_4}h^T(u)^2\nonumber
  \\
  &\quad +2\pi C_{TT_4T_4}
  u^i u^j u^k u^l h^{T_4}_{ijkp}h^T_{lp}
  +\pi C_{T_4T_4T_4}
  u^i u^j u^k u^l h^{T_4}_{ijpq}h^{T_4}_{klpq}
  +O(h^3,\varepsilon h^2).
\label{eq:tri-beta-T4}
\end{align}
Note that these too are written in terms of the polarisation vectors,
\begin{align}
& h^T(u) = u^i u^j h^T_{ij}\,, \quad h^{T_4}(u) = u^i u^j u^k u^l h^{T_4}_{ijkl}\,,\\
& \beta^T(u) = u^i u^j \beta^T_{ij}\,, \quad \beta^{T_4}(u) = u^i u^j u^k u^l \beta^{T_4}_{ijkl}\,.
\end{align}
The rank four operator must be included because of the term
\begin{equation}
    \beta^{T_4}(u) \supset \pi C_{TTT_4} h^T(u)^2,
\end{equation}
which makes $h^{T_4}=0$ inconsistent whenever $h^T\neq0$.

\subsubsection{\texorpdfstring{$O(N)$}{O(N)} symmetric solutions}
The $O(N)$ symmetric fixed points are found in the same way as before, with
\begin{equation}
  \cD_N^{\rm tri}:
  \qquad
  h^{S,\cD_N,{\rm tri}}
  =\frac{\sqrt{2N(N+2)}(N+14)}{2\pi(N+8)(3N+22)}\varepsilon,
  \qquad
  h^{T,\cD_N,{\rm tri}}(u)=0,
  \qquad
  h^{T_4,\cD_N,{\rm tri}}(u)=0.
\label{eq:DN-tri-h-couplings}
\end{equation}

\subsubsection{\texorpdfstring{$O(n)\times O(m)$}{O(n)xO(m)} symmetric solutions}
To find solutions which preserve $O(n)\times O(m)$ symmetry, 
we use an ansatz for the rank two coupling of the form 
\eqref{eq:hn-hm-ansatz}. To translate this into the 
language of polarisation vectors, we split the 
polarisation vectors as 
$u^i = (\hat u^{\hat\imath},\check u^{\check\jmath})$, 
with $\hat u^2+\check u^2=0$ and find%
\footnote{Generically, there are multiple solutions 
with the residual $O(n)\times O(m)$ symmetry. An 
extra subscript can be used to distinguish them.}
\begin{equation}
\label{eq:hTtri}
    h^{T,\mathcal{D}_{n}}(u) = h^{\mathcal{D}_{n}}_n \hat u^2+h^{\mathcal{D}_{n}}_m \check u^2.
\end{equation}
For the rank four coupling, the most general $O(n)\times O(m)$ invariant polynomial is
\begin{equation}
\label{eq:hT4tri0}
h^{T_4}(u)=a(\hat u^2)^2 + b \hat u^2 \check u^2+ c(\check u^2)^2,
\end{equation}
$O(N)$ tracelessness fixes two of the coefficients in terms of the remaining one which we take to be $b\equiv h^{\mathcal{D}_n}_{(4)}$,
\begin{equation}
h^{T_4}(u)
=
h_{(4)}^{\mathcal{D}_n}
\left(
-\frac{m}{2(n+2)}(\hat u^2)^2
+\hat u^2\check u^2
-\frac{n}{2(m+2)}(\check u^2)^2
\right).
\label{eq:rank-four-coupling-ansatz}
\end{equation}
With this ansatz, the term in the action involving the rank four operator takes the form
\begin{equation}
h^{T_4}_{ijkl}T_{4,ijkl}
=
h_{(4)}^{\mathcal{D}_n}
\frac{(N+2)(N+4)}
{2(n+2)(m+2)}
T_{4,\hat\imath\hat\imath\check\jmath\check\jmath}.
\end{equation}

After substituting \eqref{eq:hTtri} and \eqref{eq:rank-four-coupling-ansatz}, 
the fixed point equations are three coupled quadratic 
equations in three unknowns. To avoid overcomplications, 
we restrict to the special case of even $N$ with $n=m=N/2$.
As in \eqref{eq:hT-N/2}, tracelessness leaves a single rank two coupling, which we again denote
$h_{N/2}^{\cD_{N/2}}\equiv h_n^{\cD_{N/2}} = -h_m^{\cD_{N/2}}$.
Its sign is again immaterial, since the $\mathbb{Z}_2$ block exchange sends $h_{N/2}^{\cD_{N/2}}\rightarrow-h_{N/2}^{\cD_{N/2}}$ while leaving the singlet and rank four couplings invariant, the latter because it is symmetric under $\hat u \leftrightarrow \check u$ for $n=m$.
Then, the leading order fixed point equations are
\begin{align}
0={}&
\gamma_S h^{S,\mathcal{D}_{N/2}}
+\pi C_{SSS}\bigl(h^{S,\mathcal{D}_{N/2}}\bigr)^2
+\pi N C_{STT}\bigl(h_{N/2}^{\mathcal{D}_{N/2}}\bigr)^2
+\frac{\pi N^2(N+2)}{12(N+4)}
C_{ST_4T_4}
\bigl(h_{(4)}^{\mathcal{D}_{N/2}}\bigr)^2,
\label{eq:En-equal-rank-fixed-point-S}
\\
0={}&
\left(
\gamma_T
+2\pi C_{STT}h^{S,\mathcal{D}_{N/2}}
-\frac{\pi N}{3}C_{TTT_4}
h_{(4)}^{\mathcal{D}_{N/2}}
\right)
h_{N/2}^{\mathcal{D}_{N/2}},
\label{eq:En-equal-rank-fixed-point-T}
\\
0={}&
\left(
\gamma_{T_4}
+2\pi C_{ST_4T_4}h^{S,\mathcal{D}_{N/2}}
\right)h_{(4)}^{\mathcal{D}_{N/2}}
-2\pi C_{TTT_4}\frac{N+4}{N+2}
\bigl(h_{N/2}^{\mathcal{D}_{N/2}}\bigr)^2
\nonumber\\&\quad
-\pi C_{T_4T_4T_4}
\frac{(N-2)(N+8)}{18(N+4)}
\bigl(h_{(4)}^{\mathcal{D}_{N/2}}\bigr)^2.
\label{eq:En-equal-rank-fixed-point-T4}
\end{align}

Since \eqref{eq:En-equal-rank-fixed-point-T} factorises, the solutions fall into two families according to whether $h_{N/2}^{\cD_{N/2}}$ vanishes. In both cases, there is an equation which is linear in $h^S$ and eliminates it, leaving the singlet equation as a single quadratic for $h^{\cD_{N/2}}_{(4)}$.

We first study the family of solutions where the rank two coupling vanishes, $h^{\mathcal{D}_{N/2}}_{N/2}=0$.
From \eqref{eq:En-equal-rank-fixed-point-T4}, we solve for $h^{S,\mathcal{D}_{N/2}}$,
\begin{equation}
    h^{S,\mathcal{D}_{N/2}}
    =
    -\frac{\gamma_{T_4}}{2\pi C_{ST_4T_4}}
    +\frac{C_{T_4T_4T_4}}{C_{ST_4T_4}}\frac{(N-2)(N+8)}{36(N+4)}h^{\mathcal{D}_{N/2}}_{(4)}.
\end{equation}
The value of $h^{\mathcal{D}_{N/2}}_{(4)}$ is then constrained by \eqref{eq:En-equal-rank-fixed-point-S}.
Substituting the expression for $h^{S,\mathcal{D}_{N/2}}$ and multiplying by an overall nonzero factor, the constraint reads
\begin{equation}
\begin{aligned}
0={}&
\pi^2
\Bigg[
C_{SSS}C_{T_4T_4T_4}^{\,2}
\left(
\frac{(N-2)(N+8)}{18(N+4)}
\right)^2
+
\frac{N^2(N+2)}{3(N+4)}
C_{ST_4T_4}^{\,3}
\Bigg]
\bigl(h_{(4)}^{\mathcal{D}_{N/2}}\bigr)^2
\\[2mm]
&-
\pi C_{T_4T_4T_4}
\frac{(N-2)(N+8)}{9(N+4)}
\left(
C_{SSS}\gamma_{T_4}
-C_{ST_4T_4}\gamma_S
\right)
h_{(4)}^{\mathcal{D}_{N/2}}
\\[2mm]
&+
\gamma_{T_4}
\left(
C_{SSS}\gamma_{T_4}
-2C_{ST_4T_4}\gamma_S
\right).
\end{aligned}
\end{equation}
The discriminant of the quadratic is given by 
\begin{equation}
\begin{aligned}
  \Delta_{(4)}^{\mathcal{D}_{N/2}}
  ={}&
  \frac{N^2\pi^2 C_{ST_4T_4}^{\,2}}{(N+4)^2}
  \Bigg[
    C_{T_4T_4T_4}^{\,2}\gamma_S^2
    \frac{(N-2)^2(N+8)^2}{81N^2}
  \\
  &\qquad
    -\frac{4}{3}
    C_{ST_4T_4}\gamma_{T_4}(N+2)(N+4)
    \left(
      C_{SSS}\gamma_{T_4}
      -2C_{ST_4T_4}\gamma_S
    \right)
  \Bigg].
\end{aligned}
\label{eq:En-hn-zero-discriminant}
\end{equation}
After substituting the CFT data for the tricritical model, the discriminant is
\begin{equation}
    \Delta^{\mathcal{D}_{N/2},\text{tri}}_{(4)} = \frac{3072\pi^2(N+2)}{N(N+4)^2(3N+22)^2}\varepsilon^2 \left(N^4 -180 N^3 -3068 N^2 -9792 N+12544\right).
\end{equation}
This expression is negative at $N=194$ and positive at $N=196$.
Hence for even $N\leq194$ there are no real solutions, and for even $N\geq196$ there is a pair.

Note that these solutions have an additional 
$\bZ_2$ symmetry exchanging the two $N/2$ blocks. 
This exchanges $\hat u$ and $\check u$ in 
\eqref{eq:hTtri} and \eqref{eq:rank-four-coupling-ansatz}.
The defect conformal manifold for this configuration is therefore
\begin{equation}
    \widetilde{\Gr}(N/2,\mathbb{R}^N)=\frac{O(N)}{\left(O(N/2)\times O(N/2)\right)\rtimes\mathbb{Z}_2}.
\end{equation}
For $N=2$, this space is still a circle, so its fundamental group is $\pi_1\left(\widetilde{\Gr}(1,\mathbb{R}^2)\right)=\mathbb{Z}$. 
For even $N\geq4$, the fundamental group is instead
\begin{equation}
    \pi_1\left(\widetilde{\Gr}(N/2,\mathbb{R}^N)\right)
    =
    \begin{cases}
        \mathbb{Z}_2\times\mathbb{Z}_2, \quad & N\equiv 0 \mod 4,
        \\
        \mathbb{Z}_4, &N\equiv2\mod4.
    \end{cases}
\end{equation}
This allows for vortex configurations on the surface operator, 
as discussed after \eqref{eq:grassmannian}.

Now we look for solutions with $h_{N/2}^{\mathcal{D}_{N/2}}\neq0$. We solve for $h^{S,\mathcal{D}_{N/2}}$ using \eqref{eq:En-equal-rank-fixed-point-T},
\begin{equation}
h^{S,\mathcal{D}_{N/2}}
=
-\frac{\gamma_T}{2\pi C_{STT}}
+\frac{N C_{TTT_4}}{6C_{STT}}
h_{(4)}^{\mathcal{D}_{N/2}}.
\label{eq:En-hn-nonzero-hS}
\end{equation}
The third fixed point equation \eqref{eq:En-equal-rank-fixed-point-T4} allows us to determine $\left(h_{N/2}^{\mathcal{D}_{N/2}}\right)^2$ as
\begin{equation}
\begin{aligned}
\label{eq:h-N/2-squared-constraint}
\bigl(h_{N/2}^{\mathcal{D}_{N/2}}\bigr)^2
={}&
\frac{N+2}
{2\pi (N+4)C_{TTT_4}}
h_{(4)}^{\mathcal{D}_{N/2}}
\Biggl[
\gamma_{T_4}
-\frac{C_{ST_4T_4}}{C_{STT}}\gamma_T
\\
&\qquad
+\pi\left(
\frac{N C_{ST_4T_4}C_{TTT_4}}
{3C_{STT}}
-\frac{(N-2)(N+8)C_{T_4T_4T_4}}
{18(N+4)}
\right)
h_{(4)}^{\mathcal{D}_{N/2}}
\Biggr],
\end{aligned}
\end{equation}
in terms of $h^{\mathcal{D}_{N/2}}_{(4)}$.
The final fixed point equation \eqref{eq:En-equal-rank-fixed-point-S} then serves as a quadratic constraint on $h^{\mathcal{D}_{N/2}}_{(4)}$ which, up to an overall nonzero factor, is given by
\begin{equation}
\begin{aligned}
0&=
C_{TTT_4}\gamma_T
 \bigl(C_{SSS}\gamma_T-2C_{STT}\gamma_S\bigr)
+\pi^2N^2
\Bigg[
\frac{1}{9}C_{SSS}C_{TTT_4}^3
+\frac{N+2}{N+4}
 C_{STT}^2C_{ST_4T_4}C_{TTT_4}
\\
&\qquad{}-\frac{(N+2)(N-2)(N+8)}
 {9N(N+4)^2}
 C_{STT}^3C_{T_4T_4T_4}
\Bigg]
\bigl(h_{(4)}^{\mathcal{D}_{N/2}}\bigr)^2
\\&\quad
+2\pi N
\Bigg[
-\frac{1}{3}C_{TTT_4}^2
 \bigl(C_{SSS}\gamma_T-C_{STT}\gamma_S\bigr)
+\frac{N+2}{N+4}C_{STT}^2
 \bigl(C_{STT}\gamma_{T_4}
       -C_{ST_4T_4}\gamma_T\bigr)
\Bigg]
h_{(4)}^{\mathcal{D}_{N/2}}.
\end{aligned}
\end{equation}
After substituting the data for the tricritical model 
\eqref{eq:tri-uv-gammas} and \eqref{eq:tri-scalar-constants} 
the discriminant is
\begin{equation}
    \Delta^{\mathcal{D}_{N/2},\text{tri}} = 
    \frac{8192 \pi ^2 \varepsilon^2 (N+2) (N+16)^4 }{N
   (N+4)^4 (3 N+22)^2}\left(-2
   N^3-5 N^2+404 N+2468\right).
\end{equation}
The discriminant is positive for even $N\leq14$, where the quadratic has two real roots for the rank four coupling.
Since \eqref{eq:h-N/2-squared-constraint} determines the square of the rank two coupling, reality requires the right-hand side to be positive, which is a further condition on $h_{(4)}^{\cD_{N/2}}$.
For $N=2,4,6,8,10$, only one value for $h_{(4)}^{\cD_{N/2}}$ satisfies $\left(h_{N/2}^{\cD_{N/2}}\right)^2>0$, leading to a pair of solutions distinguished by the sign of the rank two coupling.
For $N=12,14$, the fixed points are not real.

\subsubsection{Operators of dimension near two}
The stability matrix is now a $3\times3$ matrix, obtained by evaluating
\begin{equation}
    \Gamma =    \frac{\partial(\beta^S,\beta^T_{ij},\beta^{T_4}_{pqrs})}{\partial(h^S,h^T_{kl},h^{T_4}_{tuvw})}
\end{equation}
at a fixed point. For the $O(N)$ symmetric fixed point 
$\mathcal{D}_{N}^{\text{tri}}$ characterised in 
\eqref{eq:DN-tri-h-couplings}, the stability matrix is given by
\begin{equation}
    \Gamma^{\mathcal{D}_N} = \begin{pmatrix}
        \gamma_S + 2\pi C_{SSS} h^{S,\mathcal{D}_N,\text{tri}} & 0 & 0
        \\
        0 & \gamma_T + 2\pi C_{STT} h^{S,\mathcal{D}_N,\text{tri}} & 0
        \\
        0 & 0 & \gamma_{T_4} + 2\pi C_{ST_4T_4} h^{S,\mathcal{D}_N,\text{tri}}
    \end{pmatrix},
\end{equation}
where we suppressed the projectors on the rank two and rank four tensor subspaces.
The unbroken $O(N)$ symmetry forbids mixing between the three representations, which is why the matrix is diagonal and why we can read off the eigenvalues.
Substituting in the values for the CFT data, we find
\begin{equation}
    \gamma_S^{\mathcal{D}_N,\text{tri}} = \frac{2(N+14)}{3N+22}\varepsilon,
    \quad 
    \gamma_T^{\mathcal{D}_N,\text{tri}} = -\frac{2(N^2-112)}{(N+8)(3N+22)}\varepsilon,
    \quad
    \gamma_{T_4}^{\mathcal{D}_N,\text{tri}} = - \frac{6(N^2+10N-8)}{(N+8)(3N+22)}\varepsilon.
\end{equation}
In the tricritical theory the current has dimension $2-\varepsilon$. 
In the $O(N)$ symmetric fixed point the current is conserved so does 
not develop an anomalous dimension. In the symmetry breaking fixed 
points the $O(n)$ and $O(m)$ adjoint components are still conserved, 
while the bifundamental components get anomalous dimensions.
Because of the transverse vector index, 
$j_{\hat\imath\check\jmath,r}^{\rm tri}$ does not mix with the tilt 
$\bt_{\hat\imath\check\jmath}^{\rm tri}$ originating from $T_{ij}$ and $T_{4,ijkl}$.

For $N=1$, the symmetric fixed point $\mathcal{D}_N$ is perturbatively stable because only the singlet operator exists for this value of $N$ and its anomalous dimension is positive. 

For $N\geq2$ where the rank four operator exists, its anomalous dimension at the symmetric fixed point $\mathcal{D}_N$ is negative. 
This means that the symmetric fixed point is perturbatively unstable for all $N\geq2$ when symmetry breaking perturbations are allowed. 
It remains stable when only symmetry preserving deformations by the singlet are allowed.

\subsubsection{Displacements and tilts}
In the tricritical theory, the displacement and tilt operators receive extra contributions from the rank four operator. 
Performing a calculation analogous to that leading to \eqref{eq:t-explicit-from-variation} and \eqref{eq:D-explicit-from-variation}, we find
\begin{align}
  \bt_{\hat\imath\check\jmath}
  &=
  2\mu^{-\gamma_T}\big(h_n-h_m\big)T_{\hat\imath\check\jmath} + 4\mu^{-\gamma_{T_4}}\left(h^{T_4}_{\hat\imath klm}T_{4,\check\jmath klm}-h^{T_4}_{\check\jmath klm}T_{4,\hat\imath klm}\right),
  \label{eq:t-tricritical-from-variation}
\\
  \bD_r&=
  -\mu^{-\gamma_S}h^S\partial_rS
  -\mu^{-\gamma_T}h^T_{ij}\partial_rT_{ij}
  -\mu^{-\gamma_{T_4}}h^{T_4}_{ijkl}\partial_rT_{4,ijkl}
\nonumber  \\
  &=
  -\sqrt{2\Delta_S}\,\mu^{-\gamma_S}h^{S} V_r
  -\sqrt{2\Delta_T}\,\mu^{-\gamma_T}h^T_{ij}U_{ij,r}
  -\sqrt{2\Delta_{T_4}}\,\mu^{-\gamma_{T_4}}h^{T_4}_{ijkl}U^4_{ijkl,r},
  \label{eq:D-tricritical-from-variation}
\end{align}
where $U^4_{ijkl,r} = \frac{1}{\sqrt{2\Delta_{T_4}}}\partial_r T_{4,ijkl}$ is the unit normalised descendant of the rank four operator.

At the $O(N)$ symmetric fixed point $\cD_N$ \eqref{eq:DN-tri-h-couplings} there are no tilts.
To determine $C^{\mathcal{D}_n}_\bt$ for the coupling ansatz in \eqref{eq:rank-four-coupling-ansatz}, we first write the combination of rank four operators in \eqref{eq:t-tricritical-from-variation} as 
\begin{equation}
    h^{T_4}_{\hat\imath klm}T_{4,\check\jmath klm}-h^{T_4}_{\check\jmath klm}T_{4,\hat\imath klm} = -\frac{(N+2)(N+4)}{2(n+2)(m+2)}h^{\mathcal{D}_n}_{(4)} T_{4,\hat\imath\check\jmath \hat k \hat k}.
\end{equation}
The coefficient of the tilt two-point function is then
\begin{equation}
\label{eq:tricritical-Ct-at-En}
    C^{\mathcal{D}_n}_\bt = 
    2(h^{\mathcal{D}_n}_n - h^{\mathcal{D}_n}_m)^2+
    \frac{(N+2)^2(N+4)}{3(n+2)(m+2)}\left(h^{\mathcal{D}_n}_{(4)}\right)^2,
\end{equation}
which for $n=m=N/2$, simplifies to
\begin{equation}
\label{eq:tricritical-Ct-N/2}
    C^{\mathcal{D}_{N/2}}_\bt = 
    8(h^{\mathcal{D}_{N/2}}_{N/2})^2+
    \frac{4(N+2)^2}{3(N+4)}\left(h^{\mathcal{D}_{N/2}}_{(4)}\right)^2.
\end{equation}

The displacement normalisation at a fixed point is
\begin{equation}
\label{eq:tricritical-CD-general}
    C_\bD = 2\Delta_S \big(h^S\big)^2 + 2\Delta_T h^T_{ij} h^T_{ij} + 2\Delta_{T_4} h^{T_4}_{ijkl} h^{T_4}_{ijkl}
\end{equation}
with the necessary values of the couplings.
For the symmetric fixed point \eqref{eq:DN-tri-h-couplings},
\begin{equation}
  C_{\bD}^{\cD_N,{\rm tri}}
  =2\Delta_S\big(h^{S,\cD_N,{\rm tri}}\big)^2
  =\frac{2N(N+2)(N+14)^2}{\pi^2(N+8)^2(3N+22)^2}\varepsilon^2
  +O(\varepsilon^3).
\label{eq:CD-DN-tri-evaluated}
\end{equation}
For the fixed points $\mathcal{D}_n$, we use the ansatz \eqref{eq:rank-four-coupling-ansatz} to calculate the norm of the rank four coupling,
\begin{equation}
    h^{T_4}_{ijkl} h^{T_4}_{ijkl} = \frac{nm(N+2)(N+4)}{12(n+2)(m+2)} \left(h^{\mathcal{D}_n}_{(4)}\right)^2,
\end{equation}
and therefore the displacement two-point coefficient is
\begin{equation}
\label{eq:tricritical-CD-at-En}
    C_{\bD}^{\mathcal{D}_n} = 4 (h^{S,\mathcal{D}_n})^2 + \frac{4nN}{m} (h_n^{\mathcal{D}_n})^2 + \frac{nm(N+2)(N+4)}{3(n+2)(m+2)} \left(h^{\mathcal{D}_n}_{(4)}\right)^2.
\end{equation}
Since the fixed point couplings are of order $\varepsilon$, in the expression for $C_\bD$ above we replaced $\Delta_S$, $\Delta_T$ and $\Delta_{T_4}$ by their leading value $2$. For $n=m=N/2$, the displacement normalisation is
\begin{equation}
\label{eq:tricritical-CD-N/2}
    C_{\bD}^{\mathcal{D}_{N/2}} = 4 (h^{S,\mathcal{D}_{N/2}})^2 + 4N (h_{N/2}^{\mathcal{D}_{N/2}})^2 + \frac{N^2(N+2)}{3(N+4)} \left(h^{\mathcal{D}_{N/2}}_{(4)}\right)^2.
\end{equation}

\section{Discussion}
\label{sec:discuss}

We presented a unified approach to studying surface defects at first order 
in conformal perturbation theory. This reproduces many of the results 
in the Wilson--Fisher $O(N)$ model found in~\cite{Trepanier:2023tvb, 
Raviv-Moshe:2023yvq, Giombi:2023dqs}, but extended the results 
to many other settings including the long range model, 
overlapping with~\cite{Bianchi:2024eqm}, the chiral $O(N)\times O(2)$ 
model and the tricritical $O(N)$ model. As an example, in the 
last case we find that the rank four symmetric traceless perturbation renders the $O(N)$ symmetric defect unstable under generic symmetry-breaking perturbations for every $N\geq2$. 
This defect is stable for $N\geq2$ when only symmetry-preserving deformations by the singlet operator are allowed.
For $N=1$, only the singlet perturbation exists and the symmetric defect is stable. 
The same techniques could be carried over to other theories, like those 
studied in~\cite{Anataichuk:2025zoq}.

In addition to studying deformation of the trivial UV defect, we 
looked at deformations away from an existing defect in 
Section~\ref{sec:deformations-of-the-defect}. In particular, 
we show that under certain assumptions, requiring an $O(N)$ 
invariant IR fixed point constrains the starting point to be equivalent to 
one of the $\cD_n$ defects that arise from a deformation of 
the bulk theory.

Results at higher loop order in the $\varepsilon$ expansion were 
presented in \cite{Diatlyk:2024ngd, deSabbata:2024xwn}. This extension 
is beyond the scope of the present work, though it should not be hard 
to adapt those calculations to the other models in $d=4-\varepsilon$. 
In particular, we present a rich family of fixed point collisions in the 
long range model in Section~\ref{sec:near2-LR}, which would 
be corrected at higher orders in perturbation theory. 
Such degenerations appear at the boundaries of stability 
regions analysed in extensive detail in a unified manner 
in Section~\ref{sec:real-and-stable} and then applied to the 
different models. Since the long-range model has a continuous 
parameter $\alpha$, the degenerations would remain, but at 
shifted values of $\alpha$. So these are defect marginal
operators that exist to all orders in the $\varepsilon$ 
expansion, but presumably are not exactly marginal.

Going to higher orders in perturbation theory may also be useful 
to try to make connections with $O(N)$ models in three dimensions, 
which possess varied universality classes of surface 
defects~\cite{Metlitski:2020cqy, Toldin:2021kun, Krishnan:2023cff}.

Tilt and displacement operators are nonperturbatively protected. 
Tilts are exactly marginal 
and move the defect along the defect conformal manifold, which in our 
cases are either the Grassmannian 
$\Gr(n,\bR^N)=O(N)/(O(n)\times O(m))$ or 
$O(N)/(U(p)\times O(N-2p))$, 
a $O(2p)/U(p)$ bundle over $\Gr(2p,\bR^N)$. In particular the second 
case is an example of a homogeneous but not symmetric space. 
While this is a natural consequence of complicated symmetry 
breaking patterns, in practice few have been studied in detail 
(but see e.g.~\cite{Drukker:2022txy}).

In particular, in this case the theory has two different types of 
tilt operators, those responsible for motion along the base 
and those along the fibre. These surface operators 
offer a tractable example of such rich defect conformal manifolds.

We also noted that the real Grassmannian is not simply connected, 
so there are natural local vortex operators on them, which warrant 
further study. In certain cases the space of defects is not connected, 
allowing for line interfaces between defects on the two 
connected components. Of course line interfaces can exist also between 
two surface operators on the same connected component, or 
between completely inequivalent surface operators. 
Yet another generalisation of connecting two semi-infinite surfaces 
is to glue them along a ``crease,'' as already considered in 
this context in \cite{Trepanier:2023tvb}. This fits within the 
more general framework of defect fusion, as discussed in
\cite{Diatlyk:2024qpr, Kravchuk:2024qoh} and it would 
be interesting to explore it in detail for our examples.

For all the flows and fixed points considered,
we evaluated the 
normalisation constants 
of the displacements and tilts: $C_\bD$, $C_\bt$ respectively.
For the scalar--tensor and scalar--tensor--antisymmetric flows analysed in Sections \ref{sec:SandT} and \ref{sec:scalar-tensor-anti},
we also 
examined their behaviour under RG flows showing in detail the 
protection of the dimension at all scales. In the 
simple one-dimensional system of a symmetry preserving 
flow we solved for the explicit 
flow. In other cases the explicit flows can generically only be 
solved numerically.

\section*{Acknowledgements}

We are grateful to E.~de Sabbata and Z.~Kong for inspiring 
discussions. 
BS's and ND's research is supported by the Science Technology 
\& Facilities Council (STFC) under the grant ST/X000753/1.

\bibliographystyle{utphys2}
\bibliography{refs}

\providecommand{\href}[2]{#2}\begingroup\raggedright\begin{thebibliography}{10}\setlength{\parskip}{1pt}\setlength{\itemsep}{0pt plus 0.3ex}

\bibitem{Billo:2016cpy}
M.~Bill\`o, V.~Gon\c{c}alves, E.~Lauria, and M.~Meineri, ``{Defects in conformal field theory},'' \href{http://dx.doi.org/10.1007/JHEP04(2016)091}{{\em JHEP} {\bfseries 04} (2016) 091}, \href{http://arxiv.org/abs/1601.02883}{{\ttfamily arXiv:1601.02883}}.

\bibitem{Padayasi:2021sik}
J.~Padayasi, A.~Krishnan, M.~A. Metlitski, I.~A. Gruzberg, and M.~Meineri, ``{The extraordinary boundary transition in the 3d $O(N)$ model via conformal bootstrap},'' \href{http://dx.doi.org/10.21468/SciPostPhys.12.6.190}{{\em SciPost Phys.} {\bfseries 12} no.~6, (2022) 190}, \href{http://arxiv.org/abs/2111.03071}{{\ttfamily arXiv:2111.03071}}.

\bibitem{Drukker:2022pxk}
N.~Drukker, Z.~Kong, and G.~Sakkas, ``{Broken global symmetries and defect conformal manifolds},'' \href{http://dx.doi.org/10.1103/PhysRevLett.129.201603}{{\em Phys. Rev. Lett.} {\bfseries 129} no.~20, (2022) 201603}, \href{http://arxiv.org/abs/2203.17157}{{\ttfamily arXiv:2203.17157}}.

\bibitem{Gabai:2025zcs}
B.~Gabai, A.~Sever, and D.-l. Zhong, ``{Universal constraints for conformal line defects},'' \href{http://dx.doi.org/10.1103/gsfg-wrps}{{\em Phys. Rev. D} {\bfseries 112} no.~6, (2025) 065004}, \href{http://arxiv.org/abs/2501.06900}{{\ttfamily arXiv:2501.06900}}.

\bibitem{Belton:2025hbu}
J.~Belton, N.~Drukker, Z.~Kong, and A.~Stergiou, ``{Fine spectrum from crude analytic bootstrap},'' \href{http://dx.doi.org/10.1088/1751-8121/adf925}{{\em J. Phys. A} {\bfseries 58} no.~34, (2025) 345401}, \href{http://arxiv.org/abs/2503.07710}{{\ttfamily arXiv:2503.07710}}.

\bibitem{Gabai:2025hwf}
B.~Gabai, V.~Gorbenko, and J.~Qiao, ``{Yang--Mills flux tube in AdS},'' \href{http://arxiv.org/abs/2508.08250}{{\ttfamily arXiv:2508.08250}}.

\bibitem{Kong:2025sbk}
Z.~Kong, ``{Integral identities from symmetry breaking of conformal defects},'' in {\em {16th International Workshop on Lie Theory and Its Applications in Physics}}.
\newblock 9, 2025.
\newblock \href{http://arxiv.org/abs/2509.23797}{{\ttfamily arXiv:2509.23797}}.

\bibitem{Girault:2025kzt}
B.~Girault, M.~F. Paulos, and P.~van Vliet, ``{Consequences of symmetry-breaking on conformal defect data},'' \href{http://arxiv.org/abs/2509.26561}{{\ttfamily arXiv:2509.26561}}.

\bibitem{Belton:2025ief}
J.~Belton and Z.~Kong, ``{There and back again: bulk-to-defect via Ward identities},'' \href{http://dx.doi.org/10.1007/JHEP05(2026)103}{{\em JHEP} {\bfseries 05} (2026) 103}, \href{http://arxiv.org/abs/2510.08519}{{\ttfamily arXiv:2510.08519}}.

\bibitem{Drukker:2025dfm}
N.~Drukker, Z.~Kong, and P.~Kravchuk, ``{Nonlinearly realised defect symmetries and anomalies},'' \href{http://arxiv.org/abs/2512.15913}{{\ttfamily arXiv:2512.15913}}.

\bibitem{whale}
N.~Drukker, Z.~Kong, and P.~Kravchuk. To appear.

\bibitem{Bianchi:2015liz}
L.~Bianchi, M.~Meineri, R.~C. Myers, and M.~Smolkin, ``{R\'enyi entropy and conformal defects},'' \href{http://dx.doi.org/10.1007/JHEP07(2016)076}{{\em JHEP} {\bfseries 07} (2016) 076}, \href{http://arxiv.org/abs/1511.06713}{{\ttfamily arXiv:1511.06713}}.

\bibitem{Herzog:2017xha}
C.~P. Herzog and K.-W. Huang, ``{Boundary conformal field theory and a boundary central charge},'' \href{http://dx.doi.org/10.1007/JHEP10(2017)189}{{\em JHEP} {\bfseries 10} (2017) 189}, \href{http://arxiv.org/abs/1707.06224}{{\ttfamily arXiv:1707.06224}}.

\bibitem{Herzog:2021spv}
C.~P. Herzog and V.~Schaub, ``{A sum rule for boundary contributions to the trace anomaly},'' \href{http://dx.doi.org/10.1007/JHEP01(2022)121}{{\em JHEP} {\bfseries 01} (2022) 121}, \href{http://arxiv.org/abs/2107.11604}{{\ttfamily arXiv:2107.11604}}.

\bibitem{Drukker:2020atp}
N.~Drukker, M.~Probst, and M.~Tr\'epanier, ``{Defect CFT techniques in the 6d $\mathcal{N} = (2,0)$ theory},'' \href{http://dx.doi.org/10.1007/JHEP03(2021)261}{{\em JHEP} {\bfseries 03} (2021) 261}, \href{http://arxiv.org/abs/2009.10732}{{\ttfamily arXiv:2009.10732}}.

\bibitem{Jensen:2017eof}
K.~Jensen, E.~Shaverin, and A.~Yarom, ``{'t~Hooft anomalies and boundaries},'' \href{http://dx.doi.org/10.1007/JHEP01(2018)085}{{\em JHEP} {\bfseries 01} (2018) 085}, \href{http://arxiv.org/abs/1710.07299}{{\ttfamily arXiv:1710.07299}}.

\bibitem{Choi:2025ebk}
Y.~Choi, H.~Ha, D.~Kim, Y.~Kusuki, S.~Ohyama, and S.~Ryu, ``{Higher structures on boundary conformal manifolds: Higher Berry phase and boundary conformal field theory},'' \href{http://dx.doi.org/10.1103/z8pw-8ckx}{{\em Phys. Rev. D} {\bfseries 113} no.~10, (2026) 106005}, \href{http://arxiv.org/abs/2507.12525}{{\ttfamily arXiv:2507.12525}}.

\bibitem{Wen:2025xka}
X.~Wen, ``{Space of conformal boundary conditions from the view of higher Berry phase: Flow of Berry curvature in parametrized BCFTs},'' \href{http://arxiv.org/abs/2507.12546}{{\ttfamily arXiv:2507.12546}}.

\bibitem{Copetti:2025sym}
C.~Copetti, ``{'t Hooft anomalies and defect conformal manifolds: topological signatures from modulated effective actions},'' \href{http://arxiv.org/abs/2507.15466}{{\ttfamily arXiv:2507.15466}}.

\bibitem{Jensen:2015swa}
K.~Jensen and A.~O'Bannon, ``{Constraint on defect and boundary renormalization group flows},'' \href{http://dx.doi.org/10.1103/PhysRevLett.116.091601}{{\em Phys. Rev. Lett.} {\bfseries 116} no.~9, (2016) 091601}, \href{http://arxiv.org/abs/1509.02160}{{\ttfamily arXiv:1509.02160}}.

\bibitem{Casini:2018nym}
H.~Casini, I.~Salazar~Landea, and G.~Torroba, ``{Irreversibility in quantum field theories with boundaries},'' \href{http://dx.doi.org/10.1007/JHEP04(2019)166}{{\em JHEP} {\bfseries 04} (2019) 166}, \href{http://arxiv.org/abs/1812.08183}{{\ttfamily arXiv:1812.08183}}.

\bibitem{Wang:2020xkc}
Y.~Wang, ``{Surface defect, anomalies and $b$-extremization},'' \href{http://dx.doi.org/10.1007/JHEP11(2021)122}{{\em JHEP} {\bfseries 11} (2021) 122}, \href{http://arxiv.org/abs/2012.06574}{{\ttfamily arXiv:2012.06574}}.

\bibitem{Shachar:2024ubf}
T.~Shachar, R.~Sinha, and M.~Smolkin, ``{The defect $b$-theorem under bulk RG flows},'' \href{http://dx.doi.org/10.1007/JHEP09(2024)057}{{\em JHEP} {\bfseries 09} (2024) 057}, \href{http://arxiv.org/abs/2404.18403}{{\ttfamily arXiv:2404.18403}}.

\bibitem{Metlitski:2020cqy}
M.~A. Metlitski, ``{Boundary criticality of the $O(N)$ model in $d = 3$ critically revisited},'' \href{http://dx.doi.org/10.21468/SciPostPhys.12.4.131}{{\em SciPost Phys.} {\bfseries 12} no.~4, (2022) 131}, \href{http://arxiv.org/abs/2009.05119}{{\ttfamily arXiv:2009.05119}}.

\bibitem{Toldin:2021kun}
F.~P. Toldin and M.~A. Metlitski, ``{Boundary criticality of the 3D $O(N)$ model: from normal to extraordinary},'' \href{http://dx.doi.org/10.1103/PhysRevLett.128.215701}{{\em Phys. Rev. Lett.} {\bfseries 128} no.~21, (2022) 215701}, \href{http://arxiv.org/abs/2111.03613}{{\ttfamily arXiv:2111.03613}}.

\bibitem{Krishnan:2023cff}
A.~Krishnan and M.~A. Metlitski, ``{A plane defect in the 3d $O(N)$ model},'' \href{http://dx.doi.org/10.21468/SciPostPhys.15.3.090}{{\em SciPost Phys.} {\bfseries 15} no.~3, (2023) 090}, \href{http://arxiv.org/abs/2301.05728}{{\ttfamily arXiv:2301.05728}}.

\bibitem{Trepanier:2023tvb}
M.~Tr\'epanier, ``{Surface defects in the $O(N)$ model},'' \href{http://dx.doi.org/10.1007/JHEP09(2023)074}{{\em JHEP} {\bfseries 09} (2023) 074}, \href{http://arxiv.org/abs/2305.10486}{{\ttfamily arXiv:2305.10486}}.

\bibitem{Raviv-Moshe:2023yvq}
A.~Raviv-Moshe and S.~Zhong, ``{Phases of surface defects in scalar field theories},'' \href{http://dx.doi.org/10.1007/JHEP08(2023)143}{{\em JHEP} {\bfseries 08} (2023) 143}, \href{http://arxiv.org/abs/2305.11370}{{\ttfamily arXiv:2305.11370}}.

\bibitem{Giombi:2023dqs}
S.~Giombi and B.~Liu, ``{Notes on a surface defect in the $O(N)$ model},'' \href{http://dx.doi.org/10.1007/JHEP12(2023)004}{{\em JHEP} {\bfseries 12} (2023) 004}, \href{http://arxiv.org/abs/2305.11402}{{\ttfamily arXiv:2305.11402}}.

\bibitem{Diatlyk:2024ngd}
O.~Diatlyk, Z.~Sun, and Y.~Wang, ``{Surprises in the ordinary: $O(N)$ invariant surface defect in the $\varepsilon$-expansion},'' \href{http://dx.doi.org/10.1007/JHEP06(2025)131}{{\em JHEP} {\bfseries 06} (2025) 131}, \href{http://arxiv.org/abs/2411.16522}{{\ttfamily arXiv:2411.16522}}.

\bibitem{Anataichuk:2025zoq}
A.~Anataichuk and S.~Harribey, ``{Note on surface defects in multiscalar critical models},'' \href{http://dx.doi.org/10.1088/1751-8121/adf26d}{{\em J. Phys. A} {\bfseries 58} no.~31, (2025) 315403}, \href{http://arxiv.org/abs/2503.05519}{{\ttfamily arXiv:2503.05519}}.

\bibitem{Bianchi:2024eqm}
L.~Bianchi, L.~S. Cardinale, and E.~de~Sabbata, ``{Defects in the long-range $O(N)$ model},'' \href{http://dx.doi.org/10.1088/1751-8121/adf788}{{\em J. Phys. A} {\bfseries 58} no.~33, (2025) 335401}, \href{http://arxiv.org/abs/2412.08697}{{\ttfamily arXiv:2412.08697}}.

\bibitem{Osborn:2017ucf}
H.~Osborn and A.~Stergiou, ``{Seeking fixed points in multiple coupling scalar theories in the $\varepsilon$ expansion},'' \href{http://dx.doi.org/10.1007/JHEP05(2018)051}{{\em JHEP} {\bfseries 05} (2018) 051}, \href{http://arxiv.org/abs/1707.06165}{{\ttfamily arXiv:1707.06165}}.

\bibitem{Henriksson:2020fqi}
J.~Henriksson, S.~R. Kousvos, and A.~Stergiou, ``{Analytic and numerical bootstrap of CFTs with $O(m)\times O(n)$ global symmetry in 3D},'' \href{http://dx.doi.org/10.21468/SciPostPhys.9.3.035}{{\em SciPost Phys.} {\bfseries 9} no.~3, (2020) 035}, \href{http://arxiv.org/abs/2004.14388}{{\ttfamily arXiv:2004.14388}}.

\bibitem{Kawamura:1988zz}
H.~Kawamura, ``{Renormalization-group analysis of chiral transitions},'' \href{http://dx.doi.org/10.1103/PhysRevB.38.4916}{{\em Phys. Rev. B} {\bfseries 38} (1988) 4916--4928}. [Erratum: Phys.Rev.B 42, 2610--2610 (1990)].

\bibitem{Pannell:2024hbu}
W.~H. Pannell, ``{A note on defect stability in $d = 4 - \varepsilon$},'' \href{http://dx.doi.org/10.1007/JHEP12(2024)187}{{\em JHEP} {\bfseries 12} (2024) 187}, \href{http://arxiv.org/abs/2408.15315}{{\ttfamily arXiv:2408.15315}}.

\bibitem{deSabbata:2024xwn}
E.~de~Sabbata, N.~Drukker, and A.~Stergiou, ``{Transdimensional defects},'' \href{http://dx.doi.org/10.1007/JHEP08(2025)187}{{\em JHEP} {\bfseries 08} (2025) 187}, \href{http://arxiv.org/abs/2411.17809}{{\ttfamily arXiv:2411.17809}}.

\bibitem{Drukker:2022beq}
N.~Drukker and M.~Tr\'epanier, ``{Ironing out the crease},'' \href{http://dx.doi.org/10.1007/JHEP08(2022)193}{{\em JHEP} {\bfseries 08} (2022) 193}, \href{http://arxiv.org/abs/2204.12627}{{\ttfamily arXiv:2204.12627}}.

\bibitem{Antunes:2021qpy}
A.~Antunes, ``{Conformal bootstrap near the edge},'' \href{http://dx.doi.org/10.1007/JHEP10(2021)057}{{\em JHEP} {\bfseries 10} (2021) 057}, \href{http://arxiv.org/abs/2103.03132}{{\ttfamily arXiv:2103.03132}}.

\bibitem{Sun:2024qhv}
X.~Sun and S.-K. Jian, ``{Holographic dual of defect conformal field theory with corner contributions},'' \href{http://dx.doi.org/10.1103/kqk3-lc64}{{\em Phys. Rev. D} {\bfseries 112} no.~4, (2025) L041902}, \href{http://arxiv.org/abs/2407.19003}{{\ttfamily arXiv:2407.19003}}.

\bibitem{Shimamori:2024yms}
S.~Shimamori, ``{Conformal field theory with composite defect},'' \href{http://arxiv.org/abs/2404.08411}{{\ttfamily arXiv:2404.08411}}.

\bibitem{Michel:1983in}
L.~Michel, ``{Renormalization-group fixed points of general $n$-vector models},'' \href{http://dx.doi.org/10.1103/PhysRevB.29.2777}{{\em Phys. Rev. B} {\bfseries 29} (1984) 2777--2783}.

\bibitem{Diatlyk:2024qpr}
O.~Diatlyk, H.~Khanchandani, F.~K. Popov, and Y.~Wang, ``{Effective field theory of conformal boundaries},'' \href{http://dx.doi.org/10.1103/PhysRevLett.133.261601}{{\em Phys. Rev. Lett.} {\bfseries 133} no.~26, (2024) 261601}, \href{http://arxiv.org/abs/2406.01550}{{\ttfamily arXiv:2406.01550}}.

\bibitem{Kravchuk:2024qoh}
P.~Kravchuk, A.~Radcliffe, and R.~Sinha, ``{Effective theory for fusion of conformal defects},'' \href{http://dx.doi.org/10.1088/1751-8121/ae14c5}{{\em J.Phys.A} {\bfseries 58} no.~46, (2025) 465402}, \href{http://arxiv.org/abs/2406.04561}{{\ttfamily arXiv:2406.04561}}.

\bibitem{Cappelli:1989yu}
A.~Cappelli and J.~I. Latorre, ``{Perturbation theory of higher spin conserved currents off criticality},'' \href{http://dx.doi.org/10.1016/0550-3213(90)90463-N}{{\em Nucl. Phys. B} {\bfseries 340} (1990) 659--691}.

\bibitem{Karateev:2024skm}
D.~Karateev and B.~Sahoo, ``{Correlation functions and trace anomalies in weakly relevant flows},'' \href{http://arxiv.org/abs/2408.16825}{{\ttfamily arXiv:2408.16825}}.

\bibitem{Baume:2024poj}
F.~Baume, A.~Miscioscia, and E.~Pomoni, ``{Constraints on RG flows from protected operators},'' \href{http://arxiv.org/abs/2409.09006}{{\ttfamily arXiv:2409.09006}}.

\bibitem{Henriksson:2022rnm}
J.~Henriksson, ``{The critical $O(N)$ CFT: Methods and conformal data},'' \href{http://dx.doi.org/10.1016/j.physrep.2022.12.002}{{\em Phys. Rept.} {\bfseries 1002} (2023) 1--72}, \href{http://arxiv.org/abs/2201.09520}{{\ttfamily arXiv:2201.09520}}.

\bibitem{Petkou:1995vu}
A.~C. Petkou, ``{$C_T$ and $C_J$ up to next-to-leading order in $1/N$ in the conformally invariant $0(N)$ vector model for $2<d<4$},'' \href{http://dx.doi.org/10.1016/0370-2693(95)00936-F}{{\em Phys. Lett. B} {\bfseries 359} (1995) 101--107}, \href{http://arxiv.org/abs/hep-th/9506116}{{\ttfamily arXiv:hep-th/9506116}}.

\bibitem{Fisher:1972zz}
M.~E. Fisher, S.-k. Ma, and B.~G. Nickel, ``{Critical exponents for long-range interactions},'' \href{http://dx.doi.org/10.1103/PhysRevLett.29.917}{{\em Phys. Rev. Lett.} {\bfseries 29} (1972) 917--920}.

\bibitem{Benedetti:2020rrq}
D.~Benedetti, R.~Gurau, S.~Harribey, and K.~Suzuki, ``{Long-range multi-scalar models at three loops},'' \href{http://dx.doi.org/10.1088/1751-8121/abb6ae}{{\em J. Phys. A} {\bfseries 53} no.~44, (2020) 445008}, \href{http://arxiv.org/abs/2007.04603}{{\ttfamily arXiv:2007.04603}}. [Erratum: J.Phys.A 58, 129401 (2025)].

\bibitem{Bianchi:2026sax}
L.~Bianchi, E.~de~Sabbata, and M.~Meineri, ``{Conformal defects and Goldstone bosons in anti-de Sitter space},'' \href{http://arxiv.org/abs/2605.13947}{{\ttfamily arXiv:2605.13947}}.

\bibitem{Qiao:2026ijh}
J.~Qiao, ``{Protected operators in non-local defect CFTs from AdS},'' \href{http://arxiv.org/abs/2605.13975}{{\ttfamily arXiv:2605.13975}}.

\bibitem{Henriksson:2021lwn}
J.~Henriksson and A.~Stergiou, ``{Perturbative and nonperturbative studies of CFTs with $MN$ global symmetry},'' \href{http://dx.doi.org/10.21468/SciPostPhys.11.1.015}{{\em SciPost Phys.} {\bfseries 11} (2021) 015}, \href{http://arxiv.org/abs/2101.08788}{{\ttfamily arXiv:2101.08788}}.

\bibitem{Calabrese:2004nt}
P.~Calabrese, P.~Parruccini, A.~Pelissetto, and E.~Vicari, ``{Critical behavior of $O(2)\times O(N)$ symmetric models},'' \href{http://dx.doi.org/10.1103/PhysRevB.70.174439}{{\em Phys. Rev. B} {\bfseries 70} (2004) 174439}, \href{http://arxiv.org/abs/cond-mat/0405667}{{\ttfamily cond-mat/0405667}}.

\bibitem{Kompaniets:2019xez}
M.~V. Kompaniets, A.~Kudlis, and A.~I. Sokolov, ``{Six-loop epsilon expansion study of three-dimensional $O(n)\times O(m)$ spin models},'' \href{http://dx.doi.org/10.1016/j.nuclphysb.2019.114874}{{\em Nucl. Phys. B} {\bfseries 950} (2020) 114874}, \href{http://arxiv.org/abs/1911.01091}{{\ttfamily arXiv:1911.01091}}.

\bibitem{Pelissetto:2001fi}
A.~Pelissetto, P.~Rossi, and E.~Vicari, ``{Large $n$ critical behavior of $O(n)\times O(m)$ spin models},'' \href{http://dx.doi.org/10.1016/S0550-3213(01)00223-1}{{\em Nucl. Phys. B} {\bfseries 607} (2001) 605--634}, \href{http://arxiv.org/abs/hep-th/0104024}{{\ttfamily hep-th/0104024}}.

\bibitem{Gracey:2002ze}
J.~A. Gracey, ``{Critical exponent $\omega$ at $O(1/N)$ in $O(N) \times O(m)$ spin models},'' \href{http://dx.doi.org/10.1016/S0550-3213(02)00818-0}{{\em Nucl. Phys. B} {\bfseries 644} (2002) 433--450}, \href{http://arxiv.org/abs/hep-th/0209053}{{\ttfamily arXiv:hep-th/0209053}}.

\bibitem{Drukker:2022txy}
N.~Drukker and Z.~Kong, ``{1/3 BPS loops and defect CFTs in ABJM theory},'' \href{http://dx.doi.org/10.1007/JHEP06(2023)137}{{\em JHEP} {\bfseries 06} (2023) 137}, \href{http://arxiv.org/abs/2212.03886}{{\ttfamily arXiv:2212.03886}}.

\end{thebibliography}\endgroup

\end{document}